\documentclass[12pt]{article}
\usepackage{graphicx, color}
\usepackage{natbib} 
\usepackage{booktabs}
\usepackage{subfigure}
\usepackage[margin=1.25in]{geometry}
\usepackage{listings}
\usepackage{amsmath}
\usepackage{amssymb} 
\usepackage{setspace}
\usepackage{rotating}
\usepackage{mathtools}
\usepackage{enumerate}
\usepackage[toc]{appendix}

\usepackage{algpseudocode}
\usepackage{algorithm}

\usepackage{theorem} 

\usepackage{thrmappendix}

\long\def\rptstmtonlyproof#1{
	\pf \cs{#1proof} \endpf
}

\newtype{result}{statement}{Result} 
\newtype{assumption}{assumption}{Assumption}
\newtype{property}{statement}{Property}
\newtype{cor}{statement}{Corollary}
\def\endpf{\hfill $\blacksquare$}

\numberwithin{statement}{section}

\usepackage{array}
\makeatletter
\newcommand*{\compress}{\@minipagetrue}
\makeatother

\newcommand{\relerr}[2]{\delta_{\widehat{\bar{#1}}_{#2}}}
\newcommand{\relcov}[4]{C_{{#1}_{#2}, {#3}_{#4}}}
\newcommand{\relcovbar}[4]{C_{\widehat{\bar{#1}}_{#2}, \widehat{\bar{#3}}_{#4}}}

\newcommand{\prc}{\eta}
\newcommand{\TPR}{\tau}
\newcommand{\Bin}{\text{Binomial}}
\newcommand{\Ber}{\text{Bernoulli}}

\newcommand{\E}{\mathbb{E}}

\newcommand{\sd}{\text{sd}}
\newcommand{\cov}{\text{cov}}
\newcommand{\correl}{\text{cor}}
\newcommand{\cv}{\text{cv}}

\newcommand{\piw}{\zeta}

\newcommand{\cprime}{c^{\prime}}
\newcommand{\wprime}{w^{\prime}}

\newcommand{\yfhhatprime}{\widehat{y}^{\prime}_{F,H}}

\newcommand{\vhabartildehatprime}{\widehat{\bar{\tilde{v}}}^{\prime}_{H,\mathcal{A} \cap F}}
\newcommand{\vhabartilde}{\bar{\tilde{v}}_{H,\mathcal{A} \cap F}}

\newcommand{\nhprimehat}{\widehat{N}_H^{\prime}}
\newcommand{\nhprime}{N_H^{\prime}}
\newcommand{\wdiff}{\epsilon}

\newcommand{\piF}{\pi^{F}}
\newcommand{\piH}{\pi^{H}}
\newcommand{\wF}{w^{F}}

\newcommand{\piprime}{\pi^{\prime}}
\newcommand{\piprimeF}{\pi^{\prime F}}
\newcommand{\piprimeH}{\pi^{\prime H}}
\newcommand{\wprimeF}{w^{\prime F}}
\newcommand{\wprimeH}{w^{\prime H}}
\newcommand{\wdiffF}{\epsilon^{F}}
\newcommand{\wdiffFbar}{\bar{\epsilon}^{F}}
\newcommand{\wdiffH}{\epsilon^{H}}

\newcommand{\leadstounbiased}{\rightarrow}

\usepackage{blkarray} 
\usepackage{multirow} 

\usepackage{comment}

\begin{document}

\title{
Generalizing the Network Scale-Up Method:\\A New Estimator for the Size of Hidden Populations\footnotemark[1]~\footnotemark[4]~\footnotemark[6]}
\author{Dennis M. Feehan\footnotemark[2]~\phantom{ }and Matthew J. Salganik\footnotemark[3]~\footnotemark[4]}
    \date{\today}

\renewcommand{\thefootnote}{\fnsymbol{footnote}}
\thispagestyle{empty}
\footnotetext[1]{The authors thank Alexandre Abdo, Francisco Bastos, Russ
    Bernard, Neilane Bertoni, Dimitri Fazito, Sharad Goel, Wolfgang Hladik,
    Jake Hofman, Mike Hout, Karen Levy, Rob Lyerla, Mary Mahy, Chris McCarty,
    Maeve Mello, Tyler McCormick, Damon Phillips, Justin Rao, Adam Slez, and
    Tian Zheng for helpful discussions.  This research was supported by The
    Joint United Nations Programme on HIV/AIDS (UNAIDS), NSF (CNS-0905086), and
    NIH/NICHD (R01-HD062366, R01-HD075666, \& R24-HD047879).  Some of this
    research was conducted while MJS was an employee Microsoft Research.  The
    opinions expressed here represent the views of the authors and not the
    funding agencies.}
\footnotetext[2]{Department of Demography, University of California, Berkeley, CA, USA}
\footnotetext[3]{Office of Population Research, Princeton University, Princeton, NJ, USA}
\footnotetext[4]{Department of Sociology, Princeton University, Princeton, NJ, USA}
\footnotetext[5]{Upon publication, we will make replication materials available through a public archive.  Many of the methods described in this paper can be implemented using our accompanying open-source R package, which is available on CRAN.}
\footnotetext[6]{Word count: main text 6,520 words; abstract 164 words}
\renewcommand{\thefootnote}{\arabic{footnote}}

\maketitle
\newpage
\begin{abstract}
The network scale-up method enables researchers to estimate the size of hidden
populations, such as drug injectors and sex workers, using sampled social
network data. The basic scale-up estimator offers advantages over other size
estimation techniques, but it depends on problematic modeling assumptions. We
propose a new generalized scale-up estimator that can be used in settings with
non-random social mixing and imperfect awareness about membership in the hidden
population. Further, the new estimator can be used when data are collected via
complex sample designs and from incomplete sampling frames. However, the
generalized scale-up estimator also requires data from two samples: one from
the frame population and one from the hidden population.  In some situations
these data from the hidden population can be collected by adding a small number
of questions to already planned studies. For other situations, we develop
interpretable adjustment factors that can be applied to the basic scale-up
estimator.  We conclude with practical recommendations for the design and
analysis of future studies.
\end{abstract}





\newpage


\section{Introduction}

Many important problems in social science, public health, and public policy
require estimates of the size of hidden populations. For example, in HIV/AIDS
research, estimates of the size of the most at-risk populations---drug
injectors, female sex workers, and men who have sex with men---are critical for
understanding and controlling the spread of the epidemic.  However, researchers
and policy makers are unsatisfied with the ability of current statistical
methods to provide these estimates~\citep{jointunitednationsprogrammeonhiv/aids_guidelines_2010}.  
We address this problem by improving the network scale-up method, a promising
approach to size estimation.  Our results are immediately applicable in many
substantive domains in which size estimation is challenging, and the framework
we develop advances the understanding of sampling in networks more generally.

The core insight behind the network scale-up method is that ordinary people
have embedded within their personal networks information that can be used to
estimate the size of hidden populations, if that information can be properly
collected, aggregated, and adjusted~\citep{bernard_estimating_1989,
bernard_counting_2010a}.  In a typical scale-up survey, randomly sampled 
adults are asked about the number of connections they have to people in a
hidden population (e.g., ``How many people do you know who inject drugs?'') and
a series of similar questions about groups of known size (e.g., ``How many
widowers do you know?''; ``How many doctors do you know?'').
Responses to these questions are called \emph{aggregate relational
data}~\citep{mccormick_surveying_2012}.

To produce size estimates from aggregate relational data, previous researchers
have begun with the \emph{basic scale-up model}, which makes three important
assumptions: (i) social ties are formed completely at random (i.e., random
mixing), (ii) respondents are perfectly aware of the characteristics of their
alters, and (iii) respondents are able to provide accurate answers to survey
questions about their personal networks. From the basic scale-up
model~\citet{killworth_estimation_1998a} derived the \emph{basic scale-up
estimator}.  This estimator, which is widely used in practice, has two main
components. For the first component, the aggregate relational data about the
hidden population are used to estimate the number of connections that
respondents have to the hidden population.  For the second component, the
aggregate relational data about the groups of known size are used to estimate
the number of connections that respondents have in total.  For example, a
researcher might estimate that members of her sample have 5,000 connections to
people who inject drugs and 100,000 connections in total.  The basic scale-up
estimator combines these pieces of information to estimate that 5\% ($5,000 /
100,000$) of the population injects drugs.  This estimate is a sample
proportion, but rather than being taken over the respondents, as would be
typical in survey research, the proportion is taken over the respondents'
alters. Researchers who desire absolute size estimates multiply the alter
sample proportion by the size of the entire population, which is assumed to be
known (or estimated using some other method).

Unfortunately, the three assumptions underlying the basic scale-up model have
all been shown to be problematic. Scale-up researchers call violations of the
random mixing assumption \emph{barrier
effects}~\citep{killworth_investigating_2006, zheng_how_2006,
maltiel_estimating_2015}; they call violations of the perfect awareness
assumption \emph{transmission error}~\citep{shelley_who_1995, shelley_who_2006,
killworth_investigating_2006, salganik_game_2011a, maltiel_estimating_2015}; and
they call violations of the respondent accuracy assumption \emph{recall
error}~\citep{killworth_two_2003, killworth_investigating_2006,
mccormick_adjusting_2007, maltiel_estimating_2015}.  

\begin{table}
\center
\scalebox{0.9}{
\begin{tabular}{lll}
\toprule
Hidden population(s) & Location & Citation\\
\midrule
Mortality in earthquake & Mexico City, Mexico & \citep{bernard_estimating_1989}\\
Rape victims & Mexico City, Mexico & \citep{bernard_estimating_1991}\\
HIV prevalence, rape, and homelessness & U.S. & \citep{killworth_estimation_1998a}\\
Heroin use & 14 U.S. cities & \citep{kadushin_scaleup_2006a}\\
Choking incidents in children & Italy & \citep{snidero_use_2007,
snidero_question_2009, snidero_scaleup_2012}\\
Groups most at-risk for HIV/AIDS & Ukraine & \citep{paniotto_estimating_2009}\\
Heavy drug users & Curitiba, Brazil & \citep{salganik_assessing_2011}\\
Groups most at-risk for HIV/AIDS & Kerman, Iran & \citep{shokoohi_size_2012}\\
Men who have sex with men & Japan & \citep{ezoe_population_2012}\\
Groups most at-risk for HIV/AIDS & Almaty, Kazakhstan & \citep{scutelniciuc_network_2012}\\
Groups most at-risk for HIV/AIDS & Moldova & \citep{scutelniciuc_network_2012a}\\
Groups most at-risk for HIV/AIDS & Thailand & \citep{aramrattan_network_2012}\\
Groups most at-risk for HIV/AIDS & Rwanda & \citep{rwandabiomedicalcenter_estimating_2012}\\
Groups most at-risk for HIV/AIDS & Chongqing, China & \citep{guo_estimating_2013}\\
Groups most at-risk for HIV/AIDS & Tabriz, Iran & \citep{khounigh_size_2014}\\
Men who have sex with men & Taiyuan, China & \citep{jing_estimating_2014}\\
Drug and alcohol users & Kerman, Iran & \citep{sheikhzadeh_comparing_2014}\\
Men who have sex with men & Shanghai, China & \citep{wang_application_2015}\\
\bottomrule
\end{tabular}
}
\caption{Network scale-up studies that have been completed.}
\label{tab:scaleup_studies}
\end{table}

In this paper, we develop a new approach to producing size estimates from
aggregate relational data.  Rather than depending on the basic scale-up model
or its variants (e.g.,~\citet{maltiel_estimating_2015}), we use a simple
identity to derive a series of new estimators.  Our new approach reveals that
one of the two main components of the basic scale-up estimator is problematic.
Therefore, we propose a new estimator---the \emph{generalized scale-up
estimator}---that combines the aggregate relational data traditionally used in
scale-up studies with similar data collected from the hidden population.
Collecting data from the hidden population is a major departure from current
scale-up practice, but we believe that it enables a more principled approach to
estimation.  For researchers who are not able to collect data from the hidden
population, we propose a series of adjustment factors that highlight the
possible biases of the basic scale-up estimator.  Ultimately, researchers must
balance the trade-offs between the basic scale-up estimator, generalized
scale-up estimator, and other size estimation techniques based on the specific
features of their research setting.

In the next section, we derive the generalized scale-up estimator, and we
describe the data collection procedures needed to use it. 
In Section~\ref{sec:relationshiptoscaleup}, we compare the generalized and
basic scale-up approaches analytically and with simulations; our comparison
leads us to propose a decomposition that separates the difference between the
two approaches into three measurable and substantively meaningful factors
(Equation~\ref{eqn:gnsum-text}).  
In Section~\ref{sec:recs-for-practice} we make practical recommendations for the
design and analysis of future scale-up studies, and in
Section~\ref{sec:conclusion}, we conclude with an discussion of next steps. Online
Appendices A - G provide technical details and supporting arguments.

\section{The generalized scale-up estimator}
\label{sec:framework}

\begin{figure}
  \centering
   \subfigure[]{%
     \label{fig:reporting-network-panel1} 
     \includegraphics[width=0.3\textwidth]{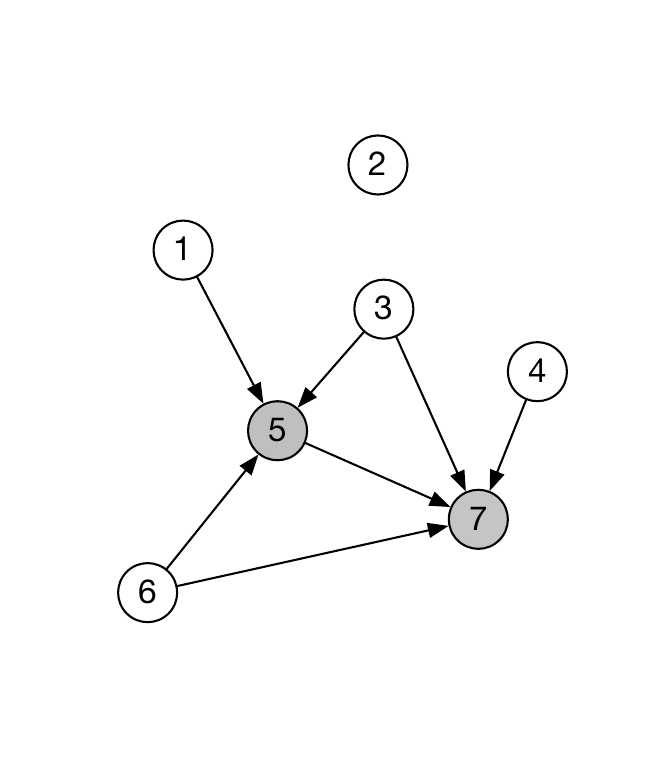}}
  \hspace{0.0in}
   \subfigure[]{%
     \label{fig:reporting-network-panel2} 
     \includegraphics[width=0.3\textwidth]{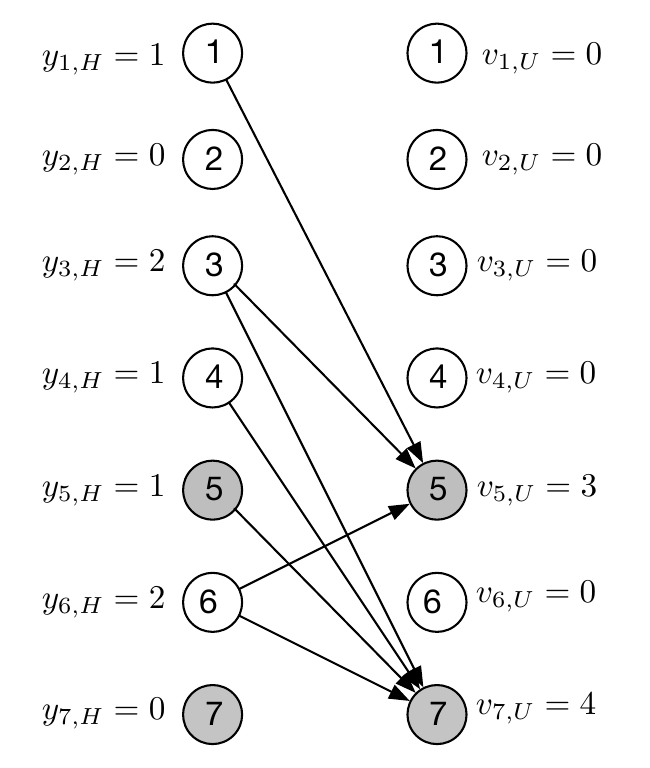}}
  \hspace{0.0in}
   \subfigure[]{%
     \label{fig:reporting-network-panel3} 
     \includegraphics[width=0.3\textwidth]{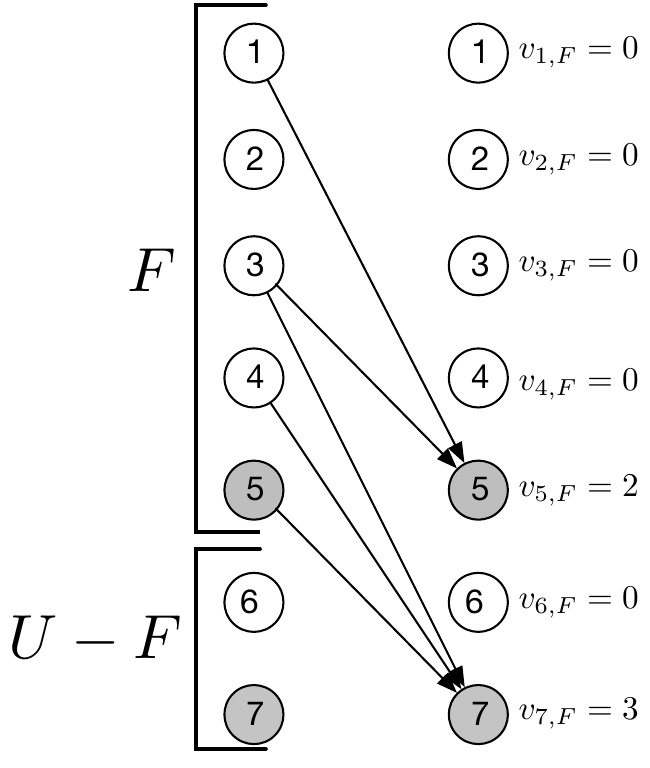}}
     \caption{Illustration of the derivation of the generalized scale-up
         estimator. Panel (a) shows a population of 7 people, 2 of whom are
         drug injectors (shown in grey).  A directed edge $i \rightarrow j$
         indicates that $i$ counts $j$ as a drug injector when answering the
         question ``How many drug injectors do you know?'' Panel (b) shows the
         same population, but redrawn so that each person now appears twice: as
         a source of out-reports, on the left, and as a recipient of
         in-reports, on the right. This arrangement shows that total
         out-reports and total in-reports must be equal. Panel (c) shows the
         same population again, but now some of the people are in the frame
         population $F$ and some are not. In real scale-up studies, we can only
         learn about out-reports from the frame population. 
     }
     \label{fig:reporting-network} 
\end{figure}

The generalized scale-up estimator can be derived from a simple accounting identity that requires no assumptions about the underlying social network structure in the population.  
Figure~\ref{fig:reporting-network} helps illustrate the derivation, which was inspired by earlier research on multiplicity estimation~\citep{sirken_household_1970} and indirect sampling~\citep{lavallee_indirect_2007}.
Consider a population of 7 people, 2 of
whom are drug injectors (Figure~\ref{fig:reporting-network-panel1}).  
In this population, two people are connected by a directed edge $i
\rightarrow j$  if person $i$ would count person $j$ as a drug injector when
answering the question ``How many drug injectors do you know?''  
Whenever $i \rightarrow j$, we say that $i$ makes an \emph{out-report} about $j$
and that $j$ receives an \emph{in-report} from $i$.%
\footnote{%
Throughout the paper, we only consider the case where $i$ never reports $j$ more than once.
}

Each person can be viewed as both a source of out-reports and a recipient of in-reports, and in order to emphasize this point, Figure~\ref{fig:reporting-network-panel2} shows the population with each person represented twice: on the left as a sender of out-reports and on the right as a receiver of in-reports. This visual representation highlights the following identity:
\begin{align}
    \label{eqn:nr-identity}
    \text{total out-reports} = \text{total in-reports}. 
\end{align}
\noindent Despite its simplicity, the identity in Equation~\ref{eqn:nr-identity} turns out to be very useful because it leads directly to the new estimator that we propose.  

In order to derive an estimator from Equation~\ref{eqn:nr-identity}, we must define some notation.  Let $U$ be the entire population, and let $H \subset U$ be the hidden population.
Further, let $y_{i,H}$ be the total number of out-reports from person $i$
(i.e., person $i$'s answer to the question ``How many drug injectors do you
know?'').  For example, Figure~\ref{fig:reporting-network-panel2} shows that
person 5 would report knowing 1 drug injector, so $y_{5,H}=1$. Let $v_{i,U}$ be
the total number of in-reports to $i$ if everyone in $U$ is interviewed; that
is, $v_{i,U}$ is the \emph{visibility} of person $i$ to people in $U$.  For
example, Figure~\ref{fig:reporting-network-panel2} shows person 5 would be
reported as a drug injector by 3 people so $v_{5,U}=3$.  Since total out-reports must
equal total in-reports, it must be the case that 
\begin{align}
    \label{eqn:nr-identity-gnsum}
    y_{U,H} = v_{U,U},
\end{align}
where $y_{U,H} = \sum_{i \in U} y_{i,H}$ and $v_{U,U} = \sum_{i \in U} v_{i,U}$. 
Multiplying both sides of Equation~\ref{eqn:nr-identity-gnsum} by $N_H$, the
number of people in the hidden population, and then rearranging terms, we get
\begin{align}
    \label{eqn:nr-id}
    N_H = \frac{y_{U,H}}{v_{U,U} / N_H}.
\end{align}
Equation~\ref{eqn:nr-id} is an expression for the size of the hidden population
that does not depend on any assumptions about network structure or reporting
accuracy; it is just a different way of expressing the identity that the total
number of out-reports must equal the total number of in-reports.  If we could estimate the two terms on
the right side of Equation~\ref{eqn:nr-id}---one term related to out-reports
($y_{U,H}$) and one term related to in-reports ($v_{U,U} / N_H$)---then we
could estimate $N_H$. 

However, in order to make the identity in Equation~\ref{eqn:nr-id} useful in
practice we need to modify it to account for an important logistical
requirement of survey research.  In real scale-up studies, researchers do not
sample from the entire population $U$, but instead they sample from a subset of
$U$ called the frame population, $F$.  For example, in almost all scale-up
studies the frame population has been adults (but note that our mathematical results hold for any frame population).  In standard survey research,
restricting interviews to a frame population does not cause problems because inference is being made about the frame population.  In other words, when respondents report about themselves it is  clear to which group inferences apply.  
However, with the scale-up
method, respondents report about others, so the group that inferences are being made about 
is not necessarily the same as the group that is being interviewed. 
As we show in Section~\ref{sec:recommendation-only-sf}, failure to consider
this fact requires the introduction of an awkward adjustment factor that had
previously gone unnoticed. 
Here, we avoid this awkward adjustment factor by deriving an identity
explicitly in terms of the frame population. Restricting our attention to
out-reports coming from people in the frame population, it must be the case
that 
\begin{equation}
    \label{eqn:nr-nofp-id}
    N_H = \frac{y_{F,H}}{v_{U,F} / N_H},
\end{equation}
where  $y_{F,H} = \sum_{i \in F} y_{i,H}$ and $v_{U,F} = \sum_{i \in U}
v_{i,F}$.  The only difference between Equation~\ref{eqn:nr-id} and
Equation~\ref{eqn:nr-nofp-id} is that Equation~\ref{eqn:nr-nofp-id} restricts
out-reports and in-reports to come from people in the frame population
(Figure~\ref{fig:reporting-network-panel3}).  The identity in
Equation~\ref{eqn:nr-nofp-id} is extremely general: it does not depend on any
assumptions about the relationship between the entire population $U$, the frame
population $F$, and the hidden population $H$.  For example, it holds if
no members of the hidden population are in the frame population, if there are
barrier effects, and if there are transmission errors. Thus, if we could
estimate the two terms on the right side of Equation~\ref{eqn:nr-nofp-id}---one
term related to out-reports ($y_{F,H}$) and one term related to in-reports
($v_{U, F} / N_H$)---then we could estimate $N_H$ under very
general conditions.  

Unfortunately, despite repeated attempts, we were unable to develop a practical
method for estimating the term related to in-reports ($v_{U, F} / N_H$).
However, if we make an assumption about respondents' reporting behavior, then
we can re-express Equation~\ref{eqn:nr-nofp-id} as an identity made up of
quantities that we can actually estimate.  Specifically, if we assume that the
out-reports from people in the frame population only include people in the
hidden population, then it must be the case that the visibility of everyone not
in the hidden population is 0: $v_{i,F} = 0 \mbox{ for all } i \notin H$.  In
this case, we can re-write Equation~\ref{eqn:nr-nofp-id} as
\begin{align}
\label{eqn:qoi-census-nofp}
N_H = \frac{y_{F,H}}{v_{H,F} / N_H}  = \frac{y_{F,H}}{\bar{v}_{H,F}} \quad \mbox{if } v_{i,F} = 0 \mbox{ for all } i \notin H,
\end{align}
where $\bar{v}_{H,F} = v_{H,F} / N_H$.  

To understand the reporting assumption substantively, consider the two possible types of reporting errors: false
positives and false negatives. Previous scale-up research on transmission
error focused on the problem of false negatives, where a respondent is
connected to a member of the hidden population but does not report this,
possibly because she is not aware that the person she is connected to is in the
hidden population~\citep{bernard_counting_2010a}.  Since hidden populations like
drug injectors are often stigmatized, it is reasonable to suspect that false
negatives will be a serious problem for the scale-up method. Fortunately, Equation~\ref{eqn:qoi-census-nofp} holds even if there are false negative
reporting errors.  However, false positives---which do not seem to have been
considered previously in the scale-up literature---are also possible.  For
example, a respondent who is not connected to any drug injectors might
report that one of her acquaintances is a drug injector. These
false positive reports are not accounted for in the identity in
Equation~\ref{eqn:qoi-census-nofp} and the estimators that we derive
subsequently.  If false positive reports exist, they will introduce a
positive bias into estimates from the generalized scale-up estimator.
Therefore, in Online Appendix~\ref{ap:generalized} we (i) formally define an
interpretable measure of false positive reports, the \emph{precision of
out-reports}; (ii) analytically show the bias in size estimates as a function
of the precisions of out-reports; and (iii) discuss two research designs that
could enable researchers to estimate the precision of out-reports. 

\subsection{Estimating $N_H$ from sampled data}

Equation~\ref{eqn:qoi-census-nofp} relates our quantity of interest, the size
of the hidden population ($N_H$), to two other quantities: the total number of
out-reports from the frame population ($y_{F,H}$) and the average number of
in-reports in the hidden population ($\bar{v}_{H,F}$).  We now show how to
estimate $y_{F,H}$ with a probability sample from the frame population and $\bar{v}_{H,F}$
with a relative probability sample from the hidden population.

The total number of out-reports ($y_{F,H}$) can be estimated from respondents' reported number of connections to the hidden population,
\begin{align}
\label{eqn:yft-ht}
\widehat{y}_{F,H} = \sum_{i \in s_F} \frac{y_{i,H}}{\pi_{i}},
\end{align}
where $s_F$ denotes the sample, $y_{i,H}$ denotes the reported number of
connections between $i$ and $H$, and $\pi_{i}$ is $i$'s probability of
inclusion from a conventional probability sampling design from the frame
population.  Because $\widehat{y}_{F,H}$ is a standard Horvitz-Thompson
estimator, it is consistent and unbiased as long as all members of $F$ have a
positive probability of inclusion under the sampling
design~\citep{sarndal_model_1992}; for a more formal statement, see Result~\ref{res:estimator-yft}.  This estimator depends only on an assumption about the sampling design for the frame population, and in Table~\ref{tab:sampling-robustness-all} we show the sensitivity of our estimator to violations of this assumption.

Estimating the average number of in-reports for the hidden population
($\bar{v}_{H,F}$) is more complicated.  First, it will usually be impossible to
obtain a conventional probability sample from the hidden population.  As we
show below, however, estimating $\bar{v}_{H,F}$ only requires a relative
probability sampling design in which hidden population members have a nonzero
probability of inclusion and respondents' probabilities of inclusion are known
up to a constant of proportionality, $c \pi_i$ (see Online
Appendix~\ref{ap:sampling-designs-from-t} for a more precise definition).  Of
course, even selecting a relative probability sample from a hidden population
can be difficult.

A second problem arises because we do not expect respondents to be able to
easily and accurately answer direct questions about their visibility
($v_{i,F}$).  That is, we do not expect respondents to be able to answer
questions such as ``How many people on the sampling frame would include you
when reporting a count of the number of drug injectors that they know?''
Instead, we propose asking hidden population members a series of questions about
their connections to certain groups and their visibility to those groups.  For
example, each sampled hidden population respondent could be asked ``How many
widowers do you know?'' and then ``How many of these widowers are
aware that you inject drugs?''  This question pattern can be repeated for many
groups (e.g., widowers, doctors, etc.).  We call data with this
structure \emph{enriched aggregate relational data} to emphasize its similarity
to the aggregate relational data that is familiar to scale-up researchers.
An interviewing procedure called the \emph{game of contacts} enables
researchers to collect enriched aggregated relational data, even in realistic
field settings~\citep{salganik_game_2011a, maghsoudi_network_2014}.  

Given a relative probability sampling design and enriched aggregate relational
data, we can now formalize our proposed estimator for $\bar{v}_{H,F}$. Let
$A_1, A_2, \ldots, A_J$, be the set of groups about which we collect enriched
aggregate relational data (e.g., widowers, doctors, etc). Here, to
keep the notation simple, we assume that these groups are all contained in the
frame population, so that $A_j \subset F$ for all $j$; in
Online Appendix~\ref{sec:estimating-vtf} we extend the results to groups that do not
meet this criterion. Let $\mathcal{A}$ be the concatenation of these groups,
which we call the \emph{probe alters}.  For example, if $A_1$ is widowers and $A_2$ is doctors, then the probe alters $\mathcal{A}$ is the
collection of all widowers and all doctors, with doctors who are widowers included
twice.  Also, let $\tilde{v}_{i,A_j}$ be respondent $i$'s report about her
visibility to people in $A_j$ and let $v_{i,A_j}$ be respondents $i$'s actual
visibility to people in $A_j$ (i.e., the number of times that this respondent
would be reported about if everyone in $A_j$ was asked about their connections
to the hidden population). 

The estimator for $\bar{v}_{H,F}$ is:
\begin{align}
    \label{eqn:vtf-inmaintext}
    \widehat{\bar{v}}_{H,F} &= 
    \frac{N_F}{N_{\mathcal{A}}}~\frac{\sum_{i \in s_H} \sum_j \widetilde{v}_{i, A_j}/ (c \pi_i)}
    {\sum_{i \in s_H} 1/(c \pi_i)},
\end{align}
\noindent where $N_\mathcal{A}$ is the number of probe alters,
$c$ is the constant of proportionality from the relative probability
sample, and $s_H$ is a relative probability sample of the hidden population.  
Equation~\ref{eqn:vtf-inmaintext} is a standard weighted sample mean
\citep[Sec. 5.7]{sarndal_model_1992} multiplied by a constant, $N_F /
N_{\mathcal{A}}$. Result~\ref{res:goc-v-estimator} shows that, 
this estimator is consistent and \emph{essentially unbiased}\footnote{%
We use the term ``essentially unbiased'' because Equation~\ref{eqn:vtf-inmaintext} is not, strictly speaking, unbiased; the ratio of two unbiased estimators is not itself unbiased.  However, a large literature confirms that the biases caused by the nonlinear form of ratio estimators are typically insignificant relative to other sources of error in estimate~\citep[e.g.][chap. 5]{sarndal_model_1992}.  Unfortunately, many of the estimators we propose are actually ratios of ratios, sometimes called ``compound ratio estimators'' or ``double ratio estimators.''  In Online Appendix~\ref{ap:ratio} we demonstrate that the bias caused the nonlinear form of our estimators is not a practical cause for concern.
}%
, when three conditions are satisfied: one about the design of the survey, one about reporting behavior and one about sampling from the hidden population. 

The first condition underlying the estimator in Equation~\ref{eqn:vtf-inmaintext} is related to the design of the survey, and we call it the \emph{probe alter condition}.  This condition describes the required relationship between the visibility of the hidden population to the probe alters and the visibility of the hidden population to the frame population:
\begin{align}
\frac{v_{H, \mathcal{A}}}{N_{\mathcal{A}}} &= \frac{v_{H,F}}{N_F},
\label{eq:probe_alter_condition_v_tf}
\end{align}
where $v_{H, \mathcal{A}}$ is the total visibility of the hidden population to
the probe alters, $v_{H, F}$ is the total visibility of the hidden population
to the frame population, $N_{\mathcal{A}}$ is the number of probe alters, and
$N_F$ is the number of people in the frame population.  In words,
Equation~\ref{eq:probe_alter_condition_v_tf} says that the rate at which the
hidden population is visible to the probe alters must be the same as the rate
at which the hidden population is visible to the frame population.  For
example, in a study to estimate the number of drug injectors in a city, drug
treatment counselors would be a poor choice for membership in the probe alters
because drug injectors are probably more visible to drug treatment counselors
than to typical members of the frame population. On the other hand, postal
workers would probably be a reasonable choice for membership in the probe
alters because drug injectors are probably about as visible to postal workers
as they are to typical members of the frame population.  Additional results
about the probe alter condition are presented in the Online Appendixes:
(i) Result~\ref{res:well_constructed_equiv} presents three other algebraically
equivalent formulations of probe alter condition, some of which offer
additional intuition; (ii) Result~\ref{res:well_constructed_test} provides a
method to empirically test the probe alter condition; and (iii) Table~\ref{tab:robustness} quantifies the bias introduced when the
probe alter condition is not satisfied.

The second condition underlying the estimator $\widehat{\bar{v}}_{H,F}$ 
(Equation~\ref{eqn:vtf-inmaintext}) is 
related to reporting behavior, and we call it \emph{accurate aggregate reports about visibility}:
\begin{equation}
\tilde{v}_{H, \mathcal{A}} = v_{H, \mathcal{A}},
\label{eq:reporting_condition_v_tf}
\end{equation}
where $\tilde{v}_{H, \mathcal{A}}$ is the total reported visibility of members
of the hidden population to the probe alters ($\sum_{i \in H} \sum_{j \in J}
\tilde{v}_{i, A_j}$) and $v_{H, \mathcal{A}}$ is the total actual visibility of
members of the hidden population to the probe alters ($\sum_{i \in H} \sum_{j
\in J} v_{i, A_j}$). In words, Equation~\ref{eq:reporting_condition_v_tf} says that hidden
population members must be correct in their reports about their visibility to probe
alters in aggregate, but Equation~\ref{eq:reporting_condition_v_tf} does not require the stronger condition
that each individual report be accurate.  In practice, we expect that there are two main ways
that there might not be accurate aggregate reports about visibility.  First,
hidden population members might not be accurate in their assessments of what
others know about them.  For example, research on the ``illusion of
transparency'' suggests that people tend to over-estimate how much others know
about them~\citep{gilovich_illusion_1998}.  Second, although we propose asking
hidden population members what other people know about them (e.g., ``How many of
these widowers know that you are a drug injector?'') what actually matters for
the estimator is what other people would report about them (e.g., ``How many of
these widowers would include you when reporting a count of the number of drug
injectors that they know?'').  In cases where the hidden population is
extremely stigmatized, some respondents to the scale-up survey might conceal
the fact that they are connected to people whom they know to be in the hidden
population, and if this were to occur, it would lead to a difference between
the information that we collect ($\tilde{v}_{i, \mathcal{A}}$) and the
information that we want ($v_{i, \mathcal{A}}$).  Unfortunately, there is
currently no empirical evidence about the possible magnitude of these two problems
in the context of scale-up studies.  
However, Table~\ref{tab:robustness} quantifies the bias introduced into estimates
if the accurate aggregate reports about visibility condition is not satisfied.

Finally, the third condition underlying the estimator $\widehat{\bar{v}}_{H,F}$ 
(Equation~\ref{eqn:vtf-inmaintext}) is that researchers have a relative probability sample from the hidden population.  Currently the most widely used method for drawing relatively probability samples from hidden populations is respondent-driven sampling~\citep{heckathorn_respondentdriven_1997};
see~\citet{volz_probabilitybased_2008} for a set of conditions under which
respondent-driven sampling leads to a relative probability sample.
Although respondent-driven sampling has been used in hundreds of studies
around the world~\citep{white_strengthening_2015}, there is active debate about
the characteristics of samples that it yields~%
\citep{%
heimer_critical_2005, scott_they_2008,
bengtsson_global_2010, goel_assessing_2010, gile_respondentdriven_2010a,
mccreesh_evaluation_2012, salganik_commentary_2012, mills_respondent_2012,
rudolph_importance_2013, yamanis_empirical_2013, li_central_2015,
gile_network_2015, 
gile_diagnostics_2015,
rohe_network_2015}.  
If other methods for sampling from hidden populations are demonstrated to be
better than respondent-driven sampling (see e.g.,~\citet{kurant_unbiased_2011,
mouw_network_2012, karon_statistical_2012}), then researchers should consider using these methods when using the generalized scale-up estimator.  
Further, researchers can use Table~\ref{tab:sampling-robustness-all} to quantify the bias that results
if the condition requiring a relative probability sample is not satisfied.

To recap, using two different data collection procedures---one with the frame
population and one with the hidden population---we can estimate the two
components of the expression for $N_H$ given in
Equation~\ref{eqn:qoi-census-nofp}.  The estimator for the numerator ($\widehat{y}_{F,H}$) depends on an assumption about the ability to select a probability sample from the frame population (see Result~\ref{res:estimator-yft}), and the estimator for the denominator ($\widehat{\bar{v}}_{H,F}$) depends on assumptions about survey construction, reporting behavior, and the ability to select a relative probability sample from the hidden population (see Result~\ref{res:goc-v-estimator}).

We can combine these component estimators to form the \emph{generalized scale-up estimator}:
\begin{align}
\label{eqn:nofpestimator}
\widehat{N}_H &= \frac{\widehat{y}_{F,H}}{\widehat{\bar{v}}_{H,F}}.
\end{align}
\noindent 
Result~\ref{res:goc-gnsum-new} proves that the generalized scale-up
estimator will be consistent and essentially unbiased if
(i) the estimator for the numerator ($\widehat{y}_{F,H}$) is consistent and essentially 
unbiased; 
(ii) the estimator for the denominator ($\widehat{\bar{v}}_{H,F}$) is consistent
and essentially unbiased;
and (iii) there are no false positive reports.

One attractive feature of the generalized scale-up estimator
(Equation~\ref{eqn:nofpestimator}) is that it is a combination of standard
survey estimators.  This structure enabled us to derive very general
sensitivity results about the impact of violations of assumptions,
either individually or jointly.  We return to the issue of assumptions and sensitivity
analysis when discussing recommendations for practice
(Section~\ref{sec:recs-for-practice}).

\section{Comparison between the generalized and basic scale-up approaches}
\label{sec:relationshiptoscaleup}

In Section~\ref{sec:framework}, we derived the generalized network scale-up
estimator by using an identity relating in-reports and out-reports as the basis
for a design-based estimator. The approach we followed differs from previous
scale-up studies, which have posited the basic scale-up model and derived
estimators conditional on that model.  In this section, we compare these two
different approaches from a design-based perspective.

We begin our comparison by reviewing the basic scale-up model, which was used
in most of the studies listed in Table~\ref{tab:scaleup_studies}. In order to
review this model, we need to define another quantity: we call $d_{i,U}$ person
$i$'s degree, the number of undirected network connections she has to everyone
in $U$.  

The basic scale-up model assumes that each person's connections are formed
independently, that reporting is perfect, and that visibility is
perfect~\citep{killworth_estimation_1998a}.  Together, these three assumptions
lead to the probabilistic model:
\begin{align}
\label{eqn:basic-scaleup-model}
y_{i,A_j} = d_{i,A_j} \sim \Bin\left(d_{i,U}, \frac{N_{A_j}}{N}\right),
\end{align}
\noindent for all $i$ in $U$ and for any group $A_j$.  In words, this model
suggests that the number of connections from a person $i$ to members of a group
$A_j$ is the result of a series of $d_{i,U}$ independent random draws, where
the probability of each edge being connected to $A_j$ is $\frac{N_{A_j}}{N}$.  

The basic scale-up model leads to what we call the basic scale-up estimator:
\begin{align}
\label{eqn:basic-scaleup-estimator-intheory}
\widehat{N}_H = 
\frac{\sum_{i \in s_F} y_{i,H}}{\sum_{i \in s_F} \widehat{d}_{i,U} } \times N,
\end{align}
\noindent where $\widehat{d}_{i,U}$ is the estimated degree of respondent $i$
from the known population method~\citep{killworth_social_1998a}.
\citet{killworth_estimation_1998a} showed that
Equation~\ref{eqn:basic-scaleup-estimator-intheory} is the maximum-likelihood
estimator for $N_H$ under the basic scale-up model, conditional on the additional
assumption that $d_{i,U}$ is known for each $i \in s_F$. 

Given this background, we can now compare the basic and generalized scale-up
approaches by comparing their estimands; that is, we compare the quantities that they
produce in the case of a census with perfectly observed degrees.  The basic
scale-up estimand can be written
\begin{align}
    \label{eqn:basic-scaleup-estimator}
    \widehat{N}_H &= \frac{y_{F,H}}{d_{F,U}} \times N 
    = \frac{y_{F,H}}{\bar{d}_{U,F}},
\end{align}
\noindent where $d_{F,U} = \sum_{i \in F} d_{i,U}$ and 
$\bar{d}_{U,F} = d_{U,F} / N = d_{F,U} / N$.  Further, as shown in Section~\ref{sec:framework}, the generalized scale-up estimand is
\begin{align}
    \label{eqn:gen-scaleup-estimator-census}
    \widehat{N}_H &= \frac{y_{F,H}}{\bar{v}_{H,F}}.
\end{align}

Comparing Equations~\ref{eqn:basic-scaleup-estimator}
and~\ref{eqn:gen-scaleup-estimator-census} reveals that both estimands have the
same numerator but they have different denominators.  The network reporting
identity from Section~\ref{sec:framework} (total out-reports = total
in-reports) shows that the appropriate way to adjust the out-reports is based
on in-reports, as in the generalized scale-up approach.  However, the basic
scale-up approach instead adjusts out-reports with the degree of respondents.
While using the degree of respondents cleverly avoids any data collection from
the hidden population, our results reveal that it will only be correct under a
very specific special case ($\bar{d}_{U,F} = \bar{v}_{H,F}$).  

In order to further clarify the relationship between the basic and generalized
scale-up approaches, we propose a decomposition that separates the difference
between the two estimands into three measurable and substantively meaningful
\emph{adjustment factors}:
\begin{align}
\label{eqn:gnsum-text}
N_H &= 
\underbrace{%
\left( \frac{y_{F,H}}{\bar{d}_{U,F}} \right)
}_{\substack{\text{basic} \\ \text{scale-up}}}
\times 
\underbrace{%
    \underbrace{\frac{1}{\bar{d}_{F,F}/{\bar{d}_{U,F}}}}_{\substack{\text{frame ratio} \\ \phi_F}} \times 
    \underbrace{\frac{1}{\bar{d}_{H,F}/{\bar{d}_{F,F}}}}_{\substack{\text{degree ratio} \\ \delta_F}} \times 
    \underbrace{\frac{1}{\bar{v}_{H,F}/{\bar{d}_{H,F}}}}_{\substack{\text{true positive rate} \\ \tau_F }} }_{\mbox{adjustment factors}}
= 
\underbrace{%
\left( \frac{y_{F,H}}{\bar{v}_{H,F}} \right)
}_{\substack{\text{generalized} \\ \text{scale-up}}}.
\end{align}

The decomposition shows that when the product of the adjustment factors is 1,
the two estimands are both correct.  However, when the product of the
adjustment factors is not 1, then the generalized scale-up estimand is correct
but the basic scale-up estimand is incorrect.   We now describe each of the
three adjustment factors in turn. 

First, we define
the frame ratio, $\phi_F$, to be
\begin{align}
\label{eqn:frameratio}
\phi_F &= 
\frac{\text{avg \# connections from a member of $F$ to the rest of $F$}}
     {\text{avg \# connections from a member of $U$ to $F$}} =
\frac{\bar{d}_{F,F}}{\bar{d}_{U,F}}.
\end{align}
\noindent $\phi_F$ can range from zero to infinity, and in most practical
situations we expect $\phi_F$ will be greater than one.
Result~\ref{res:goc-phi-estimator} shows that we can make consistent and
essentially unbiased estimates of $\phi_F$ from a sample of $F$.\footnote{%
Note that, since $\bar{d}_{U,F} = (N_F/N)~\bar{d}_{F,U}$, an equivalent
expression for the frame ratio is
\begin{align*}
\phi_F &= \frac{\bar{d}_{F,F}}{\bar{d}_{F,U}~(N_F / N)}
= \frac{\bar{d}_{F,F}}{\bar{d}_{F,U}}~\frac{N}{N_F}.
\end{align*}
}

Next, we define the degree ratio $\delta_F$ to be
\begin{align}
\label{eqn:degratio}
\delta_F &= 
\frac{\text{avg \# connections from a member of $H$ to $F$}}
     {\text{avg \# connections from a member of $F$ to the rest of $F$}} =
\frac{\bar{d}_{H,F}}{\bar{d}_{F,F}}.
\end{align}
\noindent $\delta_F$ ranges from zero to infinity, and it is less than one when
the hidden population members have, on average, fewer connections to the frame
population than frame population members.  Result~\ref{res:goc-delta-estimator}
shows that we can to make consistent and essentially unbiased estimates of
$\delta_F$ from samples of $F$ and $H$.

Finally, we define the true positive rate, $\tau_F$, to be
\begin{equation}
    \label{eqn:tau-defn}
\tau_{F} = \frac{\text{\# in-reports to $H$ from $F$}}{\text{\# edges connecting $H$ and $F$}} = \frac{v_{H,F}}{d_{H,F}} = \frac{\bar{v}_{H,F}}{\bar{d}_{H,F}}.
\end{equation}
$\tau_F$ relates network degree to network reports.\footnote{%
    Note that the fact that in-reports must equal out-reports means that
    $\tau_{F}$ can also be defined
    \begin{equation*}
        \tau_{F} = \frac{\text{\# reported edges from $F$ actually connected to
        $H$}}{\text{\# edges connecting $F$ and $H$}} =
        \frac{y_{F,H}^{+}}{d_{F,H}}.
    \end{equation*}
    Here we have written $y_{F,H}^{+}$ to mean the true positive reports among
    the $y_{F,H}$; see Online Appendix~\ref{ap:generalized} for a detailed
    explanation.
}  $\tau_F$ ranges from $0$, if none of the edges are correctly reported, to
$1$ if all of the edges are reported.  Substantively, the more stigmatized the
hidden population, the closer we would expect $\tau_F$ to be to 0.
Result~\ref{res:goc-tpr-estimator} shows that we can to make consistent and
essentially unbiased estimates of $\tau_F$ from a sample of $H$.

Further, the decomposition in Equation~\ref{eqn:gnsum-text} can be used to derive an
expression for the bias in the basic scale-up estimator when we have a census and
degrees are known:
\begin{align}
    \text{bias}(\widehat{N}_H^{\text{basic}}) &\equiv \widehat{N}_H^{\text{basic}} - N_H \\
    &= \widehat{N}_H^{\text{basic}}~
       \left[
           1 - \frac{1}{\phi_F~\delta_F~\tau_F}
       \right].
    \label{eqn:addbias}
\end{align}

The comparison between the basic and generalized scale-up approaches leads to
two main conclusions.  First, the estimand of the basic scale-up approach is
correct only in one particular situation: when the product of the three
adjustment factors is 1.  The estimand of generalized scale-up approach, in
contrast, is correct more generally.  Second, as Equation~\ref{eqn:gnsum-text}
shows, if the adjustment factors are known (or have been estimated), then they
can be used to improve basic scale-up estimates. 


\subsection{Illustrative simulation}

In order to illustrate our comparison between the basic and generalized
scale-up approaches, we conducted a series of simulation studies. The
simulations were not meant to be a realistic model of a scale-up study, but
rather, they were designed to clearly illustrate our analytic results.  More
specifically, the simulation investigated the performance of the estimators as
three important quantities vary: 
(1) the size of the frame population $F$, relative to the size of the entire
population $U$; 
(2) the extent to which people's network connections are not formed completely
at random; and 
(3) the accuracy of reporting, as captured by the true positive rate
$\tau_F$ (see Equation~\ref{eqn:tau-defn}).\footnote{Computer code to perform the simulations was written in
    R~\citep{rcoreteam_r_2014} and used the following packages:
    devtools~\citep{wickham_devtools_2013};
    functional~\citep{danenberg_functional_2013}; 
    ggplot2~\citep{wickham_ggplot2_2009};
    igraph~\citep{csardi_igraph_2006};
    networkreporting~\citep{feehan_networkreporting_2014};
    plyr~\citep{wickham_splitapplycombine_2011};
    sampling~\citep{tille_sampling_2015};
    and stringr~\citep{wickham_stringr_2012}.}

As described in detail in Online Appendix~\ref{sec:sim_overview}, we created
populations of $5,000$ people with different proportions of the population on
the sampling frame ($p_F$).  Next, we connected the people with a social network
created by a stochastic block-model~\citep{white_social_1976,
wasserman_social_1994} in which the randomness of the mixing was
controlled by a parameter $\rho$ such that $\rho=1$ is equivalent to random
mixing (i.e., an Erdos-Reyni random graph) and the mixing becomes more
non-random as $\rho \rightarrow 0$.  Then, for each combination of parameters,
we drew 10 populations, and within each of these populations, we simulated 500
surveys.  For each survey, we drew a probability sample of 500 people from the
frame population, a relative probability sample of 30 people from the hidden
population, and simulated responses with a specific level of reporting accuracy
($\tau_F$).  Finally, we used these reports and the appropriate sampling
weights to calculate the basic and generalized scale-up estimates.

\begin{figure}
\centering
\includegraphics[width=\textwidth]{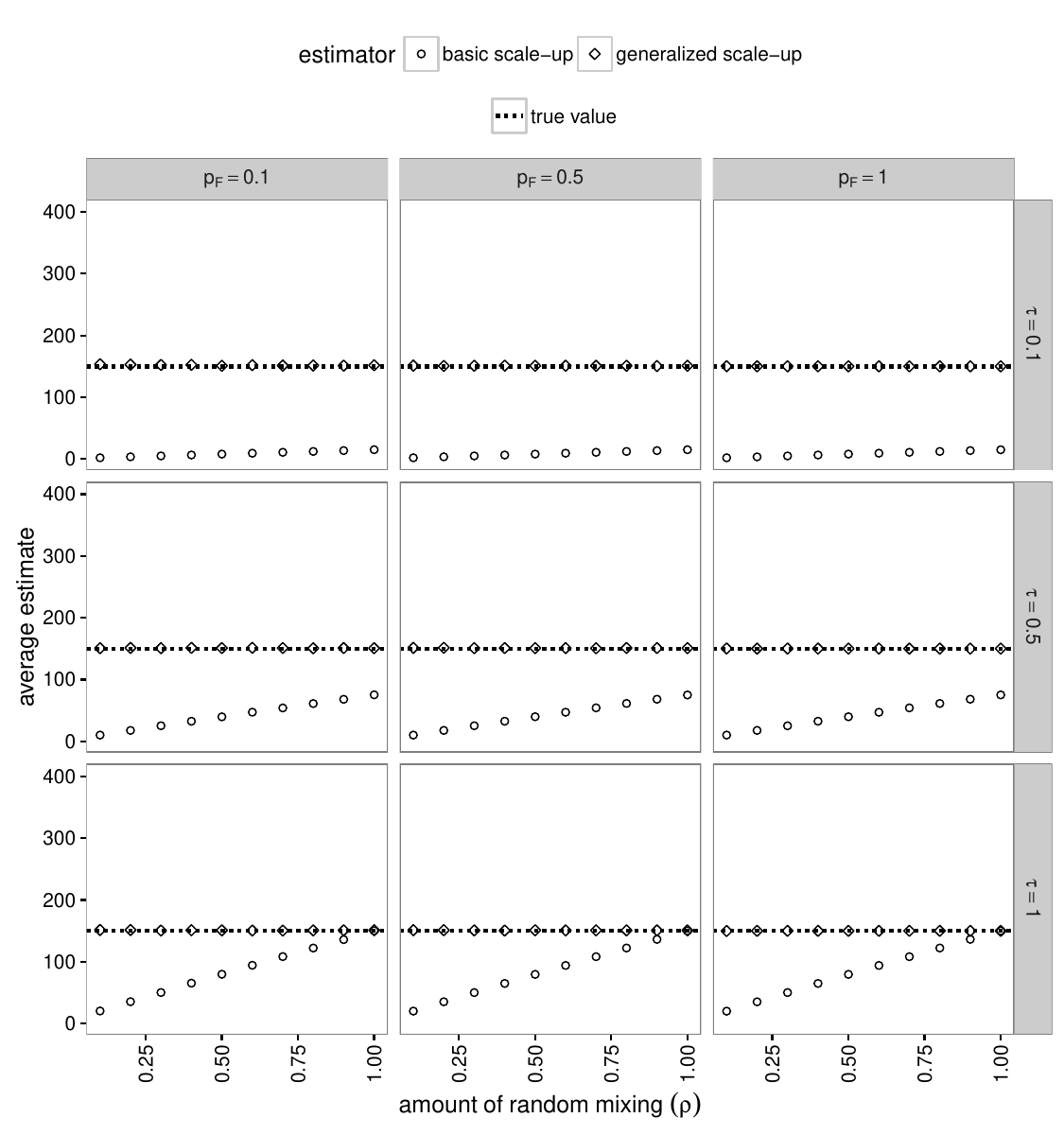}
\caption{
    Estimated size of the hidden population for the generalized and
    basic scale-up estimators. Each panel shows how the two
    estimators change as the amount of random mixing is varied from low
    ($\rho=0.1$; members of the hidden population are relatively unlikely to
    form contacts with nonmembers) 
    to high 
    ($\rho=1$; members of the hidden population form contacts independent of
    other people's hidden population membership).  
    The columns show results for different sizes of the frame population, from
    small (left column, $p_F = 0.1$), to large (right column, $p_F = 1$).
    The rows show results for different levels of reporting accuracy, from a
    small amount of true positives (top row, $\tau_F = 0.1$), to 
    perfect reporting (bottom row, $\tau_F = 1$). 
    For example, looking at the middle of the center
    panel, when $p_F = 0.5$, $\tau_F = 0.5$, and $\rho = 0.5$, we see that the
    average basic scale-up estimate is about 50, while the 
    average generalized scale-up estimate is 150 (the true value).  
    The generalized scale-up estimator is unbiased
    for all parameter combinations, while the basic scale-up estimator
    is only unbiased for certain special cases (e.g., when $\rho=1$, $\tau_F=1$, and $p_F =1$). Full details of the simulation are presented in Online Appendix~\ref{sec:sim_overview}.
}
\label{fig:sim_results}
\end{figure}

\begin{figure}
\centering
\includegraphics[width=\textwidth]{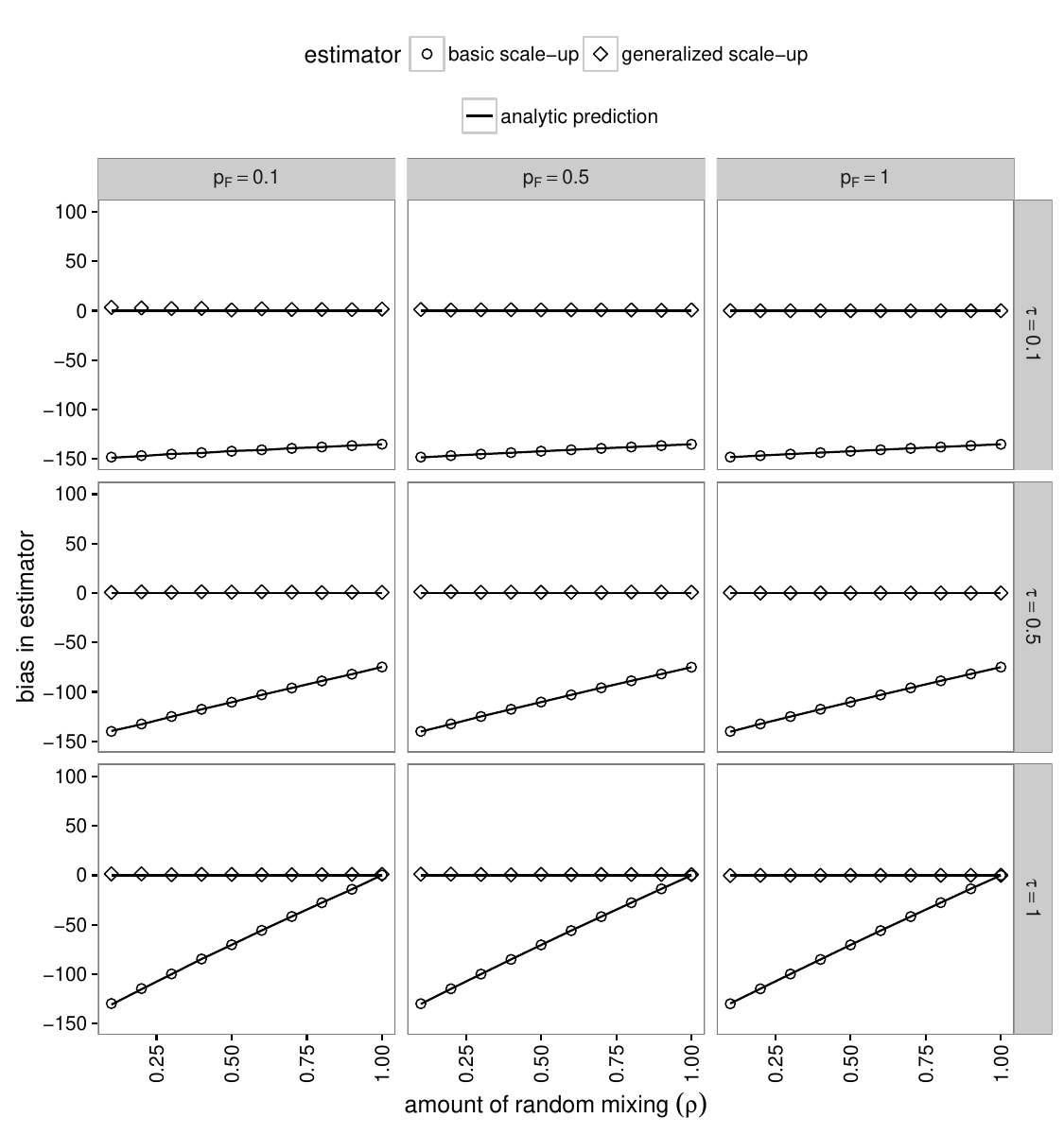}
\caption{
    Bias (open circles and diamonds) and predicted bias (solid lines) in the
    basic scale-up estimates and generalized scale-up estimates for the same
    parameter configurations depicted in Figure~\ref{fig:sim_results}. Our
    analytical results (Equation~\ref{eqn:addbias}) accurately predict the bias
    observed in our simulation study.
}
\label{fig:sim_bias}
\end{figure}

Figure~\ref{fig:sim_results} shows that the simulations support our analytic
results.  First, the simulations show that the generalized scale-up estimator
is unbiased even in the presence of incomplete sampling frames, non-random
mixing, and imperfect reporting.  Second, they show that the basic scale-up
estimator is unbiased in a much smaller set of situations.  More concretely,
the basic scale-up estimator is unbiased in situations where the basic scale-up
model holds---when everyone is in the frame population ($p_F=1$), there is
random mixing ($\rho=1$), and respondents' reports are perfect
($\tau_F=1$).\footnote{In addition to the settings where the basic scale-up
    model holds, the basic scale-up estimator can also be unbiased when its
different biases cancel (e.g., when the product of the adjustment factors is
1).}   Further, Figure~\ref{fig:sim_bias} illustrates that our analytic
approach (Equation~\ref{fig:sim_bias}) can correctly predict the bias of the
basic scale-up estimator.

\section{Recommendations for practice}
\label{sec:recs-for-practice}


\begin{table}[!p]
\centering
\scalebox{0.7}{
\begin{tabular}{>{\compress}p{3cm} >{\compress}p{8cm} >{\compress} p{4cm}> {\compress}p{1.8cm}}
\toprule
Quantity & Conditions required & Condition type & Result \\
\midrule
reported\par
connections to $H$\par
($\widehat{y}_{F,H}$) &
\begin{enumerate}
\item probability sample from $F$
\end{enumerate} 
&
\begin{enumerate}
\item[] sampling
\end{enumerate} 
&
\ref{res:estimator-yft}
\\
\midrule
average personal network size of $F$\par
($\widehat{\bar{d}}_{F,F}$) &
\begin{enumerate}
\item probability sample from $F$
\item groups of known size total is accurate $N_\mathcal{A}$
\item probe alter condition ($\bar{d}_{\mathcal{A},F}=\bar{d}_{F,F}$)
\item accurate reporting condition ($y_{F,\mathcal{A}} = d_{F,\mathcal{A}}$)
\end{enumerate} 
&
\begin{enumerate}
\item[] sampling
\item[] survey construction
\item[] survey construction
\item[] reporting behavior
\end{enumerate} 
&
\ref{res:kpestimator-dff}
\\
\midrule
average\par 
visibility of $H$\par
($\widehat{\bar{v}}_{H,F}$) &
\begin{enumerate}
\item relative probability sample from $H$
\item groups of known size total is accurate $N_{\mathcal{A}_H \cap F}$
\item probe alter condition 
    ($\frac{v_{H, \mathcal{A}\cap F}}{N_{\mathcal{A}\cap F}} = 
    \frac{v_{H,F}}{N_F}$)
\item accurate aggregate reports about visibility \par
    ($\tilde{v}_{H, \mathcal{A}_H \cap F} = v_{H, \mathcal{A}_H \cap F}$)
\end{enumerate} 
&
\begin{enumerate}
\item[] sampling
\item[] survey construction
\item[] survey construction
\item[] reporting behavior
\end{enumerate} 
&
\ref{res:goc-v-estimator}
\\
\midrule
\midrule
generalized\par
scale-up\par
($\widehat{N}_{H} = \frac{\widehat{y}_{F,H}}{\widehat{\bar{v}}_{H,F}}$)&
\begin{enumerate}
    \item conditions needed for $\widehat{y}_{F,H}$
    \item conditions needed for $\widehat{\bar{v}}_{H,F}$ ~\par ~\par ~\par
    \item no false positive reports about connections to $H$ ($\eta_F = 1$)
\end{enumerate} 
&
\begin{enumerate}
    \item[] sampling
    \item[] sampling, \par survey construction, \par reporting behavior
    \item[] reporting behavior
\end{enumerate} 
&
\ref{res:goc-gnsum-new}
\\
\midrule
modified\par
basic scale-up\par
($\widehat{N}_{H} = \frac{\widehat{y}_{F,H}}{\widehat{\bar{d}}_{F,F}}$) &
\begin{enumerate}
    \item conditions needed for $\widehat{y}_{F,H}$
    \item condition needed for $\widehat{\bar{d}}_{F,F}$ \par ~ \par ~ \par
    \item no false positive reports about connections to $H$ ($\eta_F = 1$)
    \item members of $H$ and members of $F$ have same average personal network size ($\delta_F=1$)
    \item no false negative reports about connections to $H$ ($\tau_F=1$)
\end{enumerate} 
&
\begin{enumerate}
    \item[] sampling
    \item[] sampling, \par survey construction, \par reporting behavior
    \item[] reporting behavior \par ~
    \item[] network structure \par ~
    \item[] reporting behavior \par ~
\end{enumerate} 
&
Sections~\ref{sec:framework}-\ref{sec:relationshiptoscaleup}
\\
\bottomrule
\end{tabular}
}
\caption{%
    Summary of the conditions needed for the generalized and modified basic
    network scale-up estimators, and their components, to produce estimates
    that are consistent and essentially unbiased. 
    This table uses the version of the basic scale-up estimator we recommend in
    Section~\ref{sec:recommendation-only-sf}. 
}
\label{tab:overall-assumptions}
\end{table}



\begin{table}[!p]
\centering
\scalebox{.7}{
\begin{tabular}{>{\compress}p{3cm} >{\compress}p{9cm} >{\compress}p{5cm}}
\toprule
Quantity & Conditions required & Adjusted estimand \par for sensitivity analysis\\
\midrule
generalized\par 
scale-up\par
($\widehat{N}_{H} = \frac{\widehat{y}_{F,H}}{\widehat{\bar{v}}_{H,F}}$)&
\begin{enumerate}
    \item probability sample from $F$ \par
          with accurate weights\par
          ($K_{F_2} = 0$ and $\bar{\wdiff}_F = 1$)
    \item relative probability sample from $H$ \par
          with accurate weights\par
          ($K_{H} = 0$)
    \item conditions needed for $\widehat{\bar{v}}_{H,F}$ \par
          ($\widehat{\bar{v}}_{F,H} = \kappa \bar{v}_{F,H}$)
    \item no false positive reports about connections to $H$ \par
          ($\eta_F = 1$)
\end{enumerate} 
&
$\widehat{N}_H \cdot \frac{(1 + K_H)}{\bar{\wdiff}_F (1 + K_{F_2})} \cdot \kappa \cdot \eta_F \leadsto N_H$
\\
\midrule
modified\par
basic scale-up\par
($\widehat{N}_{H} = \frac{\widehat{y}_{F,H}}{\widehat{\bar{d}}_{F,F}}$) &
\begin{enumerate}
    \item probability sample from $F$ \par
          with accurate weights for $y_{F, H}$\par
          ($K_{F_2} = 0$)
    \item probability sample from $F$ \par
          with accurate weights for $y_{F, \mathcal{A}}$\par
          ($K_{F_1} = 0$)
    \item condition needed for $\widehat{\bar{d}}_{F,F}$ \par
          ($\widehat{\bar{d}}_{F,F} = \kappa \bar{d}_{F,F}$)
    \item no false positive reports about connections to $H$ \par
          ($\eta_F = 1$)
    \item members of $H$ and members of $F$ have same average personal network size \par ($\delta_F = 1$)
    \item no false negative reports about connections to $H$ \par ($\tau_F = 1$)
\end{enumerate} 
&
$\widehat{N}_H \cdot \frac{(1 + K_{F_1})}{(1 + K_{F_2})} \cdot \kappa \cdot \frac{\eta_F}{\delta_F \tau_F} \leadsto N_H$
\\
\bottomrule
\end{tabular}
}
\caption{ 
    Analytical expressions that researchers can use to perform sensitivity
    analysis for estimates made using scale-up estimators (see
    Online Appendix~\ref{ap:sensitivity} for more detail).
    $K_{F_1}$, $K_{F_2}$, and $K_{H}$ are indices that reflect how
    imperfect the sampling weights researchers use to make estimates are;
    when these $K$ values are 0, the weights are exactly correct; the
    farther they are from 0, the more imperfect the weights are.
    %
    (NB: we use the symbol $\leadsto$ as a shorthand for `is consistent and
    essentially unbiased for'.) 
}
\label{tab:sensitivity-guide}
\end{table}

The results in Sections~\ref{sec:framework} and~\ref{sec:relationshiptoscaleup}
lead to us to recommend a major departure from current scale-up practice.   In
addition to collecting a sample from the frame population, we recommend that
researchers consider collecting a sample from the hidden population so that
they can use the generalized scale-up estimator.  As our results clarify,
researchers using the scale-up method face a decision: they can collect data
from the hidden population or they can make assumptions about the
adjustment factors described in Section~\ref{sec:relationshiptoscaleup}.  The
appropriate decision depends on a number of factors, but
we think that two are most important: (i) the difficulty of sampling from the
hidden population and (ii) the availability of high-quality estimates of the
adjustment factors in Section~\ref{sec:relationshiptoscaleup}.  For example, if
it is particularly difficult to sample from a specific hidden population and
high-quality estimates of the adjustment factors are already available, then
a basic scale-up estimator may be appropriate.  If however, it is possible to
sample from the hidden population and there are no high-quality estimates of
adjustment factors, then the generalized scale-up estimator may be appropriate.
Many realistic situations will be somewhere between these two extremes, and the
trade-offs must be weighed on a case-by-case basis.

In order to aid researchers deciding between basic and generalized
scale-up approaches, we collected the conditions needed for consistent and
essentially unbiased estimates into
Table~\ref{tab:overall-assumptions}; formal proofs of these results are
presented in Online Appendicies~\ref{ap:sample-from-f} and~\ref{ap:goc}.  We
find it helpful to group these conditions into four broad categories: sampling, survey construction, network
structure, and reporting behavior.  

A review of the conditions in Table~\ref{tab:overall-assumptions} necessarily
raises practical concerns.  In situations where researchers are trying to make
estimates about real hidden populations, they probably won't know how close
they are to meeting these conditions.  Therefore, researchers may wonder how
their estimates will be impacted by violations of these assumptions, both
individually (e.g., ``How would my estimates be impacted if there was a problem
with the survey construction?'') and jointly (e.g., ``How would my estimate be
impacted if there was a problem with my survey construction and reporting
behavior?'').  To address this concern, in Online
Appendix~\ref{ap:sensitivity}, we develop a framework for sensitivity analysis
that shows researchers exactly how estimates will be impacted by violations of
all assumptions, either individually or jointly.  Table~\ref{tab:sensitivity-guide}
summarizes the results of our sensitivity framework.  

Another problem that researchers face in practice is putting appropriate
confidence intervals around estimates.  The procedure currently used
in scale-up studies was proposed in~\citet{killworth_estimation_1998a}, but it
has a number of conceptual problems, and in practice, it produces intervals that are anti-conservative (e.g., the actual coverage rate is lower than the desired coverage rate).  Both of these problems---theoretical and empirical---do not seem to be
widely appreciated in the scale-up literature.  Therefore, instead of the
current procedure, we recommend that researchers use the rescaled bootstrap
procedure~\citep{rao_resampling_1988, rao_recent_1992a, rust_variance_1996},
which has strong theoretical foundations; does not depend on the basic scale-up
model; can handle both simple and complex sample designs; and can be used for
both the basic scale-up estimator and the generalized scale-up estimator.  In
Online Appendix~\ref{ap:variance_estimation} we review the current scale-up
confidence interval procedure and the rescaled bootstrap,
highlighting the conceptual advantages of the rescaled bootstrap.
Further, we show that the rescaled bootstrap produces slightly better
confidence intervals in three real scale-up datasets: one collected via simple
random sampling~\citep{mccarty_comparing_2001a} and two collected via complex
sample designs~\citep{salganik_assessing_2011,
rwandabiomedicalcenter_estimating_2012}.  Finally, and somewhat
disappointingly, our results show that none of the confidence interval procedures work very well in an absolute sense, a finding that highlights
an important problem for future research.

We now provide more specific guidance for researchers based on the data they
decide to collect.  In Section~\ref{sec:recommendation-with-st} we present
recommendations for researchers who collect a sample from both the frame
population, $F$, and the hidden population, $H$; and, in
Section~\ref{sec:recommendation-only-sf}, we present recommendations for
researchers who only select a sample from the frame population.

\subsection{Estimation with samples from $F$ and $H$}
\label{sec:recommendation-with-st}

We recommend that researchers who have samples from $F$ and $H$ use a
generalized scale-up estimator to produce estimates of $N_H$ (see
Section~\ref{sec:framework}):
\begin{align}
\widehat{N}_H &= \frac{\widehat{y}_{F,H}}{\widehat{\bar{v}}_{H,F}}.
\end{align} 

\noindent For researchers using the generalized scale-up estimator we have
three specific recommendations.  Of all the conditions needed for consistent
and essentially unbiased estimation, the ones most under the control of the
researcher are those related to survey construction, and so we recommend that
researchers focus on these during the study design phase.  In particular, we
recommend that the probe alters be designed so that the rate at which the
hidden population is visible to the probe alters is the same as the rate at
which the hidden population is visible to the frame population (see
Result~\ref{res:goc-v-estimator} for a more formal statement, and see
Section~\ref{ap:v-probe-alter-rec} for more advice about choosing probe
alters).  Second, when presenting estimates, we recommend that researchers use
the results in Table~\ref{tab:sensitivity-guide} to also present sensitivity
analyses highlighting how the estimates may be impacted by assumptions that are
particularly problematic in their setting.   Finally, we recommend that
researchers produce confidence intervals around their estimate using the
rescaled bootstrap procedure, keeping in mind that this will likely produce
intervals that are anti-conservative. 

\begin{figure}[t]
  \centering
     \includegraphics[width=0.9\textwidth]{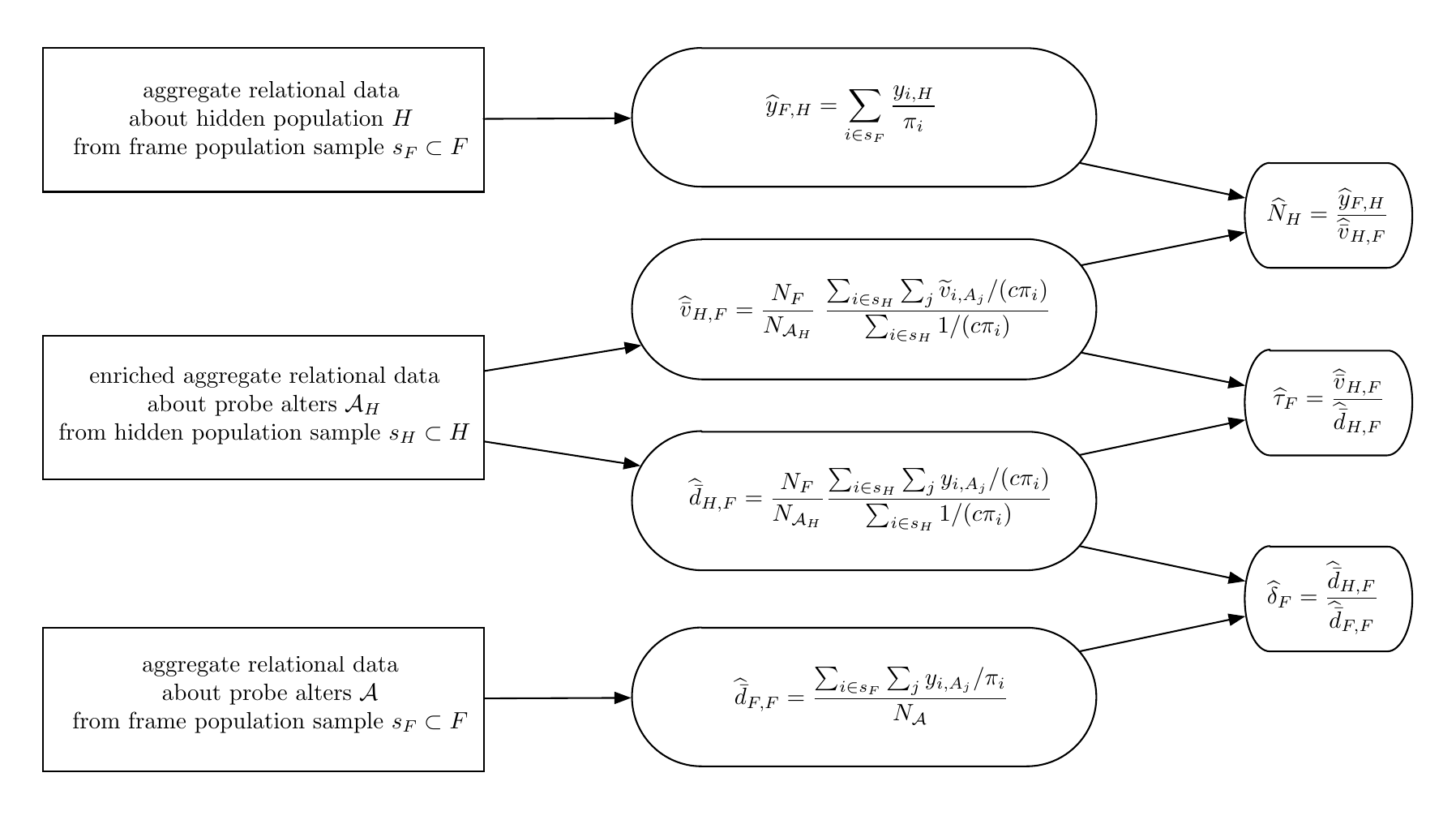}
     \caption{Recommended schematic of inputs and outputs for a study using the
         generalized scale-up estimator.  We recommend that researchers produce
         size estimates using the generalized scale-up estimator, and that
         researchers produce estimates of the adjustment factors $\delta_F$ and
     $\tau_F$ in order to aid other researchers.}
     \label{fig:estimator-data-collection}
\end{figure}

We also have three additional recommendations that will facilitate the
cumulation of knowledge about the scale-up method.  First, although the
generalized scale-up estimator does not require aggregate relational data from
the frame population about groups of known size, we recommend that researchers
collect this data so that the basic and generalized estimators can be compared.
Second, we recommend that researchers publish estimates of $\delta_F$ and
$\tau_F$, although these quantities play no role in the generalized scale-up
estimator (Fig.~\ref{fig:estimator-data-collection}). As a body of evidence
about these adjustment factors accumulates
(e.g.,~\citet{salganik_assessing_2011, maghsoudi_network_2014}), studies that
are not able to collect a sample from the hidden population will have an
empirical foundation for adjusting basic scale-up estimates, either by
borrowing values directly from the literature, or by using published values as
the basis for priors in a Bayesian model. Finally, we recommend that
researchers design their data collections---both from the frame population and
the hidden population---so that size estimates from the generalized scale-up
method can be compared to estimates from other methods (see
e.g.,~\citet{salganik_assessing_2011}).  For example, if respondent-driven
sampling is used to sample from the hidden population, then researchers could
use methods that estimate the size of a hidden
population from recruitment patterns in the respondent-driven sampling
data~\citep{berchenko_modeling_2013, handcock_estimating_2014,
handcock_estimating_2015, crawford_hidden_2015,
wesson_if_2015, johnston_estimating_2015}.

\subsection{Estimation with only a sample from $F$}
\label{sec:recommendation-only-sf}

If researchers cannot collect a sample from the hidden population, we have
three recommendations.  First, we recommend two simple changes to the basic
scale-up estimator that remove the need to adjust for the frame ratio,
$\phi_F$. Recall, that the basic scale-up estimator that has been used in
previous studies is (see Section~\ref{sec:relationshiptoscaleup}):
\begin{align}
\widehat{N}_H &= \frac{\widehat{y}_{F,H}}{\widehat{d}_{F,U}} \times N = \frac{\widehat{y}_{F,H}}{\widehat{d}_{F,U} / N}.
\label{eq:current_basic_scaleup}
\end{align} 
Instead of Equation \ref{eq:current_basic_scaleup}, we suggest a new estimator,
called the modified basic scale-up estimator,
that more directly deals with the fact that researchers sample from the frame
population $F$ (typically adults), and not from the entire population $U$
(adults and children):
\begin{align}
\widehat{N}_H &= \frac{\widehat{y}_{F,H}}{\widehat{d}_{F,F}} \times N_F = \frac{\widehat{y}_{F,H}}{\widehat{d}_{F,F} / N_F} 
\label{eq:future_basic_scaleup}
\end{align} 
There are two differences between the modified basic scale-up estimator
(Equation~\ref{eq:future_basic_scaleup}) and the basic scale-up estimator
(Equation~\ref{eq:current_basic_scaleup}).  First, we recommend that
researchers estimate $\widehat{d}_{F,F}$ (i.e., the total number of connections
between adults and adults) rather than $\widehat{d}_{F,U}$ (i.e., the total
number of connections between adults and everyone).  In order to do so,
researchers should design the probe alters for the frame population
so that they have similar personal networks to the frame
population; in Online Appendix~\ref{ap:kp} we define this requirement formally,
and in Section~\ref{ap:probe-alter-guidance} we provide guidance for
choosing the probe alters.  
Second, we recommend that researchers use $N_F$ rather than $N$.\footnote{In some cases this difference between $N_F$ and $N$ can be substantial.  
For example, if $F$ is adults, then in many developing countries, $N \approx 2
N_F$.}  
These two simple changes remove the need to adjust for the frame ratio
$\phi_F$, and thereby eliminate an assumption about an unmeasured quantity. 
An improved version of the basic scale-up estimator would then be:
\begin{align}
\label{eqn:gnsum-text-recs}
\widehat{N}_H &= 
\underbrace{%
\frac{\widehat{y}_{F,H}}{(\widehat{d}_{F,F} / N_F)} 
}_{\substack{\text{modified basic} \\ \text{scale-up}}}
\times 
\underbrace{%
    \frac{1}{\widehat{\delta}_{F}} \times 
    \frac{1}{\widehat{\tau}_F}}
    _{\substack{\text{adjustment} \\ \text{factors}}}
\end{align}

Our second recommendation is that researchers using the modified basic scale-up
estimator (Equation~\ref{eq:future_basic_scaleup}) perform a sensitivity
analysis using the results in Table~\ref{tab:sensitivity-guide}.  In
particular, we think that researchers should be explicit about the values that
they assume for the adjustment factors $\delta_F$ and $\tau_F$.  Our third
recommendation is that researchers construct confidence intervals using the
rescaled bootstrap procedure, while explicitly accounting for the fact that
there is uncertainty around the assumed adjustment factors and bearing in mind
that this procedure will likely produce intervals that are anti-conservative. 

\section{Conclusion and next steps} 
\label{sec:conclusion}

In this paper, we developed the generalized network scale-up estimator. This
new estimator improves upon earlier scale-up estimators in several ways: it
enables researchers to use the scale-up method in populations with non-random
social mixing and imperfect awareness about membership in the hidden
population, and it accommodates data collection with complex sample designs and
incomplete sampling frames. We also compared the generalized and basic scale-up
estimators, leading us to introduce a framework that makes the design-based 
assumptions of the basic scale-up estimator precise. 
Finally, researchers who use either the basic or generalized scale-up estimator can use
our results to assess the sensitivity of their size estimates to assumptions.

The approach that we followed to derive the generalized scale-up estimator has
three elements, and these elements may prove useful in other problems related
to sampling in networks.  First, we distinguished between the network of
reports and the network of relationships. Second, using the network of reports,
we derived a simple identity that permitted us to develop a design-based
estimator free of any assumptions about the structure of the network of
relationships. Third, we combined data from different types of samples.
Together, these three elements may help other researchers in other situations
derive relatively simple, design-based estimators that are an important
complement to complex, model-based techniques.

Although the generalized scale-up estimator has many attractive features, it
also requires that researchers obtain two different samples, one from the frame
population and one from the hidden population.  In cases where studies of the
hidden population are already planned (e.g., the behavioral surveillance
studies of the groups most at-risk for HIV/AIDS), the necessary information for
the generalized scale-up estimator could be collected at little additional cost
by appending a modest number of questions to existing questionnaires. In cases
where these studies are not already planned, researchers can either collect their own
data from the hidden population, or they can use the modified basic scale-up estimator and borrow estimated adjustment
factors from other published studies.

The generalized scale-up estimator, like all estimators, depends on a number of
assumptions and we think three of them will be most problematic in practice.
First, the estimator depends on the assumption that there are no false positive
reports, which is unlikely to be true in all situations. Although we have
derived an estimator that works even in the presence of false positive reports
(Online Appendix~\ref{ap:generalized}), we were not able to design a practical
data collection procedure that would allow us to estimate one of the terms it
requires.  Second, the generalized scale-up estimator depends on the assumption
that hidden population members have accurate aggregate awareness about
visibility (Equation~\ref{eq:reporting_condition_v_tf}). That is, researchers
have to assume that hidden population respondents can accurately report whether
or not their alters would report them, and we expect this assumption will be
difficult to check in most situations.  Third, the generalized scale-up
estimator depends on having a relative probability sample from the hidden
population.  Unfortunately, we cannot eliminate any of these assumptions, but
we have stated them clearly and we have derived the sensitivity of the
estimates to violations of these assumptions, individually and jointly.

Our results and their limitations highlight several directions for further
work, in terms of both of improved modeling and improved data collection.  We
think the most important direction for future modeling is developing estimators
in a Bayesian framework, and a recent paper by~\citet{maltiel_estimating_2015}
offers some promising steps in this direction.  We see two main advantages of
the Bayesian approach in this setting.  First, a Bayesian approach would allow
researchers to propagate the uncertainty they have about the many assumptions
involved in scale-up estimates, whereas our current approach only captures
uncertainty introduced by sampling.  Further, as more empirical studies produce
estimates of the adjustment factors ($\tau_F$ and $\delta_F$), a Bayesian
framework would permit researchers to borrow values from other studies in a
principled way.  In terms of future directions for data collection, researchers
need practical techniques for estimating the rate of false positive reporting.
These estimates, combined with the estimator in Online
Appendix~\ref{ap:generalized}, would permit the relaxation of one of the most
important remaining assumptions made by all scale-up studies to date.    We
hope that the framework introduced in this paper will provide a basis for these
and other developments.


\clearpage
\newpage

\bibliographystyle{apalike}
\bibliography{gnsum}

\newpage
\part*{Online Appendices}
\appendix

\setcounter{figure}{0} \renewcommand{\thefigure}{A.\arabic{figure}}
\numberwithin{figure}{section}
\setcounter{table}{0} \renewcommand{\thetable}{A.\arabic{table}}
\numberwithin{table}{section}
\setcounter{equation}{0} \renewcommand{\theequation}{A.\arabic{equation}}
\numberwithin{equation}{section}

\clearpage
\pagenumbering{arabic}
\renewcommand*{\thepage}{A\arabic{page}}

\newpage


\section{Estimation in the presence of false positive reports}
\label{ap:generalized}

In the main text, we follow all previous scale-up studies to date in assuming
that there are never any false positive reports.  In this appendix, we
generalize our analysis to the situation where false positive reports are
possible.

In Section~\ref{sec:framework}, Equation~\ref{eqn:qoi-census-nofp}, we
discussed false positive reports in terms of in-reports: we explained that if
there are no false positive reports, then $v_{i,F} = 0$ for all $i \notin H$.
In this appendix, we will re-orient the analysis and focus on how false positives affect
out-reports. Each individual $i$'s out-reports can be divided into two groups:
true positives, which actually connect to the hidden population
($y^{+}_{i,H}$); and false positives, which do not connect to the hidden
population ($y^{-}_{i,H}$). Therefore,
\begin{align}
y_{i,H} &= y^{+}_{i,H} + y^{-}_{i,H}.
\end{align}
\noindent We can also define the aggregate quantities $y^{+}_{F,H} = \sum_{i \in F} y^{+}_{i,H}$ and $y^{-}_{F,H} = \sum_{i \in F} y^{-}_{i,H}$, so that
\begin{align}
y_{F,H} &= y^{+}_{F,H} + y^{-}_{F,H}.
\end{align}
\noindent Because the total number of true-positive out-reports must equal the total number of true-positive in-reports, it is the case that 
\begin{align}
\label{eqn:nr-id-withfp}
y^{+}_{F,H} &= v_{H,F}
\end{align}
where $y^{+}_{F,H}$ is the total number of true-positive out-reports and $v_{H,F}$ is the total number of true positive in-reports.  Dividing both sides by $v_{H,F}$, and then multiplying both sides by $N_H$ produces
\begin{align}
\label{eqn:nt-id-tp}
N_H &= \frac{y^{+}_{F,H}}{\bar{v}_{H,F}}.
\end{align}

In the main text, we introduce a strategy for estimating $\bar{v}_{H,F}$.  If
there was also a strategy for estimating $y^{+}_{F,H}$, then we could use
Equation~\ref{eqn:nt-id-tp} to estimate $N_H$, even if some reports are false
positives. Unfortunately, we cannot typically estimate $y^{+}_{F,H}$ directly
from $F$, since any attempt to do so would learn about $y_{F,H}$ instead.
Therefore, we propose that researchers collect information about $y_{F,H}$ and
then estimate an adjustment factor that relates $y_{F,H}$ to $y^{+}_{F,H}$.
This approach leads us to introduce a new quantity called the \emph{precision
of out-reports},
$\prc_F$: 
\begin{align} 
    \prc_F &= \frac{y^{+}_{F,H}}{y_{F,H}}.
\end{align} 
The precision is useful because it relates the observed out-reports, $y_{F,H}$
to the true positive out-reports, $y^{+}_{F,H}$.  It varies from 0, when none
of the out-reports are true positives, to 1, when the out-reports are perfect.
The precision allows us to derive an identity that relates out-reports to
$N_H$:
\begin{align}
\label{eqn:nt-tp}
N_H &= \frac{\prc_F~y_{F,H}}{\bar{v}_{H,F}}.
\end{align}
\noindent 
Equation~\ref{eqn:nt-tp} then suggests the estimator:
\begin{align}
\label{eqn:nt-hat-tp}
\widehat{N}_H &= \frac{\widehat{\prc}_F~\widehat{y}_{F,H}}{\widehat{\bar{v}}_{H,F}}.
\end{align}
\noindent If we could find a consistent and essentially unbiased estimator for
$\prc_F$, then we could use Equation \ref{eqn:nt-hat-tp} to form a consistent
and essentially unbiased estimator for $N_H$, even in the presence of false
positive reports.

Unfortunately, we are not aware of a practical strategy for estimating
the precision of out-reports.  The most direct approach would be to
interview each alter that a respondent reports as being in the hidden
population.  In other words, if a respondent reports knowing 3 drug injectors,
researchers could try to interview these three people and see if they are
actually drug injectors. \citet{killworth_investigating_2006} attempted a
version of this procedure, which they called an ``alter-chasing'' study, but
they later abandoned it because of the numerous logistical challenges that
arose; see also~\citet{laumann_friends_1969} for a related attempt.  A second
possible approach would be to conduct a census of a networked population where
respondents are asked about themselves and specific people to whom they are
connected.  For example,~\citet{goel_real_2010} collected responses about the
political attitudes of thousands of interconnected people on Facebook,
including respondents' attitudes as well as their beliefs about specific
alters' attitudes.  For a subset of respondents, they could compare $i$'s
belief about $j$'s attitude with $j$'s report of her own attitude in order to
measure the precision.  Unfortunately, we think it would be difficult to
include a sufficiently large number of members of a stigmatized hidden
population in this type of study.  

We expect that the measurement of the precision of out-reports will pose a major challenge for future scale-up research, and we hope that practical solutions to this problem can be found. For the time being, we recommend that researchers show the impact that different values of the precision of out-reports would have on size estimates (Equation~\ref{eqn:nt-hat-tp}).


\section{Estimates with a sample from $F$}
\label{ap:sample-from-f}

In this appendix, we present the full results for all of the estimators that
require a sample from the frame population. First, we describe the general
requirements that our sampling design for $F$ must satisfy
(Section~\ref{ap:sampling-designs-from-f}). Then we describe how to estimate
the total number of out-reports, $y_{F,H}$ (Section~\ref{ap:y-ft}). Next we
turn to some background material on multisets (Section~\ref{ap:multisets}),
which is needed for the following section on the known population method
for estimating network degree (Section~\ref{ap:kp}). Finally, we present an
estimator for the frame ratio, $\phi_F$, which makes use of the known
population method results (Section~\ref{sec:hat_phi_f}).

\subsection{Requirements for sampling designs from F}
\label{ap:sampling-designs-from-f}

We follow \citet{sarndal_model_1992}'s definition of a probability sampling
design, which we repeat here for convenience.  Suppose that we have a set of
possible samples $\{s_1, \ldots, s_j, \dots, s_{\text{max}} \}$, with each $s_j
\subset F$. Furthermore, suppose $p(s_j)$ gives the probability of selection
for each possible sample $s_j$. If we select a sample $s_F$ at random using a
process that will produce each possible sample $s_j$ with probability $p(s_j)$,
and if every element $i \in F$ has a nonzero probability of inclusion $\pi_i >
0$, then we will say that we have selected a \emph{probability sample} and we
call $p(\cdot)$ the \emph{sampling design}.  

\subsection{Estimating the total number of out-reports, $y_{F,H}$}
\label{ap:y-ft}

If we have a probability sample from the frame then estimating the total number
of out-reports is a straightforward application of a standard survey estimator.

\stmt{result}{res:estimator-yft}{%
Suppose we have a sample $s_F$ taken from the frame population using a
probability sampling design with probabilities of inclusion given by $\pi_{i}$
(Sec.~\ref{ap:sampling-designs-from-f}).  Then the estimator given by
\begin{align}
\label{eqn:estimator-y-ft}
\widehat{y}_{F,H} &= \sum_{i \in s_F} y_{i,H} / \pi_i
\end{align}
\noindent is consistent and unbiased for $y_{F,H}$.
}
\stmtproof{res:estimator-yft}{%
This follows from the fact that Equation~\ref{eqn:estimator-y-ft} is a
Horvitz-Thompson estimator \citep[Section 2.8]{sarndal_model_1992}.
}
\rptstmtonlyproof{res:estimator-yft}


\subsection{Reporting about multisets}
\label{ap:multisets} 

Appendix~\ref{ap:kp} and Appendix~\ref{ap:goc} both describe strategies that
involve asking respondents to answer questions about their network alters in
specific groups. In this section, we develop the notation and some basic
properties of responses generated this way; these properties will be then be
used in the subsequent sections.

Suppose we have several groups $A_1, \dots, A_J$ with $A_j \subset U$ for all
$j$, and also a frame population $F$ of potential interviewees.  (Note that we
do not require $A_j \subset F$.) Imagine concatenating all of the people in
populations $A_1, \dots, A_J$ together, repeating each individual once for each
population she is in.  The result, which we call the \emph{probe alters},
$\mathcal{A}$, is a multiset. The size of $\mathcal{A}$ is $N_\mathcal{A} =
\sum_j N_{A_j}$.

Let $y_{i, A_j}$ be the number of members of group $A_j$ that respondent $i$
reports having among the members of her personal network.  We also write
$y_{i, \mathcal{A}} = \sum_j y_{i, A_j}$ for the sum of the responses for
individual $i$ across all of $A_1, \dots, A_J$, and $y_{F, \mathcal{A}} =
\sum_{i \in F} \sum_{j} y_{i, A_j}$ to denote the total number of reports from
$F$ to $\mathcal{A}$. Similarly, we write $d_{i, \mathcal{A}} = \sum_j d_{i,
A_j}$ for the sum of the network connections from individual $i$ to each $A_1,
\ldots, A_J$, and $d_{F, \mathcal{A}} = \sum_{i \in F} \sum_j d_{i, A_j}$ for
the total of the individual $d_{i, \mathcal{A}}$ taken over all $i$.  As
always, we will write averages with respect to the first subscript so that, for
example, $\bar{d}_{\mathcal{A}, F} = d_{\mathcal{A}, F} / N_\mathcal{A}$.

We now derive a property of estimation under multisets that will be useful
later on.  Roughly, this property says that we can estimate the total number of
reports from the entire frame population to the entire multiset of probe alters using
only a sample from the frame population with known probabilities of inclusion
(Section~\ref{ap:sampling-designs-from-f}).  While this property might seem
intuitive, we state it formally for two reasons.  First, by stating it
explicitly, we show that this property is very general: it does not
require any assumptions about the contact pattern between the frame population
and probe alters, nor does it require any assumptions about the probe alters.
Second, it will turn out to be useful in several later proofs, and so we state
it for compactness.

\stmt{property}{prop:htestimator} {%
Suppose we have a sample $s_F$ from $F$ taken using a probability sampling
design with probabilities of inclusion $\pi_i$
(Section~\ref{ap:sampling-designs-from-f}).  Then
\begin{align}
\label{eqn:multiset-ht}
\widehat{y}_{F,\mathcal{A}} &= \sum_{i \in s_F} y_{i,\mathcal{A}}/\pi_i
\end{align}
\noindent is a consistent and unbiased estimator for $y_{F,\mathcal{A}}$.
}
\stmtproof{prop:htestimator}{%
If we define $a_i = \sum_{j} y_{i,A_j}$, the sum of the responses to each $A_j$
for individual $i$, then we can write our estimator as
\begin{align}
\widehat{y}_{F,\mathcal{A}} &= \sum_{i \in s_F} a_i/\pi_i.
\end{align}
\noindent This is a Horvitz-Thompson esimator
\citep[see, e.g.,][chap. 2]{sarndal_model_1992}; it is unbiased and consistent
for the total $\sum_{i \in F} a_i = y_{F,\mathcal{A}}$.
}
\rptstmtonlyproof{prop:htestimator}


\subsection{Network degree and the known population method for estimating
    $\bar{d}_{F,F}$, $\bar{d}_{F,U}$, and $\bar{d}_{U,F}$}
\label{ap:kp}

In order to conduct a scale-up study, we need a definition of the network
that we will ask respondents to tell us about; that is, we need to define what
it will mean for two members of the population to be connected by an edge.  To
date, most scale-up studies have used slight variations of the same definition:
the respondent is told that she should consider someone a member of her network
if she ``knows'' the person, where to know someone means (i) you know her and she
knows you; (ii) you have been in contact in the past 2 years; and, (iii), if
needed, you could get in touch with her \citep{bernard_counting_2010a}.  Of
course, many other definitions are possible, and an investigation of this issue
is a matter for future study.  The only restriction on the tie definition we
impose here is that it be reciprocal; that is, the definition must imply that
if the respondent is connected to someone, then that person is also connected
to the respondent.  

For a particular definition of a network tie an individual $i$'s degree, $d_{i,U}$
may not be very easy to directly
observe, even if the network is conceptually well-defined.  For the basic scale-up estimator, the most commonly used technique
for estimating respondents' network sizes is called the known population
method \citep{killworth_social_1998a,bernard_counting_2010a}.\footnote{%
There are other techniques for estimating personal network size, including the
summation method \citep{mccarty_comparing_2001a,bernard_counting_2010a}, which
could be used in conjunction with many of our results.  We focus on the known
population method here because it is relatively easy to work with from a
statistical perspective, and also because there is some evidence that it works
better in practice
\citep{salganik_assessing_2011,rwandabiomedicalcenter_estimating_2012}
}
The known population method is based on the idea that we can estimate a
respondent's network size by asking how many connections she has to a 
number of different groups whose sizes are known.  The more connections a
respondent reports to these groups, the larger we estimate her
network to be.  Current standard practice is to ask a respondent about her connections to approximately 20 groups of known size in order to estimate her degree~\citep{bernard_counting_2010a}, although the exact number of groups used has no impact on the bias of the estimates as we show in Results~\ref{res:kpestimator-dff} and~\ref{res:kpestimator-duf}.

The known population estimator was originally introduced to estimate
the personal network size of each respondent individually
\citep{killworth_social_1998a}, but in Sections~\ref{sec:relationshiptoscaleup} and
\ref{sec:recommendation-only-sf} we showed that for the scale-up method the quantity of interest is actually the average
number of connections from a member of the frame population $F$ to the rest of
the frame population $F$ ($\bar{d}_{F,F}$),
or the average number of connections from a member of the entire population $U$ to
the frame population $F$ ($\bar{d}_{U,F}$).\footnote{%
    Although we have framed our discussion here in terms of $\bar{d}_{F,F}$,
    the same ideas apply to $\bar{d}_{U,F}$ and $\bar{d}_{F,U}$. 
}  This is fortunate, because it is easier to estimate an average degree over
all respondents than it is to estimate the individual degree for each
respondent.

\subsubsection{Guidance for choosing the probe alters, $\mathcal{A}$}
\label{ap:probe-alter-guidance}

Result~\ref{res:kpestimator-dff}, below, shows that the known population estimator will produce consistent and unbiased estimates of average network degree if (i) $y_{F,\mathcal{A}} = d_{F, \mathcal{A}}$ (\emph{reporting condition}); and (ii) $\bar{d}_{\mathcal{A}, F} = \bar{d}_{F,F}$ (\emph{probe alter condition}).  Stating these conditions precisely enables us to provide guidance about how the groups of known size ($A_1, A_2, \ldots A_J$) should be selected such that the probe alters $\mathcal{A}$ will enable consistent and unbiased estimates.  

First, the reporting condition ($y_{F, \mathcal{A}} = d_{F, \mathcal{A}}$) in Result~\ref{res:kpestimator-dff} shows that researchers should select probe alters such that reporting will be accurate in aggregate.  One way to make the reporting condition more likely to hold is to select groups that are unlikely to suffer from transmission error~\citep{shelley_who_1995, shelley_who_2006, killworth_investigating_2006, salganik_game_2011a, maltiel_estimating_2015}.  Another way to make the reporting condition more likely to hold is to avoid selecting groups that may lead to recall error~\citep{killworth_two_2003, zheng_how_2006, mccormick_adjusting_2007, mccormick_how_2010, maltiel_estimating_2015}.  That is, previous work suggests that respondents seem to under-report the number of connections they have to large groups, although the precise mechanism behind this pattern is unclear~\citep{killworth_two_2003}.  Researchers who have data that may include recall error can consider some of the empirically-calibrated adjustments that have been used in earlier studies~\citep{zheng_how_2006, mccormick_adjusting_2007, mccormick_how_2010, maltiel_estimating_2015}.

Second, the probe alter condition ($\bar{d}_{\mathcal{A}, F} = \bar{d}_{F,F}$) in Result~\ref{res:kpestimator-dff} shows that researchers should select groups to be typical of $F$ in terms of their connections to $F$. In most applied situations, we expect that $F$ will consist of adults, so that researchers should choose groups of known size that are composed of adults, or that are typical of adults in terms of their connections to adults.  Further, when trying to choose groups that satisfy the probe alter condition, it is useful to understand how connections from the individual known populations to the frame ($\bar{d}_{A_1, F}, \dots, \bar{d}_{A_J, F}$) aggregate up into connections from the probe alters to the frame ($\bar{d}_{\mathcal{A},F}$).  Basic algebraic manipulation shows that the probe alter condition can be written as:
\begin{align}
\label{eqn:kpweightedavg}
\frac{\sum_j \bar{d}_{A_j,F}~N_{A_j}}{\sum_j N_{A_j}}
&= \bar{d}_{F,F}.
\end{align} 
Equation \ref{eqn:kpweightedavg} reveals that the probe alter condition requires that $\bar{d}_{F,F}$ is equal to a weighted average of the average number of connections between each individual known population $A_j$ and the frame population $F$ ($\bar{d}_{A_j, F}$).  The weights are given by the size of each known population, $N_{A_j}$.  The simplest way that this could be satisfied is if $\bar{d}_{A_j, F} = \bar{d}_{F,F}$ for every known population $A_j$. If this is not true, then the probe alter condition can still hold as long as groups for  which $\bar{d}_{A_j, F}$ is too high are offset by other groups for which $\bar{d}_{A_{j^\prime},F}$  is too low.  

In practice it may be difficult to determine if the reporting condition and probe alter condition will be satisfied.  Therefore, we recommend that researchers assess the sensitivity of their size estimates using the procedures described in Online Appendix~\ref{ap:sensitivity}.  Further, we note that in many realistic situations, $N_{A_j}$ might not be known exactly.  Fortunately, researchers only need to know  $\sum_j N_{A_j}$, and they can assess the sensitivity of their estimates to errors in the size of known populations 	using the procedures described in Online Appendix~\ref{ap:sensitivity}.

\subsubsection{The known population estimators}

Given that background about selecting the probe alters, we present the formal results for the known population estimators for $\bar{d}_{F,F}$, $\bar{d}_{U,F}$, and $\bar{d}_{F,U}$.

\stmt{result}{res:kpestimator-dff} {%
Suppose we have a sample $s_F$ taken from the frame population using a
probability sampling design with probabilities of inclusion given by $\pi_{i}$
(see Section~\ref{ap:sampling-designs-from-f}).  Suppose also that we have a
multiset of known populations, $\mathcal{A}$. Then the known population estimator
given by
\begin{align}
\label{eqn:kpestpf-dff}
\widehat{\bar{d}}_{F,F} &= \frac{\sum_{i \in s_F} \sum_j
y_{i,A_j}/\pi_{i}}{N_{\mathcal{A}}}&&\mbox{ \phantom{(estimator)} }
\end{align}
\noindent is consistent and unbiased for $\bar{d}_{F,F}$ if
\begin{align}
\label{eqn:kpcondition-dff-reporting}
y_{F, \mathcal{A}} &= d_{F, \mathcal{A}}, &&\mbox{ (reporting condition) }
\end{align}
\noindent and if
\begin{align}
\label{eqn:kpcondition-dff}
\bar{d}_{\mathcal{A},F} &= \bar{d}_{F,F}.&&\mbox{ (probe alter condition) }
\end{align}
}
\stmtproof{res:kpestimator-dff}{
By Property~\ref{prop:htestimator}, we know that our estimator is unbiased and consistent for $y_{F,{\mathcal{A}}}/N_{\mathcal{A}}$.  
By the reporting condition in Equation~\ref{eqn:kpcondition-dff-reporting},
this means it is unbiased and consistent for
$d_{F,{\mathcal{A}}}/N_{\mathcal{A}}$. Then, by the probe alter condition in
Equation~\ref{eqn:kpcondition-dff}, it is also unbiased and consistent for
$\bar{d}_{F,F}$.
}
\rptstmtonlyproof{res:kpestimator-dff}

\stmt{result}{res:kpestimator-duf} {
Suppose we have a sample $s_F$ taken from the frame population using a
probability sampling design with probabilities of inclusion given by $\pi_{i}$
(see Section~\ref{ap:sampling-designs-from-f}).  Suppose also that we have a
multiset of known populations, $\mathcal{A}$.  Then the known population estimator
given by
\begin{align}
\label{eqn:kpestpf-uf}
\widehat{\bar{d}}_{U,F} &= \frac{\sum_{i \in s_F} \sum_j y_{i,A_j}/\pi_{i}}{N_{\mathcal{A}}}&&\mbox{ \phantom{(estimator) }}
\end{align}
\noindent is consistent and unbiased for $\bar{d}_{U,F}$ if
\begin{align}
\label{eqn:kpcondition-duf-reporting}
y_{F, \mathcal{A}} &= d_{F, \mathcal{A}},&&\mbox{ (reporting condition) }
\end{align}
\noindent and if
\begin{align}
\label{eqn:kpcondition-duf}
\bar{d}_{\mathcal{A},F} &= \bar{d}_{U,F}.&&\mbox{ (probe alter condition) }
\end{align}
}
\stmtproof{res:kpestimator-duf}{
By Property~\ref{prop:htestimator}, we know that our estimator is unbiased and consistent for $y_{F,{\mathcal{A}}}/N_{\mathcal{A}}$.  
By the reporting condition in Equation~\ref{eqn:kpcondition-duf-reporting},
this means it is unbiased and consistent for $d_{F,{\mathcal{A}}} /
N_{\mathcal{A}}$. Then, by the probe alter condition in
Equation~\ref{eqn:kpcondition-duf}, it is also unbiased and consistent for
$\bar{d}_{U,F}$.
}
\rptstmtonlyproof{res:kpestimator-duf}

Since $\bar{d}_{F,U} = \frac{N}{N_F} \bar{d}_{U,F}$, as a direct consequence of
Result~\ref{res:kpestimator-duf} we have the following corollary.

\stmt{cor}{res:kpestimator-dfu} {%
If the conditions described in Result~\ref{res:kpestimator-duf} hold,
\begin{align}
\widehat{\bar{d}}_{F,U} &= 
\widehat{\bar{d}}_{U,F}~\frac{N}{N_F}
\end{align}
\noindent is consistent and unbiased for $\bar{d}_{F,U}$.
}


\subsection{Estimating the frame ratio, $\phi_F$}
\label{sec:hat_phi_f}

Given our estimator of $\bar{d}_{F,F}$ (Result~\ref{res:kpestimator-dff}) and
our estimator of $\bar{d}_{U,F}$ (Result~\ref{res:kpestimator-duf}), we can
estimate the frame ratio, $\phi_F$.

\stmt{result}{res:goc-phi-estimator} {%
The estimator
\begin{align}
\label{eqn:goc-phi-estpf}
\widehat{\phi}_F &= \frac{\widehat{\bar{d}}_{F,F}}{\widehat{\bar{d}}_{U,F}}
\end{align}
\noindent is consistent and essentially unbiased for $\phi_F$ if
$\widehat{\bar{d}}_{F,F}$ is consistent and essentially unbiased for
$\bar{d}_{F,F}$ and $\widehat{\bar{d}}_{U,F}$ is consistent and essentially
unbiased for $\bar{d}_{U,F}$.
}
\stmtproof{res:goc-phi-estimator}{%
This follows from the properties of a ratio estimator \cite[chap. 5]{sarndal_model_1992}.}
\rptstmtonlyproof{res:goc-phi-estimator}

More concretely, combining the estimator for $\bar{d}_{F,F}$
(Result~\ref{res:kpestimator-dff}) and the estimator for $\bar{d}_{U,F}$
(Result~\ref{res:kpestimator-duf}), and assuming that we have known
populations $\mathcal{A}_{F_1}$ for $\bar{d}_{F,F}$, and $\mathcal{A}_{F_2}$
for $\bar{d}_{U,F}$, we obtain
\begin{align}
\widehat{\phi}_F &= 
\frac{N_{\mathcal{A}_{F_2}}}
     {N_{\mathcal{A}_{F_1}}}~
\frac{\sum_{i \in s_F} \sum_{A_j \in \mathcal{A}_{F_1}} y_{i, A_j}/\pi_i}
     {\sum_{i \in s_F} \sum_{A_k \in \mathcal{A}_{F_2}} y_{i, A_k} /\pi_i}.
\end{align}
\noindent In our discussion of $\widehat{\bar{d}}_{F,F}$
(Result~\ref{res:kpestimator-dff}) and $\widehat{\bar{d}}_{U,F}$
(Result~\ref{res:kpestimator-duf}), we concluded that we want the known
populations $\mathcal{A}_{F_1}$ used for $\widehat{\bar{d}}_{F,F}$ to be
typical of members of $F$ in their connections to $F$.  An analogous argument
shows that we want the known populations $\mathcal{A}_{F_2}$ used for
$\widehat{\bar{d}}_{U,F}$ to be typical of members of $U$ in their connections
to $F$.  In general, we expect that it will not be appealing to assume that $F$
and $U$ are similar to each other in terms of their connections to $F$ meaning
that, unfortunately, it will not make sense to use the same set of known
populations for $\widehat{\bar{d}}_{F,F}$ and $\widehat{\bar{d}}_{U,F}$. If
researchers wish to estimate $\phi_F$ directly, one approach would be to choose
$\mathcal{A}_{F_2}$ to be typical of $U$ in such a way that some of the
individual known populations are more typical of $F$, while others more typical of
$U-F$. The multiset formed from only the ones that are more typical of $F$
could then be our choice for $\mathcal{A}_{F_1}$.  In this case, researchers would also
want $\frac{N_{\mathcal{A}_{F_1}}}{N_{\mathcal{A}_{F_2}}} \approx
\frac{N_F}{N}$. This complication is one of the reasons we recommend in
Section~\ref{sec:recs-for-practice} that future scale-up studies estimate
$\bar{d}_{F,F}$ directly, thus avoiding the need to estimate $\phi_F$ entirely.


\section{Estimates with samples from $F$ and $H$}
\label{ap:goc}

In this appendix, we present the full results for all of the estimators that
require a sample from the hidden population.  First, we define the general
requirements that our sampling design for $H$ must satisfy
(Section~\ref{ap:sampling-designs-from-t}). Then we describe a flexible data
collection procedure called the game of contacts
(Section~\ref{ap:data_collection}).  Next, we introduce some background
material on estimation using questions about multisets
(Section~\ref{ap:weighted_mean}) and present an estimator for $\bar{v}_{H,F}$,
the average number of in-reports among the members of the hidden population
(Section~\ref{sec:estimating-vtf}). Then, we present estimators for the two
adjustment factors introduced in Section~\ref{sec:relationshiptoscaleup}: the
degree ratio, $\delta_F$, and the true positive rate, $\tau_F$
(Section~\ref{ap:term-by-term}).  Finally, we present formal results for four
different estimators for $N_H$ (Section~\ref{sec:hat_N_T}). 

\subsection{Requirements for sampling designs from H}
\label{ap:sampling-designs-from-t}  

For the results that involve a sample from the hidden population $s_H$, we do
not need a probability sample (Appendix~\ref{ap:sample-from-f}); instead, we
need a weaker type of design. We require that every element $i \in H$ have
a nonzero probability of selection $\pi_i > 0$, and that we can
determine the probability of selection up to a constant factor $c$; that is, we
only need to know $c \pi_i$.  We are not aware of any existing name for this
situation, so we will call it a \emph{relative probability sample}.
Because of the challenges involved in sampling hard-to-reach populations, the
two most likely sampling designs for $s_H$ will probably be time-location
sampling~\citep{karon_statistical_2012} and respondent-driven
sampling~\citep{heckathorn_respondentdriven_1997}.  A relative probability
sample allows us to use weighted sample means to estimate averages, but not
totals. See \citet[Section 5.7]{sarndal_model_1992} for more details on
weighted sample means, also sometimes called H\'{a}jek estimators, which is what we
use to estimate averages from a sample of hidden population members.

\subsection{Data collection}
\label{ap:data_collection}

In order to make estimates about the hidden population's visibility to the
frame population, researchers will need to collect what we call \emph{enriched
aggregate relational data} from each respondent, and a
procedure called the \emph{game of contacts} has produced promising
results from a study of heavy drug users in Brazil~\citep{salganik_game_2011a}.  In the main text, we assumed that the groups in the probe alters $A_1, \dots, A_J$ were all contained in the frame population ($A_j \subset F$ for all $j$).  However, the estimators in this Online Appendix are more general because they allow for the possibility that some of the groups $A_1, \dots A_J$ may not be contained entirely in $F$.  For example, if the frame population is adults, then this flexibility enables researchers to use groups based on names, such as Michael, even though not all people named Michael are adults. 

In order to allow for this flexibility, we need to introduce some new notation: let $A_1 \cap F, A_2 \cap F, \ldots, A_J \cap F$ be the
intersection of these groups and the frame population, and let $\mathcal{A}
\cap F$ be the concatenation of these intersected groups.  For example, if the
frame population is adults, $A_1$ is people named Michael, and $A_2$ is
doctors, then $A_1 \cap F$ is adults named Michael, $A_2 \cap F$ is adult
doctors, and $\mathcal{A} \cap F$ is the collection of all adult Michaels and
all adult doctors, with adult doctors named Michael included twice.  (In the
special case discussed in the main text, $A_1 \cap F, \dots A_J \cap F =
A_1, \dots, A_J$.)

\begin{figure}
\centering
\includegraphics[width=0.6\textwidth]{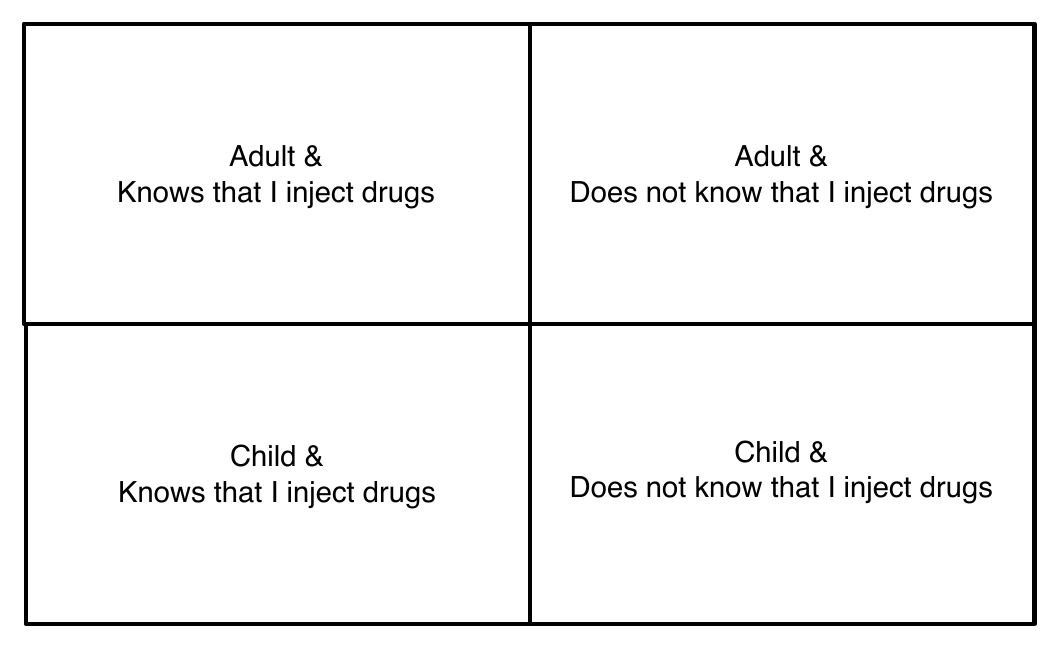}
\caption{Example of a game board that could be used in the game of contacts
    interviewing procedure if the hidden population was people who inject drugs
    and the frame was made up of adults.  This board is a variation of the
    board used in~\citet{salganik_game_2011a}.}
\label{fig:board}
\end{figure}

The data collection begins with a relative probability sample (Section~\ref{ap:sampling-designs-from-t})
from the hidden population. For a set of groups, $A_1, A_2, \ldots A_J$, each respondent in the hidden
population is asked, ``How many people do you know in group $A_j$?''  We call
the response $y_{i, A_j}$.  Next for each of the $y_{i, A_j}$ alters, the
respondent picks up a token and places it on a game board like the one in
Figure~\ref{fig:board}.  From the location of the tokens on the board, the
researcher can record whether each alter is in the frame population (or not)
and whether the alter is aware that the respondent is in the hidden population
(or not) (Table~\ref{tab:goc_adjustment}).  This process is then repeated until
the respondent has been asked about all groups.  

If all of the probe alters are in the frame population, then the process is much easier for respondents and the game board can be modified to collect alternative information.  If all of the probe alters are not in the frame population, then it is important for the researcher to define the frame population as clearly as possible.  If the respondents are not able to correctly indicate whether the alters are in the frame population or not, it could lead to biased estimates of $\bar{v}_{H,F}$.  For more on the operational implementation of this procedure, see~\citet{salganik_game_2011a}.

\begin{table}
\centering
\begin{tabular}{lccc}
\toprule
 & aware & not aware & total \\
\midrule
frame population & $\widetilde{v}_{i, A_j \cap F}$ & $\widetilde{h}_{i, A_j \cap F}$  & $y_{i, A_j \cap F}$\\
not frame population & $\widetilde{v}_{i, A_j \cap (U - F)}$ & $\widetilde{h}_{i, A_j \cap (U - F)}$ & $y_{i, A_j \cap (U - F)}$ \\
total & $\widetilde{v}_{i, A_j}$ & $\widetilde{h}_{i, A_j}$ & $y_{i, A_j}$\\
\bottomrule
\end{tabular}
\label{tab:goc_adjustment}
\caption{Responses collected during the game of contacts for each respondent
    $i$ and each group $A_j$.  We use $\quad \widetilde{ } \quad$ to indicate
    reported values.  For example, $\widetilde{v}_{i, A_j}$ is the respondent's
    reported visibility to people in $A_j$ and $v_{i, A_j}$ is respondent's
    actual visiblility to people in $A_j$.  Also, using this notational
    convention, it is the case that $y_{i, A_j} = \widetilde{d}_{i, A_j}$, but
    we have written $y_{i, A_j}$ in order to be consistent with the rest of the
paper.}
\end{table}


\subsection{Estimation using aggregated relational data from the hidden population}
\label{ap:weighted_mean}

In this section, we follow Section~\ref{ap:multisets} and present another
useful property about estimates made using aggregate relational data from the
hidden population.  Roughly, this property says that we can estimate the
average number of reports from the entire hidden population to the probe alters
using only a relative probability sample from the hidden population
(Section~\ref{ap:sampling-designs-from-t}). Similar to
Property~\ref{prop:htestimator}, the result we present below does not require
any assumptions about the contact pattern between the hidden population and the
probe alters, nor about the probe alters themselves.

\stmt{property}{prop:hajekest} {
Suppose we have a sample $s_H$ from $H$ taken using a relative probability
design, allowing us to compute the relative probabilities of inclusion $c
\pi_i$ for all sampled elements (Sec.~\ref{ap:sampling-designs-from-t}). Then
\begin{align}
\label{eqn:multiset-hajek}
\widehat{\bar{y}}_{H,\mathcal{A}} &= 
\frac{\sum_{i \in s_H} y_{i,\mathcal{A}}/(c\pi_i)}
     {\sum_{i \in s_H} 1/(c\pi_i)}
\end{align}
\noindent is a consistent and essentially unbiased estimator for
$\bar{y}_{H,\mathcal{A}} = y_{H,\mathcal{A}}/N_H$.
}
\stmtproof{prop:hajekest}{
Note that the $c$ in the relative probabilities of inclusion $c \pi_i$ cancel, so that
\begin{align}
\widehat{\bar{y}}_{H,\mathcal{A}} &= 
\frac{\sum_{i \in s_H} y_{i,\mathcal{A}}/(\pi_i)}
     {\sum_{i \in s_H} 1/(\pi_i)}.
\end{align}
If we define $a_i = \sum_{j} y_{i,A_j}$, the sum of the responses to each $A_j$
for individual $i$, then we can write our estimator as
\begin{align}
\widehat{\bar{y}}_{H,\mathcal{A}} &= 
\frac{\sum_{i \in s_H} a_i/\pi_i}{\sum_{i \in s_H} 1/\pi_i}.
\end{align}
\noindent Now we have a standard weighted mean estimator \citep[e.g.][chap.
5]{sarndal_model_1992}; it is consistent and essentially unbiased for the
average $\frac{1}{N_H} \sum_{i \in H} a_i = y_{H,\mathcal{A}}/N_H$.
}
\rptstmtonlyproof{prop:hajekest}

\subsection{Estimating the average visibility, $\bar{v}_{H,F}$}
\label{sec:estimating-vtf}

Given the data collection procedure described in Sec.~\ref{ap:data_collection},
we can estimate the average visibility ($\bar{v}_{H,F}$) as long as three
conditions are satisfied: one about reporting, one about the visibility of
the hidden population to the probe alters, and one about sampling. 

\stmt{result}{res:goc-v-estimator} {
Assume that we have a sample $s_H$ taken from the hidden population using a
relative probability design with relative probabilities of inclusion $c \pi_i$
for all sampled elements (Sec.~\ref{ap:sampling-designs-from-t}).  Then
\begin{align}
\label{eqn:goc-v-estpf}
\widehat{\bar{v}}_{H,F} &= 
\frac{N_F}{N_{\mathcal{A} \cap F}}~
\frac{\sum_{i \in s_H} \sum_j \widetilde{v}_{i, A_j \cap F}/ (c \pi_i)}
     {\sum_{i \in s_H} 1/(c \pi_i)} && \mbox{}
\end{align}
\noindent is consistent and essentially unbiased for $\bar{v}_{H,F}$ if
\begin{align}
\widetilde{v}_{H, \mathcal{A} \cap F} &= v_{H, \mathcal{A} \cap F}, 
&& \mbox{ (reporting condition) }
\label{eq:v_tf_reporting}
\end{align}
and
\begin{align}
\frac{v_{H, \mathcal{A} \cap F}}{N_{\mathcal{A} \cap F}} &= \frac{v_{H,F}}{N_F}.
&& \mbox{ (probe alter condition) }
\label{eq:v_tf_probealters}
\end{align}
}

\stmtproof{res:goc-v-estimator}{
Property~\ref{prop:hajekest} holds for estimating
$\overline{\widetilde{v}}_{F,\mathcal{A} \cap F}$ from
$\widetilde{v}_{i,\mathcal{A} \cap F}$, just as it holds for estimating
$\bar{y}_{H,\mathcal{A} \cap F}$ from $y_{i,\mathcal{A} \cap F}$. Applying
Property~\ref{prop:hajekest} here, we conclude that the estimator is consistent
and essentially unbiased for
\begin{align}
\frac{N_F}{N_{\mathcal{A} \cap F}} \overline{\widetilde{v}}_{H,\mathcal{A} \cap F} =
\frac{N_F}{N_{\mathcal{A} \cap F}} \frac{\widetilde{v}_{H,\mathcal{A} \cap F}}{N_H}.
\end{align}

\noindent Next, by applying the reporting condition in
Equation~\ref{eq:v_tf_reporting} we can conclude that 
\begin{equation}
\frac{N_F}{N_{\mathcal{A} \cap F}} \frac{\widetilde{v}_{H,\mathcal{A} \cap F}}{N_H} =
\frac{N_F}{N_{\mathcal{A} \cap F}} \frac{v_{H,\mathcal{A} \cap F}}{N_H}.
\end{equation}

\noindent Finally, by applying the probe alter condition in
Equation~\ref{eq:v_tf_probealters} and rearranging terms, we conclude that 
\begin{align}
\frac{N_F}{N_{\mathcal{A} \cap F}} \frac{v_{H,\mathcal{A} \cap F}}{N_H} &= 
\frac{N_F}{N_H} \frac{v_{H,F}}{N_F} \\
 & = \bar{v}_{H,F}
\end{align}
}
\rptstmtonlyproof{res:goc-v-estimator}

Note that Result~\ref{res:goc-v-estimator} requires us to know the size of the
probe alters in the frame population, $N_{\mathcal{A} \cap F}$. In some cases,
this may not be readily available, but it may be reasonable to assume that
\begin{align}
N_{\mathcal{A} \cap F} &= \frac{N_F}{N}~N_{\mathcal{A}}.
\end{align}
\noindent Furthermore, if $\mathcal{A}$ is chosen so that all of its members
are in $F$, then $N_{\mathcal{A} \cap F} = N_{\mathcal{A}}$ and $v_{i, A_j
\cap F} = v_{i, A_j}$. In this situation, we do not need to specifically ask
respondents about connections to $\mathcal{A} \cap F$; we can just ask about
connections to $\mathcal{A}$.

The reporting condition required for Result~\ref{eq:v_tf_reporting} states that
the hidden population's total reported visibility from the probe alters on the
frame must be correct.  This might not be the case, if for example, respondents
systematically over-estimate how much others know about them (see
e.g.,~\citet{gilovich_illusion_1998}).  

The required condition for the probe
alters is slightly more complex. It needs to be the case that the rate at which the hidden population is visible to the probe alters is the same as the rate at which the hidden population is visible to the frame population. There are several equivalent ways of stating this
condition, as we show in a moment. First, we need to
define two new quantities: the individual-level true positive rate and the
average of the individual-level true positive rates.
\stmt{defn}{def:tpr-indiv} {
We define the individual-level true positive rate for respondent $i \in F$ to be
\begin{align}
\TPR_{i} = \frac{v_{H,i}}{d_{i,H}},
\end{align}
\noindent where $v_{H,i} = \sum_{j \in H} v_{j,i}$.
}
\stmt{defn}{def:tpr-indiv-avg} {
We define the average of the individual true positive rates over a set $F$ of respondents as
\begin{align}
\overline{\TPR}_F = \frac{1}{N_F} \sum_{i \in F} \TPR_{i}.
\end{align}
}
In general, $\overline{\TPR}_F \neq \TPR_F$. To see this, note that while
$\overline{\TPR}_F$ is the average of the individual-level true positive rates
with each individual weighted equally, $\TPR_F$ can be written as the weighted
average of the individual true positive rates, with the weights given by each
individual's degree.  We can see the exact relationship between the two by
expressing $\TPR_F$ in terms of the $\TPR_i$:
\begin{align}
\label{eqn:tpr-indiv-tmp1}
\TPR_F = \frac{ \sum_{i \in F} \TPR_i~d_{i,H}}{ \sum_{i \in F} d_{i,H}},
\end{align}
\noindent since multiplying each $\TPR_i$ by $d_{i,H}$ and summing is the same
as summing the $v_{H,i}$. 

\stmt{result}{res:well_constructed_equiv} {
The following conditions are all equivalent.
\begin{enumerate}[(i)]
    \item $\frac{v_{H, \mathcal{A} \cap F}}{N_{\mathcal{A} \cap F}} = \frac{v_{H,F}}{N_F}$ 
\item $\TPR_{\mathcal{A}  \cap F}~\bar{d}_{\mathcal{A} \cap F,H} = \TPR_F~\bar{d}_{F,H}$
\item $\overline{\TPR}_{\mathcal{A} \cap F}~\bar{d}_{\mathcal{A} \cap F, H} + \cov_{\mathcal{A} \cap F}(\TPR_i, d_{i,H}) =  \overline{\TPR}_{F}~\bar{d}_{F, H} + \cov_F(\TPR_i, d_{i,H})$
    \item $\bar{y}^{+}_{F,H} = \frac{\sum_j \bar{y}^{+}_{A_j \cap F, H}~N_{A_j \cap F}}{\sum_j N_{A_j \cap F}}$,
\end{enumerate}
\noindent where $\cov_F$ is the finite-population covariance taken over the set
$F$.%
\footnote{ We define the finite-population covariance to have a denominator of
    $N_F$; this differs from some other authors, who define the
    finite-population covariance to have $N_F - 1$ in the denominator.
}
}

\stmtproof{res:well_constructed_equiv}{
First, we show that 
\begin{equation}
    \TPR_{\mathcal{A} \cap F}~\bar{d}_{\mathcal{A} \cap F,H} = 
    \TPR_F~\bar{d}_{F,H} 
    \Longleftrightarrow  
    \frac{v_{H, \mathcal{A} \cap F}}{N_{\mathcal{A}\cap F}} = 
    \frac{v_{H,F}}{N_F}.
\end{equation}
By definition, $\TPR_{F}~\bar{d}_{F,H} = (v_{H,F} / d_{F,H}) \times (d_{F,H} / N_F) = 
v_{H,F} / N_F$. The same argument demonstrates that $\TPR_{\mathcal{A} \cap F}~\bar{d}_{\mathcal{A} \cap F,H} = v_{H,\mathcal{A} \cap F} / N_{\mathcal{A}}$. We conclude that $(i) \Longleftrightarrow (ii)$.

Next, we show that $(ii)$ is equivalent to $(iii)$.  We can use the relationship between $\TPR_F$ and the $\TPR_i$, Equation~\ref{eqn:tpr-indiv-tmp1}, to deduce that
\begin{align}
\TPR_F~d_{F,H} = \sum_{i \in F} \TPR_i~d_{i,H} &= N_F~[\overline{\TPR}_F~\bar{d}_{F,H} + \cov_F(\TPR_i, d_{i,H}) ].
\end{align}
Dividing the left-most and right-most sides by $N_F$, we conclude that
\begin{align}
\TPR_{F}~\bar{d}_{F,H} = \overline{\TPR}_{F}~\bar{d}_{F, H} + \cov_F(\TPR_i, d_{i,H}).
\end{align}
\noindent The same argument shows that 
\begin{equation}
\bar{d}_{\mathcal{A} \cap F,H}~\TPR_{\mathcal{A} \cap F} =
\overline{\TPR}_{\mathcal{A} \cap F}~\bar{d}_{\mathcal{A} \cap F, H} +
\cov_{\mathcal{A} \cap F}(\TPR_i, d_{i,H}).
\end{equation}
\noindent So we conclude that $(ii) \Longleftrightarrow (iii)$. 

Finally, we show that $(iv)$ is equivalent to $(i)$.  In Appendix~\ref{ap:generalized},
showed that $y^{+}_{F,H} = v_{H,F}$ (Equation~\ref{eqn:nr-id-withfp}). Dividing both sides by $N_F$, we have $\bar{y}^{+}_{F,H} = v_{H,F}/N_H$, which is the right-hand side of
the identity in $(i)$. Similarly, starting with the left-hand side of the identity in $(i)$, we have
\begin{align}
\frac{v_{H, \mathcal{A} \cap F}}{N_{\mathcal{A} \cap F}} =
 \frac{\sum_j v_{H, A_j \cap F}}{\sum_j N_{A_j \cap F}} = 
 \frac{\sum_j y^{+}_{A_j \cap F, H}}{\sum_j N_{A_j \cap F}} =
 \frac{\sum_j \bar{y}^{+}_{A_j \cap F,H}~N_{A_j \cap F}}{\sum_j N_{A_j \cap F}}.
\end{align}
\noindent So we conclude that $(i) \iff (iv)$.

Since $(i) \Longleftrightarrow (ii)$ and $(ii) \Longleftrightarrow (iii)$, it
follows that $(i) \Longleftrightarrow (iii)$. Furthermore, since 
$(i) \Longleftrightarrow (iv)$, it follows that $(iv)$ is equivalent to
$(ii)$ and $(iii)$. 
}

\rptstmtonlyproof{res:well_constructed_equiv}

Result~\ref{res:well_constructed_equiv} shows that the probe alter condition
can be expressed in many equivalent ways.  One of these alternate expressions
is especially useful because it leads to an empirical check of the probe alter
condition that future scale-up studies can implement. This empirical check is a
direct consequence of Result~\ref{res:well_constructed_test}, below.
Intuitively, Result~\ref{res:well_constructed_test} and the empirical check are
a consequence of the identity in Equation~\ref{eqn:nr-identity}, which says
that in-reports from the perspective of $H$ are also out-reports from the
perspective of $F$.

\stmt{result}{res:well_constructed_test}{%
    Suppose that the precision of out-reports from the frame population
    is the same as the precision of the out-reports from $\mathcal{A} \cap F$:
\begin{equation}
    \label{eqn:pa_f_same_fp_rate}
    \frac{y^{+}_{F,H}}{y_{F,H}} = 
    \frac{y^{+}_{\mathcal{A} \cap F,H}}{y_{\mathcal{A} \cap F, H}}
\end{equation}
Then the probe alter condition (\ref{eq:v_tf_probealters}) is satisfied if
and only if
\begin{equation}
    \label{eqn:pa_equiv_samefp}
    \bar{y}_{F, H} = \bar{y}_{\mathcal{A} \cap F, H}. 
\end{equation}
}

\stmtproof{res:well_constructed_test}{%
    First, note that, by Result~\ref{res:well_constructed_equiv}, the probe alter
    condition is equivalent to
\begin{equation}
    \label{eqn:probe_alter_v4}
    \bar{y}^{+}_{F,H} = \frac{\sum_j \bar{y}^{+}_{A_j \cap F, H}~N_{A_j \cap F}}{\sum_j N_{A_j \cap F}}.
\end{equation}
Since $\bar{y}^{+}_{A_j \cap F, H} = y^{+}_{A_j \cap F, H} / N_{A_j \cap F}$ for all
$j$, the right-hand side of Equation~\ref{eqn:probe_alter_v4} is equal to
$\bar{y}^{+}_{\mathcal{A} \cap F, H}$, meaning that the probe alter condition is
also equivalent to
\begin{equation}
    \label{eqn:probe_alter_v5}
    \bar{y}^{+}_{F,H} = \bar{y}^{+}_{\mathcal{A} \cap F, H}.
\end{equation}
Second, note that the assumption in Equation~\ref{eqn:pa_f_same_fp_rate} can be
re-written as
\begin{equation}
    \label{eqn:pa_f_same_fp_meanrate}
    \frac{\bar{y}^{+}_{F,H}}{\bar{y}_{F,H}} = 
    \frac{\bar{y}^{+}_{\mathcal{A} \cap F, H}}{\bar{y}_{\mathcal{A} \cap F, H}},
\end{equation}
\noindent by multiplying the left-hand side by $\frac{N_F}{N_F}$ and the
right-hand side by $\frac{N_{\mathcal{A} \cap F}}{N_{\mathcal{A} \cap F}}$. So
we are left with the task of showing that if
Equation~\ref{eqn:pa_f_same_fp_meanrate} is true, then
Equation~\ref{eqn:probe_alter_v5} is satisfied if and only if
Equation~\ref{eqn:pa_equiv_samefp} is satisfied. But this is the case, since
Equation~\ref{eqn:probe_alter_v5} equates the numerators of the two fractions
in Equation~\ref{eqn:pa_f_same_fp_meanrate} and
Equation~\ref{eqn:pa_equiv_samefp} equates the denominators of the two
fractions in Equation~\ref{eqn:pa_f_same_fp_meanrate}. Two fractions that are
equal will have equal numerators if and only if they have equal denominators.
(Formally, if $a/b=c/d$ then $a=c$ if and only if $b=d$.) 
}

\rptstmtonlyproof{res:well_constructed_test}

The implication of Result~\ref{res:well_constructed_test} is that if (i)
researchers design the probe alters so that the frame population sample $s_F$
can be used to estimate $\bar{y}_{\mathcal{A} \cap F, H}$; and (ii) researchers
assume that the precision of out-reports from the frame population is the same
as the precision of out-reports from $\mathcal{A} \cap F$, then they can
evaluate how well the probe alter condition is satisfied empirically by
comparing $\widehat{\bar{y}}_{F,H}$ and $\widehat{\bar{y}}_{\mathcal{A} \cap
F, H}$. 


Finally, we can foresee four practical problems that might arise when researchers try to estimate $\bar{v}_{H,F}$.  First,
researchers might not be able to choose the probe alters to satisfy the probe
alter condition (Equation~\ref{eq:v_tf_probealters}) because of limited
information about the true visibility of the hidden population with respect to
different social groups.  A second problem might arise if researchers are not
able to choose the probe alters to satisfy the reporting condition
(Equation~\ref{eq:v_tf_reporting}) because of limited information about the
hidden population's awareness about visibility. A third problem might
arise due to errors in administrative records which would cause researchers to
have incorrect information about the size of the multiset of probe alters on
the frame ($N_{\mathcal{A} \cap F}$).  
Finally, a fourth problem might arise due to errors in the sampling method researchers
use.
Fortunately, as we show in
Online Appendix~\ref{ap:sensitivity} (Result~\ref{res:vhf_combined_sensitivity}), 
it is possible to quantify the 
effect of these problems on the resulting estimates.  
In some cases they can cancel out, but in other cases they magnify each other.

\subsection{Guidance for choosing the probe alters for the game of contacts, $\mathcal{A}$}
\label{ap:v-probe-alter-rec}

Turning the results in Online Appendix~\ref{ap:goc} into easy to
follow steps for selecting the probe alters for the game of contacts is an open
and important research problem.  Here, we briefly offer three recommendations
for selecting the probe alters for the game of contacts.  We realize that these
recommendations may be difficult to follow exactly in practice.  Therefore, we
also discuss the sensitivity of the estimators to errors in the construction of
the probe alters.  Finally, we discuss one type of data that should be
collected from the frame population in order to help the researchers evaluate
their choice of probe alters for the game of contacts.

First, we recommend that probe alters for the game of contacts be in the frame
population.  For example, if the frame population is adults, we recommend that
all members of the probe alters be adults.  This choice will simplify the data
collection task in the game of contacts, and for all the advice listed below,
we assume that it has been followed.  If it is not possible, researchers can
still use the more general procedures developed in this Online Appendix.

Second, we recommend that the probe alters be selected such that the probe
alter condition in Result~\ref{res:goc-v-estimator} is satisfied.  That is, the
probe alters as a whole should be typical of the frame population in the
following way: it should be the case that the rate at which the hidden
population is visible to the probe alters is the same as the rate at which the
hidden population is visible to the frame population 
($\frac{v_{H, \mathcal{A}}}{N_{\mathcal{A}}} = \frac{v_{H,F}}{N_F}$).   
For example, in a study to estimate the number of drug injectors in a city,
drug treatment counselors would be a poor choice for membership in the probe
alters because drug injectors are probably more visible to drug treatment
counselors than to typical members of the frame population. On the other hand,
postal workers would probably be a reasonable choice for membership in the
probe alters because drug injectors are probably about as visible to postal
workers as they are to typical members of the frame population.  

Third, we recommend that the probe alters be selected so that the reporting
condition in Result~\ref{res:goc-v-estimator} is satisfied ($\widetilde{v}_{H,
    \mathcal{A}} = v_{H, \mathcal{A}}$).  One way to help ensure that this
    condition holds is to avoid selecting large groups that may cause recall
    error~\citep{killworth_two_2003, zheng_how_2006, mccormick_adjusting_2007,
    mccormick_how_2010, maltiel_estimating_2015}.  
In practice it might be difficult to meet each of these three conditions
exactly, therefore we recommend a sensitivity analysis using the results in
Online Appendix~\ref{ap:sensitivity}.

Finally, the choice of probe alters for the game of contacts also has two
implications for the design of the survey of the frame population. First, if
researchers wish to estimate the degree ratio, $\delta_F$, then they should
design the probe alters $\mathcal{A}$ so that they can be asked of both members
of the hidden population sample and members of the frame population sample
(see Result~\ref{res:goc-delta-estimator}). Second, if researchers wish to test
the probe alter condition using the approach in
Result~\ref{res:well_constructed_test}, then additional information needs to be
collected from each member of the frame population sample. For example, if one
group in the probe alters for the game of contacts is postal workers, then
members of the frame population sample should be asked if they are
postal workers.

\subsection{Term-by-term: $\delta_F$ and $\TPR_F$}
\label{ap:term-by-term}

\begin{figure}
\centering
\includegraphics[width=0.6\textwidth]{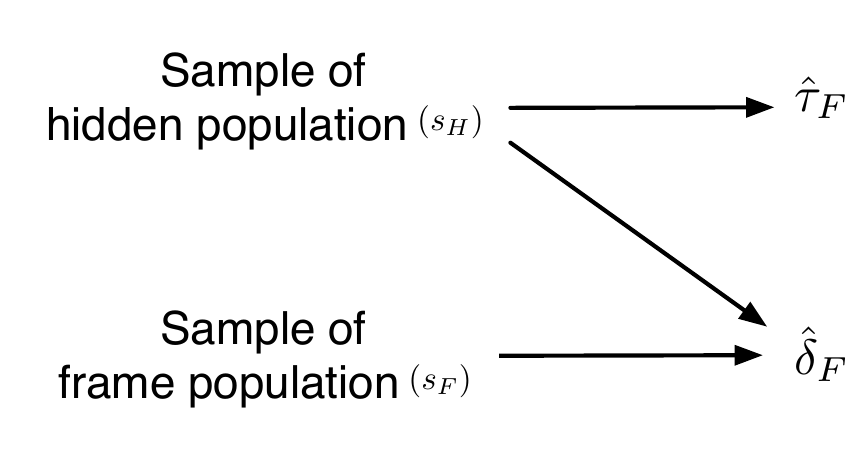}
\caption{We estimate the true positive rate $\widehat{\TPR}_F$ using data from
the survey of the hidden population, and we estimate the degree ratio
$\widehat{\delta}_F$ using the sample of the hidden population and the
sample of the frame population.}
\label{fig:adjustment_schema}
\end{figure}

In this section we describe how to estimate two adjustment factors: the degree ratio,
\begin{equation}
\delta_F = \frac{\bar{d}_{H,F}}{\bar{d}_{F,F}}
\label{eqn:delta_f_review}
\end{equation}
and the true positive rate,
\begin{equation}
\TPR_F = \frac{\bar{v}_{H,F}}{\bar{d}_{H,F}}.
\label{eqn:tpr_f_review}
\end{equation}
Estimating the degree ratio requires information from the survey of the hidden
population and the survey of the frame population, while estimating the true
positive rate only requires information from the survey of the hidden
population (Fig.~\ref{fig:adjustment_schema}).
As Equations~\ref{eqn:delta_f_review} and~\ref{eqn:tpr_f_review} make clear, both
adjustment factors involve $\bar{d}_{H, F}$ so we first present an estimator
for that quantity.

\stmt{result}{res:goc-df-estimator} { Suppose we have a sample $s_H$ taken from
    the hidden population using a relative probability sampling design with
    relative probabilities of inclusion denoted $c \pi_i$
    (Sec~\ref{ap:sampling-designs-from-t}).  Then the estimator given by
\begin{align}
\label{eqn:goc-df-estpf}
\widehat{\bar{d}}_{H,F} &= \frac{N_F}{N_{\mathcal{A} \cap F}} \frac{\sum_{i \in s_H} \sum_{j} y_{i, (A_j \cap F)} / (c\pi_i)}{\sum_{i \in s_H} 1 / (c\pi_i)}&&\mbox{ \phantom{(estimator) }}
\end{align}
\noindent is consistent and essentially unbiased for $\bar{d}_{H,F}$ if:
\begin{align}
\label{eqn:goc-df-reporting}
y_{H, \mathcal{A} \cap F} &= d_{H, \mathcal{A} \cap F},&&\mbox{ (reporting condition) }
\end{align}
and
\begin{align}
\label{eqn:goc-df-condition}
\bar{d}_{\mathcal{A} \cap F, H} &= \bar{d}_{F,H}.&&\mbox{ (probe alter condition) }
\end{align}
}

\stmtproof{res:goc-df-estimator}{
From Property~\ref{prop:hajekest}, we can see that our estimator is consistent and essentially unbiased for 
\begin{equation}
\frac{N_F}{N_{\mathcal{A} \cap F}} \frac{y_{H,\mathcal{A} \cap F}}{N_H} = 
\frac{N_F}{N_H} \frac{y_{H,\mathcal{A} \cap F}}{N_{\mathcal{A} \cap F}}. 
\end{equation}

Under the reporting condition (Equation~\ref{eqn:goc-df-reporting}) this becomes
\begin{align}
\frac{N_F}{N_H} \frac{y_{H,\mathcal{A} \cap F}}{N_{\mathcal{A} \cap F}} =
\frac{N_F}{N_H} \frac{d_{H,\mathcal{A} \cap F}}{N_{\mathcal{A} \cap F}}
\end{align}
\noindent Finally, applying the probe alter condition in Equation~\ref{eqn:goc-df-condition}, we have
\begin{align}
\frac{N_F}{N_H} \frac{d_{H,\mathcal{A} \cap F}}{N_{\mathcal{A} \cap F}} & = \frac{N_F}{N_H} \frac{d_{F,H}}{N_F} \\
 &= \bar{d}_{H,F}.
\end{align}
}

\rptstmtonlyproof{res:goc-df-estimator}

Result~\ref{res:goc-df-estimator} requires that reports are, in total, correct
(Equation~\ref{eqn:goc-df-reporting}).  Like Result~\ref{res:goc-v-estimator},
Result~\ref{res:goc-df-estimator} also requires us to know the size of the
probe alters on the frame, $N_{\mathcal{A} \cap F}$. In some cases, this may
not be readily available, but it may be reasonable to assume that
\begin{align}
N_{\mathcal{A} \cap F} &= \frac{N_F}{N}~N_{\mathcal{A}}.
\end{align}
\noindent Furthermore, if $\mathcal{A}$ is chosen so that all of its members
are in $F$, then $N_{\mathcal{A} \cap F} = N_{\mathcal{A}}$ and $y_{i, A_j \cap
F} = y_{i, A_j}$. In this situation, we do not need to specifically ask
respondents about connections to $\mathcal{A} \cap F$; we can just ask about
connections to $\mathcal{A}$. Result~\ref{res:goc-df-estimator} also requires a
specific rate of connectivity between the probe alters and the hidden
population (Equation~\ref{eqn:goc-df-condition}).   We discussed some of the
consequences of these assumption in the main text, where we made
recommendations for practice (Section~\ref{sec:recs-for-practice}).


\subsubsection{Estimating the degree ratio, $\delta_F$}
\label{sec:hat_delta_f}

We can combine our estimator for $\bar{d}_{H,F}$
(Result~\ref{res:goc-df-estimator}) and our estimator for $\bar{d}_{F,F}$
(Result~\ref{res:kpestimator-dff}), to estimate the degree ratio, $\delta_F$.

\stmt{result}{res:goc-delta-estimator} {
The estimator
\begin{align}
\label{eqn:goc-delta-estpf}
\widehat{\delta}_F &= \frac{\widehat{\bar{d}}_{H,F}}{\widehat{\bar{d}}_{F,F}}
\end{align}
\noindent is consistent and essentially unbiased for $\delta_F$ if
$\widehat{\bar{d}}_{H,F}$ is consistent and essentially unbiased for
$\bar{d}_{H,F}$ and $\widehat{\bar{d}}_{F,F}$ is consistent and essentially
unbiased for $\bar{d}_{F,F}$.
}
\stmtproof{res:goc-delta-estimator}{
This follows from the properties of a compound ratio estimator 
(Online Appendix~\ref{ap:ratio}).}

\rptstmtonlyproof{res:goc-delta-estimator}

More concretely, combing the estimators in Result~\ref{res:goc-df-estimator}
and Result~\ref{res:kpestimator-dff}, results in an estimator for
$\widehat{\delta}_F$ with the following form:
\begin{equation}
\label{eqn:delta-f-inpractice}
\widehat{\delta}_{F} = 
\frac{\frac{N_F}{N_{\mathcal{A}_H \cap F }} 
      \frac{\sum_{i \in s_H} \sum_{A_j \in \mathcal{A}_H} y_{i, (A_j \cap F)} / (c\pi_{i}^H)}
           {\sum_{i \in s_H} 1 / (c\pi_{i}^H)} } 
{\frac{1}{N_{\mathcal{A}_F}} \sum_{i \in s_F} \sum_{A_k \in \mathcal{A}_F} y_{i, A_k} / \pi_{i}^F  }.
\end{equation}

If the probe alters for the frame population and the hidden population are the
same, so that $\mathcal{A}_H = \mathcal{A}_F = \mathcal{A}$, and if the probe
alters are randomly distributed in the frame population in the sense that
\begin{align}
\label{eqn:delta-kp-prop}
N_{\mathcal{A} \cap F} = N_{\mathcal{A}}~\frac{N_F}{N},
\end{align}
then we can reduce the constants in front of Equation~\ref{eqn:delta-f-inpractice} to 
\begin{align}
\frac{\frac{N_F}{N_{\mathcal{A} \cap F}}}{\frac{1}{N_\mathcal{A}}} 
= \frac{\frac{N}{N_\mathcal{A}}}{\frac{1}{N_\mathcal{A}}} = N.
\end{align}
In other words, when the probe alters for the frame and hidden population are
the same, and when the probe alters are randomly distributed in the frame
population,  all of the factors involving the size of
$\mathcal{A}$ drop out.  This fact allows researchers to use groups defined by
first names (e.g., people named Michael) in the probe alters $\mathcal{A}$,
even if the size of these groups is not known, as long as it is reasonable to
assume that $\mathcal{A}$ satisfies Equation~\ref{eqn:delta-kp-prop}
(c.f.,~\citet{salganik_assessing_2011}).

\subsubsection{Estimating the true positive rate, $\TPR_F$}
\label{sec:hat_tpr_f}

We can combine our estimator for $\bar{v}_{H,F}$
(Result~\ref{res:goc-v-estimator}) and our estimator for $\bar{d}_{H,F}$
(Result~\ref{res:goc-df-estimator}) to estimate the true positive rate $\TPR_F$.


\stmt{result}{res:goc-tpr-estimator} {%
The estimator
\begin{align}
\label{eqn:goc-tpr-estpf}
\widehat{\TPR}_{F} &= \frac{\widehat{\bar{v}}_{H,F}}{\widehat{\bar{d}}_{H,F}}
\end{align}
\noindent is consistent and essentially unbiased for $\TPR_{F}$ if
$\widehat{\bar{v}}_{H,F}$ is a consistent and essentially unbiased estimator of
$\bar{v}_{H,F}$ and if $\widehat{\bar{d}}_{H,F}$ is a consistent and
essentially unbiased estimator of $\bar{d}_{H,F}$.
}
\stmtproof{res:goc-tpr-estimator}{%
This follows directly from the properties of a compound ratio estimator
(Online Appendix~\ref{ap:ratio}).
}
\rptstmtonlyproof{res:goc-tpr-estimator}

More concretely, combing the estimator in Result~\ref{res:goc-v-estimator} and
Result~\ref{res:goc-df-estimator} yields an estimator for $\widehat{\TPR}_F$
with the following form:
\begin{equation}
\label{eqn:goc-tpr-tmp1}
\widehat{\TPR}_{F} = \frac{ \sum_{i \in s_H} \widetilde{v}_{i, \mathcal{A}_H} / (c \pi_i) }{ \sum_{i \in s_H} y_{i, \mathcal{A}_H} / (c \pi_i) }.
\end{equation}
All of the factors involving the size of $\mathcal{A}$ drop out of
Equation~\ref{eqn:goc-tpr-tmp1}.  This fact allows researchers to use groups
defined by first names (e.g., people named Michael) in the probe alters
$\mathcal{A}$, even if the size of these groups is not known
(c.f.,~\citet{salganik_game_2011a}).

\subsection{Estimating the size of the hidden population, $N_H$}
\label{sec:hat_N_T}

We now make use of all of the results for the individual terms we derived above to present four different estimators for the size of the hidden population, $N_H$.


\stmt{result}{res:goc-gnsum-new} {
The generalized scale-up estimator given by
\begin{align}
\label{eqn:res:goc-newgnsum-estimator}
\widehat{N}_H &= \frac{\widehat{y}_{F,H}}{\widehat{\bar{v}}_{H,F}}
\end{align}
\noindent is consistent and essentially unbiased for $N_H$ if there are no false positive reports, if $\widehat{y}_{F,H}$ is consistent and unbiased for $y_{F,H}$, and if $\widehat{\bar{v}}_{H,F}$ is consistent and essentially unbiased for $\bar{v}_{H,F}$.
}
\stmtproof{res:goc-gnsum-new}{
From the properties of a compound ratio estimator, we know that our estimator is
consistent and essentially unbiased for $y_{F,H} / \bar{v}_{H,F}$
(Appendix~\ref{ap:ratio}). By the argument in the main text given in
Section~\ref{sec:framework}, leading to Equation~\ref{eqn:qoi-census-nofp},
this quantity is equal to $N_H$.
}
\rptstmtonlyproof{res:goc-gnsum-new}


\stmt{result}{res:goc-ansum} {
The adjusted basic scale-up estimator given by
\begin{align}
\label{eqn:goc-ansum-estimator}
\widehat{N}_H &= \frac{\widehat{y}_{F,H}}{\widehat{\bar{d}}_{U,F}}
~\frac{1}{\widehat{\phi}_F}
~\frac{1}{\widehat{\delta}_F}
~\frac{1}{\widehat{\TPR}_F}
\end{align}
\noindent is consistent and essentially unbiased for $N_H$ if there are no
false positive reports, and if each of the individual estimators is consistent
and essentially unbiased.
}
\stmtproof{res:goc-ansum}{
From the results in Online Appendix~\ref{ap:ratio}, we know that this compound ratio
estimator will be consistent and essentially unbiased for $y_{F,H} /
(\bar{d}_{U,F}~\phi_F~\delta_F~\TPR_F)$.  The denominator is
$\bar{v}_{H,F}$ by construction, leaving us with $y_{F,H} / \bar{v}_{H,F}$.  By
the argument in the main text given in Section~\ref{sec:framework}, leading to
Equation~\ref{eqn:qoi-census-nofp}, this quantity is equal to $N_H$.
}
\rptstmtonlyproof{res:goc-ansum}

\stmt{result}{res:goc-gnsum} {
The adjusted scale-up estimator
\begin{align}
\label{eqn:goc-gnsum-estimator}
\widehat{N}_H &= \frac{\widehat{y}_{F,H}}{\widehat{\bar{d}}_{F,F}}
~\frac{1}{\widehat{\delta}_F}
~\frac{1}{\widehat{\TPR}_F}
\end{align}
\noindent is consistent and essentially unbiased for $N_H$ if there are no
false positives, and if each of the individual estimators is consistent and
essentially unbiased.
}
\stmtproof{res:goc-gnsum}{
From the results in Online Appendix~\ref{ap:ratio}, we know that this compound ratio
estimator will be consistent and essentially unbiased for $y_{F,H} /
(\bar{d}_{F,F}~\delta_F~\TPR_F)$.  The denominator is
$\bar{v}_{H,F}$ by construction, leaving us with $y_{F,H} / \bar{v}_{H,F}$.  By
the argument in the main text given in Section~\ref{sec:framework}, leading to
Equation~\ref{eqn:qoi-census-nofp}, this quantity is equal to $N_H$.
}
\rptstmtonlyproof{res:goc-gnsum}

\stmt{result}{res:goc-asu2} {
The adjusted scale-up estimator
\begin{align}
\label{eqn:goc-gnsum-estimator2}
\widehat{N}_H &= \frac{\widehat{y}_{F,H}}{\widehat{\bar{d}}_{F,F}}
~\frac{1}{\widehat{\delta}_F}
~\frac{1}{\widehat{\TPR}_F}
~\widehat{\eta}_F
\end{align}
\noindent is consistent and essentially unbiased for $N_H$ if each of the individual estimators is consistent and
essentially unbiased.
}
\stmtproof{res:goc-asu2}{
From the results in Online Appendix~\ref{ap:ratio}, we know that this compound ratio
estimator will be consistent and essentially unbiased for $(y_{F,H}~\eta_F) /
(\bar{d}_{F,F}~\delta_F~\TPR_F)$.  The numerator is $y^{+}_{F,H}$ by
construction and the product of the denominators is $\bar{v}_{H,F}$ by
construction, leaving us with $y^{+}_{F,H} / \bar{v}_{H,F}$.  By
the argument in Online Appendix~\ref{ap:generalized} this quantity is equal to $N_H$.
}
\rptstmtonlyproof{res:goc-asu2}


\section{Sensitivity analysis}
\label{ap:sensitivity}

All of the estimators that we propose require that specific conditions hold in
order to produce consistent and essentially unbiased estimates.  
These conditions can be divided into four groups: 
survey construction, reporting behavior, network structure, and sampling.
In many practical settings, we expect that researchers may not be confident 
that these conditions hold perfectly.
Therefore, in this appendix, we derive results that enable researchers
to assess the sensitivity of their estimates to violations of all four types of conditions.
First, in Section~\ref{ap:sensitivity-nonsamp}, we develop a results that help 
researchers assess sensitivity to survey construction, reporting, and network structure;
then, in Section~\ref{sec:sampling_sensitivity}, we turn to results that help researchers 
assess sensitivity to sampling problems. 
Finally, in Section~\ref{ap:sensitivity-combined}, we combine all of the sensitivity results 
to derive expressions that enable researchers to conduct sensitivity analyses
that simultaneously account for all of the conditions.

\subsection{Sensitivity to non-sampling conditions: survey construction, reporting behavior, and network structure}
\label{ap:sensitivity-nonsamp}

Most estimators that we consider depend on conditions related to survey
construction 
(for example, choosing the probe alters for the known population method) 
and to reporting 
(for example, the assumption that respondents make accurate aggregate
reports about the probe alters);
furthermore, the basic scale-up estimator is sensitive to conditions
about network structure (for example, the relative size of hidden population 
and frame population members' personal networks).
In this section, we develop sensitivity results for these nonsampling conditions.
First, Result~\ref{res:vtf_sensitivity} shows how one of these estimators 
($\widehat{\bar{v}}_{H,F}$) is impacted by violations of the conditions it depends upon.
Next, using Result~\ref{res:vtf_sensitivity} as a template, Table~\ref{tab:robustness}
provides similar expressions for all of the estimators we discuss in the main
text.

\stmt{result}{res:vtf_sensitivity}{%
Suppose that $\widehat{N}_{\mathcal{A} \cap F}$, the researcher's estimate of
$N_{\mathcal{A} \cap F}$, is incorrect, so that $\widehat{N}_{\mathcal{A} \cap
F} = c_1 \cdot N_{\mathcal{A} \cap F}$.  Suppose also that the reporting
condition (Equation~\ref{eq:v_tf_reporting}) of Result~\ref{res:goc-v-estimator} is
incorrect, so that $\tilde{v}_{H, \mathcal{A} \cap F} = c_2 \cdot
v_{H,\mathcal{A} \cap F}$. Finally, suppose that the probe alter condition
is incorrect, so that $\frac{v_{H,{A \cap F}}}{N_{\mathcal{A} \cap F}} = c_3
\cdot \frac{v_{H,F}}{N_F}$. Call the estimator under these imperfect conditions
$\widehat{\bar{v}}_{H,F}^{\star}$. Then $\widehat{\bar{v}}_{H,F}^{\star}$ is
consistent and essentially unbiased for $\frac{c_3~c_2}{c_1} \bar{v}_{H,F}$
instead of $\bar{v}_{H,F}$. 
}

\stmtproof{res:vtf_sensitivity}{
Under the assumptions listed above, we can write the new estimator as
\begin{align}
\widehat{\bar{v}}_{F,H}^{\star} &= \frac{1}{c_1}~\frac{N_F}{N_{\mathcal{A} \cap F}}~
\frac{\sum_{i \in s_H} \sum_j \widetilde{v}_{i, A_j \cap F}/ (c \pi_i)}{\sum_{i
\in s_H} 1/(c \pi_i)}.
\end{align}
We follow the same steps as the proof of Result~\ref{res:goc-v-estimator}, but
each time we use one of our assumptions, the associated error is carried with
it.  So our estimator $\widehat{\bar{v}}^{\star}_{F,H}$ is consistent and
essentially unbiased for 
\begin{align}
    \frac{1}{c_1}~\frac{N_F}{N_{\mathcal{A} \cap F}}~
    \frac{\tilde{v}_{H, \mathcal{A} \cap F}}{N_H} =
    \frac{c_2}{c_1}~\frac{N_F}{N_{\mathcal{A} \cap F}}~
    \frac{v_{H, \mathcal{A} \cap F}}{N_H} =
    \frac{c_3~c_2}{c_1}~\frac{N_F}{N_{\mathcal{A} \cap F}}~
    \frac{v_{H, F}}{N_H}.
\end{align}
In words, the estimand is now incorrect by $\frac{c_3~c_2}{c_1}$. Since
$\widehat{\bar{v}}_{F,H}$ is consistent and essentialy unbiased for
$\bar{v}_{F,H}$, we conclude that $\widehat{\bar{v}}_{F,H}^{\star}$ is
consistent and essentially unbiased for $\frac{c_3~c_2}{c_1} \bar{v}_{F,H}$.
Note that if the assumptions needed for Result~\ref{res:goc-v-estimator} hold,
then $c_1 = 1$, $c_2=1$, and $c_3=1$, giving us the original result.  
}

\rptstmtonlyproof{res:vtf_sensitivity}

Table~\ref{tab:robustness} shows results analogous to
Result~\ref{res:vtf_sensitivity} for all of the estimators we propose. We do not
prove each one individually, since the derivations all follow the pattern of
Result~\ref{res:vtf_sensitivity} very closely.  Researchers who wish to
understand the how their estimates are affected by the assumptions they make
can use Table~\ref{tab:robustness} to conduct a sensitivity analysis.
Note that any problems with the sampling design could result in problems with
the estimates that are not captured by the results in
Table~\ref{tab:robustness}.
These sampling problems are the subject of the next section.



\begin{table}
\centering
\scalebox{.7}{
\begin{tabular}{l >{\compress}p{4cm} p{3.5cm}}
\toprule
Estimator & Imperfect assumptions & Effective estimand\\
\midrule

$\widehat{\bar{d}}_{F,F}$ (Result \ref{res:kpestimator-dff}) &
\begin{enumerate}[(i)]
\item $\widehat{N}_\mathcal{A} = c_1 ~N_\mathcal{A}$
\item $\bar{d}_{\mathcal{A}, F} = c_2 ~\bar{d}_{F,F}$
\item $y_{F, \mathcal{A}} = c_3 ~ d_{F, \mathcal{A}}$
\end{enumerate} 
&
$\frac{c_2~c_3}{c_1} ~\bar{d}_{F,F}$\\
\midrule
$\widehat{\bar{d}}_{U,F}$ (Result \ref{res:kpestimator-duf}) &
\begin{enumerate}[(i)]
\item $\widehat{N}_\mathcal{A} = c_1 ~N_\mathcal{A}$
\item $\bar{d}_{\mathcal{A}, F} = c_2 ~\bar{d}_{U,F}$
\item $y_{F, \mathcal{A}} = c_3 ~ d_{F, \mathcal{A}}$
\end{enumerate} 
&
$\frac{c_2~c_3}{c_1} ~\bar{d}_{U,F}$\\
\midrule
$\widehat{\phi}_F$ (Result \ref{res:goc-phi-estimator}) &
\begin{enumerate}[(i)]
\item $\widehat{\bar{d}}_{F,F} \leadsto c_1 ~\bar{d}_{F,F}$
\item $\widehat{\bar{d}}_{U,F} \leadsto c_2 ~\bar{d}_{U,F}$
\end{enumerate} 
&
$\frac{c_1}{c_2} ~\phi_F$\\
\midrule
$\widehat{\bar{v}}_{H,F}$ (Result \ref{res:goc-v-estimator}) &
\begin{enumerate}[(i)]
\item $\widehat{N}_{\mathcal{A} \cap F} = c_1~N_{\mathcal{A} \cap F}$
\item  $\tilde{v}_{H, \mathcal{A} \cap F} = c_2~v_{H, \mathcal{A} \cap F}$
\item  $\frac{v_{H, \mathcal{A} \cap F}}{N_{\mathcal{A} \cap F}} = c_3~\frac{v_{H,F}}{N_F}$
\end{enumerate} 
&
$\frac{c_3~c_2}{c_1} ~\bar{v}_{H,F}$\\
\midrule
$\widehat{\delta}_F$ (Result \ref{res:goc-delta-estimator}) &
\begin{enumerate}[(i)]
\item $\widehat{\bar{d}}_{H,F} \leadsto c_1 ~\bar{d}_{H,F}$
\item $\widehat{\bar{d}}_{F,F} \leadsto c_2 ~\bar{d}_{F,F}$
\end{enumerate} 
&
$\frac{c_1}{c_2} ~\delta_F$\\

\midrule
$\widehat{\TPR}_F$ (Result \ref{res:goc-tpr-estimator}) &
\begin{enumerate}[(i)]
\item $\widehat{\bar{v}}_{H,F} \leadsto c_1 ~\bar{v}_{H,F}$
\item $\widehat{\bar{d}}_{H,F} \leadsto c_2 ~\bar{d}_{H,F}$
\end{enumerate} 
&
$\frac{c_1}{c_2} ~\TPR_F$\\

\midrule
$\widehat{N}_H$ (Result \ref{res:goc-gnsum-new}) &
\begin{enumerate}[(i)]
\item $\widehat{\bar{v}}_{H,F} \leadsto c_1 ~\bar{v}_{H,F}$
\end{enumerate} 
&
$\frac{1}{c_1}~N_H$\\

\midrule
$\widehat{N}_H$ (Result \ref{res:goc-gnsum}) &
\begin{enumerate}[(i)]
\item $\widehat{\bar{d}}_{F,F} \leadsto c_1 ~\bar{d}_{F,F}$
\item $\widehat{\delta}_F \leadsto c_2 ~\delta_F$
\item $\widehat{\TPR}_F \leadsto c_3 ~\TPR_F$
\end{enumerate} 
&
$\frac{1}{c_1~c_2~c_3}~N_H$\\

\bottomrule

\end{tabular}
}
\caption{Sensitivity of estimators to nonsampling assumptions. 
    The first column lists the most important estimators we discuss in the main
    text and appendixes.
    The consistency and approximate unbiasedness of each estimator relies upon
    nonsampling conditions being satisfied. These conditions are given in the
    second column, with a modification: we add a constant to each condition; if
    the constant is $1$, then the original condition is satisfied. The estimand
    is then effectively changed to the quantity listed in the third column.
    (NB: we use the symbol $\leadsto$ as a shorthand for `is consistent and
    essentially unbiased for'.) For example, the first row shows
    $\widehat{\bar{d}}_{F,F}$ and the three conditions that the estimator in
    Result \ref{res:kpestimator-dff} relies upon. Suppose that the first and third hold,
    so that $c_1 = 1$ and $c_3=1$, but that the second does not; instead, the probe alters
    $\mathcal{A}$ have been chosen so that $\bar{d}_{\mathcal{A},F} =
    1.1~\bar{d}_{F,F}$. Then $c_2 = 1.1$. Looking at the third column, we can
    see that our estimator will then be consistent and essentially unbiased for
    $1.1 \times \bar{d}_{F,F}$ instead of $\bar{d}_{F,F}$.}
\label{tab:robustness}
\end{table}

\subsection{Sensitivity to sampling problems}
\label{sec:sampling_sensitivity}

All of the estimators we discuss throughout this paper rely upon assumptions
about the sampling procedure that researchers use to obtain their data.
In this section, we develop sensitivity results that enable researchers to assess
how violations of these sampling assumptions will impact the resulting estimates.
First, we investigate the sensitivity of the estimator $\widehat{y}_{F,H}$ from a
probability sample (Online Appendix~\ref{ap:sampling-designs-from-f}), and,
next, we investigate the estimator $\widehat{\bar{\tilde{v}}}_{H,\mathcal{A} \cap F}$ 
from relative probability sample (Online Appendix~\ref{ap:sampling-designs-from-t}).

For both estimators, we investigate how estimates are affected by differences
between the inclusion probabilities that researchers use to analyze their data
and the true inclusion probabilities that come from the sampling mechanism.
These problems could arise if the sampling design is not perfectly executed, or
if there is a problem with the information underlying the sampling design.

\subsubsection{Probability samples}
\label{sec:sampling_sensitivity_prob}

First, we must define \emph{imperfect sampling weights}.

\paragraph{Imperfect sampling weights.}
Suppose a researcher obtains a probability sample $s_F$ from the frame population $F$ 
(Online Appendix~\ref{ap:sampling-designs-from-f}).
Let $I_i$ be the random variable that assumes the value $1$ when unit $i \in F$
is included in the sample $s_F$, and $0$ otherwise.
Let $\pi_i = \E[I_i]$ be the true probability of inclusion for unit $i \in F$,
and let $w_i = \frac{1}{\pi_i}$ be the corresponding design weight for
unit $i$.
We say that researchers have \emph{imperfect sampling weights} when
researchers use imperfect estimates of the inclusion probabilities $\piprime_i$
and the corresponding design weights
$\wprime_i = \frac{1}{\piprime_i}$.
Note that we assume that both the true and the imperfect weights satisfy 
$\pi_i > 0$ and $\piprime_i > 0$ for all $i$.

The first result, Result~\ref{res:yfh_sampling_sensitivity}, concerns
researchers who obtain a probability sample, but who estimate $y_{F,H}$
imperfect sampling weights. 

Result~\ref{res:yfh_sampling_sensitivity} shows the impact that imperfect
sampling weights have on estimates of $y_{F,H}$ from a probability sample.

\stmt{result}{res:yfh_sampling_sensitivity}{%
    Suppose researchers have obtained a probability sample $s_F$, but that they
    have imperfect sampling weights. 
    Call the imperfect sampling weights
    $\wprime_i = \frac{1}{\piprime_i}$,
    call the true weights $w_i = \frac{1}{\pi_i}$, 
    and define $\wdiff_i = \frac{\wprime_i}{w_i} = \frac{\pi_i}{\piprime_i}$.
    Call $\yfhhatprime = \sum_{i \in s_F} y_{i,H} \wprime_i$ the estimator
    for $y_{F,H}$ using the imperfect weights.
    Then
    \begin{align}
        \text{bias}[\yfhhatprime] &= 
        N_F [ \bar{y}_{F,H} (\bar{\wdiff} - 1) + \cov_F(y_{i,H}, \wdiff_i)].
        \label{eqn:ht-total-bias}
    \end{align}
    where $\bar{\wdiff} = \frac{1}{N_F} \sum_{i \in F} \wdiff_i$,
    and $\cov_F(\cdot,\cdot)$ is the finite population unit covariance.
}

\stmtproof{res:yfh_sampling_sensitivity}{
We can write the bias in the estimator $\yfhhatprime$ as
\begin{align}
    \text{bias}[\yfhhatprime] &= \E[\yfhhatprime] - y_{F,H}\\
    &= \sum_{i \in F} \wprime_i \E[I_i] y_{i,H} - \sum_{i \in F} y_{i,H}\\
    &= \sum_{i \in F} \frac{\pi_i}{\piprime_i} y_{i,H} - \sum_{i \in F} y_{i,H}\\
    &= \sum_{i \in F} y_{i,H} (\wdiff_i - 1).
    \label{eqn:ht-intermed-line1}
\end{align}

Now, recall that, for any $a_i$ and $b_i$,
\begin{align}
    \sum_{i \in F} a_i~b_i &= N_F \left[ \bar{a}\bar{b} + \cov_F(a_i, b_i)\right],
\end{align}
\noindent where $\bar{a}$ and $\bar{b}$ are the mean values of $a$ and $b$, and 
$\cov_F(a_i, b_i)$ is the finite population unit covariance between
$a_i$ and $b_i$.
Applying this fact to Equation~\ref{eqn:ht-intermed-line1}, we have
\begin{align}
    \text{bias}[\yfhhatprime] &= \sum_{i \in F} y_{i,H} (\wdiff_i - 1)\\
    &= N_F \left[
        \bar{y}_{F,H}(\overline{\wdiff-1}) + \cov_F(y_{i,H}, \wdiff_i-1)
           \right],\\
    &= N_F \left[
        \bar{y}_{F,H}(\bar{\wdiff}-1) + \cov_F(y_{i,H}, \wdiff_i)
           \right].
           \label{eqn:ht-intermed-line2}
\end{align}
}

\rptstmtonlyproof{res:yfh_sampling_sensitivity}

In order to further understand Result~\ref{res:yfh_sampling_sensitivity}, 
it is helpful to use the identity
\begin{align}
\cov_F(y_{i,H}, \wdiff_i) &= 
\correl_F(y_{i,H},\wdiff_i)~ \sd_F{(y_{i,H})~ \sd_F(\wdiff_i)},
\label{eqn:yfh_covar_id}
\end{align}
where $\sd_F(\cdot)$ is the unit finite-population standard deviation, and 
$\correl_F(y_{i,H},\wdiff_i)$ is the correlation between the $y_{i,H}$ and the $\wdiff_i$.
Substituting this identity into Equation~\ref{eqn:ht-total-bias} yields
\begin{align}
    \text{bias}[\yfhhatprime] &=
    N_F \left[
        \bar{y}_{F,H}(\bar{\wdiff}-1) + 
           \correl_F(y_{i,H},\wdiff_i)~\sd_F{(y_{i,H})~\sd_F(\wdiff_i)}
           \right].
    \label{eqn:ht-total-bias-cov-bound}
\end{align}
\noindent Equation~\ref{eqn:ht-total-bias-cov-bound} provides a qualitative
understanding for when errors in the weights will be more or less problematic.
Several of the terms will typically be beyond the researcher's control: 
$N_F$, $\bar{y}_{F,H}$, and $\sd_F(y_{i,H})$ are all properties of the population
being studied.
The remaining terms, however, are related to errors in the weights.
The $\bar{\wdiff}-1$ term says that the bias will be minimized when
$\frac{\pi_i}{\piprime_i}$ is close to 1 for all $i$. 
The $\sd_F(\wdiff_i)$ term says that the bias will be reduced when the
$\frac{\pi_i}{\piprime_i}$ values have low variance---i.e., when deviations from
the correct weight value do not vary between units. 
And, finally, the 
$\correl_F(y_{i,H},\wdiff_i)$ 
term says that bias is lower in absolute value when errors in the weights are
not related to the quantity being measured.

As we will see, it will be helpful to re-express
Result~\ref{res:yfh_sampling_sensitivity} in one additional way. 
This re-expression highlights the similarities
between several of the sensitivity results we derive in this section.
This final version of Result~\ref{res:yfh_sampling_sensitivity} relies upon
a quantity, $K_F$, which serves as an index for the amount of error in the
weights.
First, note that $\sd_F(\wdiff_i) = \bar{\wdiff}~\cv_F(\wdiff_i)$, where
$\cv(\wdiff_i)$ is the coefficient of variation (i.e., the standard deviation
divided by the mean), and, likewise, $\sd_F(y_{i,H}) = \bar{y}_{F,H}~\cv_F(y_{i,H})$. 
Now, define the index
$K_F = \correl_F(y_{i,H}) \cv_F(y_{i,H}) \cv_F(\wdiff_i)$.
$K_F$ can be positive, negative, or zero.
When the weights are exactly correct (i.e., $\piprime_i = \pi_i$ for all $i$), 
$K_F = 0$; on the other hand, when there are
large errors in the weights, $K_F$ will be far from 0.%
\footnote{%
$K_F$ is similar to the identity in Equation~\ref{eqn:yfh_covar_id}, except that
it involves the coefficient of variation instead of the standard deviation.
This is convenient, because the coefficient of variation is unitless, making
$K_F$ unitless (i.e., it does not depend on the scale of the particular quantity
being estimated).
}

Using $K_F$ enables us to re-write Equation \ref{eqn:ht-total-bias-cov-bound} as 
\begin{align}
    \text{bias}[\yfhhatprime] = \E[\yfhhatprime] - y_{F,H} &=
    N_F \left[
        \bar{y}_{F,H}(\bar{\wdiff}-1) + 
        \bar{y}_{F,H}~\bar{\wdiff}~K_F
           \right]\\
\iff
    \E[\yfhhatprime] &=
    y_{F,H} +
    y_{F,H}(\bar{\wdiff}-1) + 
    y_{F,H}~\bar{\wdiff}~K_F \\
    &=
    y_{F,H}~\bar{\wdiff}~(1 + K_F)
\end{align}
\noindent Therefore, Result~\ref{res:yfh_sampling_sensitivity} directly implies
Corollary~\ref{res:yfh_sampling_sensitivity_estimand}.

\stmt{cor}{res:yfh_sampling_sensitivity_estimand}{%
    From Result~\ref{res:yfh_sampling_sensitivity}, we also have 
    \begin{align}
        \yfhhatprime &\leadstounbiased
        y_{F,H}\cdot\bar{\wdiff}\cdot(1 + K_F),
    \end{align}
    where $\leadstounbiased$ means `is consistent and unbiased for,'
    and $K_F = \correl_F(y_{i,H}, \wdiff_i) \cv_F(y_{i,H}) \cv_F(\wdiff_i)$.
}

\subsubsection{Relative probability samples}
\label{sec:sampling_sensitivity_relprob}

We now turn to the estimator for the average visibility of hidden population
members ($\bar{v}_{H,F}$). 
This estimator turns out to be more complex than the estimator we investigated
in the previous section.
In order to derive complete sensitivity results for the estimator
$\widehat{\bar{v}}_{H,F}$,
it is useful to first understand the sensitivity of the estimator for the average
reported visibility of hidden population members to the probe alters,
$\bar{\tilde{v}}_{H, \mathcal{A} \cap F}$ (see Online Appendix \ref{sec:estimating-vtf}).
$\widehat{\bar{\tilde{v}}}_{H, \mathcal{A} \cap F}$ 
turns out to be the only part of estimating $\bar{v}_{H,F}$ that
is sensitive to imperfections in sampling.

Since visibility will typically be estimated from
a relative probability sample, Result~\ref{res:vhftilde_sampling_sensitivity} concerns
researchers who obtain a relative probability sample but make estimates of
$\bar{\tilde{v}}_{H,\mathcal{A} \cap F}$ using what we call 
\emph{imperfect relative sampling weights}.
We define imperfect relative sampling weights precisely in the next paragraph, and
then we present Result~\ref{res:vhftilde_sampling_sensitivity}.

\paragraph{Imperfect relative sampling weights.}
Suppose a researcher obtains a relative probability sample $s_H$ from a population $H$ 
(Online Appendix~\ref{ap:sampling-designs-from-t}).
Let $I_i$ be the random variable that assumes the value $1$ when unit $i \in H$
is included in the sample $s_H$, and $0$ otherwise, and
let $\pi_i = \E[I_i]$.
We say that researchers have \emph{imperfect relative sampling weights} when
the true $\pi_i$ are not known and, instead, researchers use
imperfect estimates of the relative inclusion probabilities $\cprime\piprime_i$,
where $\cprime$ is some unknown constant,
and the corresponding imperfect relative probability design weights
$\wprime_i = \frac{1}{\cprime\piprime_i}$.
Note that we assume that both the true and the imperfect weights satisfy 
$\pi_i > 0$ and $\piprime_i > 0$ for all $i$.

\stmt{result}{res:vhftilde_sampling_sensitivity}{%
    Suppose researchers have obtained a relative probability sample $s_H$,
    but that the researchers
    have imperfect relative sampling weights. 
    Call the imperfect sampling weights
    $\wprime_i = \frac{1}{\cprime \piprime_i}$, 
    and define 
    $\wdiff_i = \frac{\pi_i}{\piprime_i}$.
    Call the estimator for $\bar{\tilde{v}}_{H,\mathcal{A} \cap F}$ 
    (the reported visibilities; see Section~\ref{ap:data_collection})
    using the imperfect relative sampling weights $\vhabartildehatprime$:
    \begin{align}
        \label{eqn:impweight-vhf-estimator}
        \vhabartildehatprime &= 
        \frac{\sum_{i \in s_H} \sum_j \tilde{v}_{i, A_j \cap F} / (\cprime \piprime_i)}
        {\sum_{i \in s_H} 1 / (\cprime \piprime_i)}.
    \end{align}
Then 
\begin{align}
    \text{bias}(\vhabartildehatprime) &= 
       \underbrace{
           \frac{\cov_H(\tilde{v}_{i,\mathcal{A} \cap F}, \wdiff_i)}{\bar{\wdiff}}
       }_{\substack{\text{bias from incorrect weights}}}
    -
    \underbrace{
         \frac{\cov(\vhabartildehatprime, \nhprimehat)}{\nhprime}
     }_{\text{bias from ratio estimator}},
\end{align}
\noindent 
where $\bar{\wdiff} = \frac{1}{N_H} \sum_{i \in H} \wdiff_i$;
$\nhprimehat = \sum_{i \in s_H} \wprime_i$; 
$\nhprime = \frac{1}{\cprime} \sum_{i \in H} \wdiff_i$; 
$\cov(\cdot)$ is the covariance taken with respect to the sampling distribution;
and $\cov_H(\cdot)$ is the finite population unit covariance among hidden population members.
}

\stmtproof{res:vhftilde_sampling_sensitivity}{
    The classic result of \citet{hartley_unbiased_1954}
    \citep[see also][Result 5.6.1]{sarndal_model_1992} 
    shows that 
    the expected value of the estimator in Equation~\ref{eqn:impweight-vhf-estimator} is
\begin{align}
    \E[\vhabartildehatprime] 
    &= 
    \frac{\E[\sum_{i \in s_H} \wprime_i \tilde{v}_{i,\mathcal{A} \cap F}]}
         {\E[\sum_{i \in s_H} \wprime_i]}
    - \frac{\cov(\vhabartildehatprime, \nhprimehat)}
         {\E[\sum_{i \in s_H} \wprime_i]},
         \label{eqn:vbar-samp-intermed1}
\end{align}
\noindent where the covariance is taken with respect to the sampling distribution.
Now, note that 
\begin{align}
\E[\sum_{i \in s_H} \wprime_i] &= 
\E[\sum_{i \in H} I_i \wprime_i] =
\E[\sum_{i \in H} I_i \frac{1}{\cprime \piprime_i}] =
\sum_{i \in H} \frac{\pi_i}{\cprime \piprime_i} = 
\frac{1}{\cprime} \sum_{i \in H} \wdiff_i = 
\nhprime.
\end{align}
Therefore, we substitute $\nhprime$ for the denominator of the second term
of Equation~\ref{eqn:vbar-samp-intermed1}, which
produces
\begin{align}
    \E[\vhabartildehatprime] 
    &= 
    \frac{\E[\sum_{i \in s_H} \wprime_i \tilde{v}_{i,\mathcal{A} \cap F}]}
         {\E[\sum_{i \in s_H} \wprime_i]}
    - \frac{\cov(\vhabartildehatprime, \nhprimehat)}
         {\nhprime}.
\end{align}
We do not substitute $\nhprime$ for the denominator of the first term, because
we will now see that we can instead produce a simpler expression.

The remainder of the proof focuses on the first term. Note that
\begin{align}
    \E[\sum_{i \in s_H} \wprime_i \tilde{v}_{i, \mathcal{A} \cap F}]
    &= \E[\sum_{i \in H} I_i \wprime_i \tilde{v}_{i, \mathcal{A} \cap F}]
    = \E[\sum_{i \in H} I_i \frac{1}{\cprime \piprime_i} \tilde{v}_{i, \mathcal{A} \cap F}] 
    = \sum_{i \in H} \frac{\pi_i}{\cprime \piprime_i} \tilde{v}_{i, \mathcal{A} \cap F} 
    = \frac{1}{\cprime} \sum_{i \in H} \wdiff_i \tilde{v}_{i, \mathcal{A} \cap F},
\end{align}
and also that
\begin{align}
    \E[\sum_{i \in s_H} \wprime_i ] 
    &= \E[\sum_{i \in H} I_i \wprime_i ] 
    = \E[\sum_{i \in H} I_i \frac{1}{\cprime \piprime_i}]
    = \sum_{i \in H} \frac{\pi_i}{\cprime \piprime_i} 
    = \frac{1}{\cprime} \sum_{i \in H} \wdiff_i.
\end{align}

The bias of the estimator in Equation~\ref{eqn:impweight-vhf-estimator}
is therefore
\begin{align}
    \text{bias}(\vhabartildehatprime) &= 
    \E[\vhabartildehatprime] - \vhabartilde \\
    &= 
       \frac{\sum_{i \in H} \wdiff_i \tilde{v}_{i,\mathcal{A} \cap F}}
            {\sum_{i \in H} \wdiff_i}
       - \frac{\cov(\vhabartildehatprime, \nhprimehat)}
              {\nhprime}
       - \frac{\sum_{i \in H} \tilde{v}_{i, \mathcal{A} \cap F}}{N_H}\\
    &= 
       \left(
       \frac{\sum_{i \in H} \wdiff_i \tilde{v}_{i,\mathcal{A} \cap F}}
            {\sum_{i \in H} \wdiff_i} -
            \frac{\sum_{i \in H} \tilde{v}_{i, \mathcal{A} \cap F}}{N_H}
       \right) 
       - \frac{\cov(\vhabartildehatprime, \nhprimehat)}
              {\nhprime}\\
    &= 
       \left(
              \frac{\sum_{i \in H} \wdiff_i \tilde{v}_{i, \mathcal{A} \cap F} - 
              \frac{1}{N_H}
              \sum_{i \in H} \tilde{v}_{i, \mathcal{A} \cap F} 
              \sum_{i \in H} \wdiff_i}
                   {\sum_{i \in H} \wdiff_i}
       \right)
       - \frac{\cov(\vhabartildehatprime, \nhprimehat)}
              {\nhprime}
              \\
    &= 
       \left(
        \frac{\cov_H(\tilde{v}_{i, \mathcal{A} \cap F}, \wdiff_i)}{\bar{\wdiff}}
       \right)
    - \frac{\cov(\vhabartildehatprime, \nhprimehat)}
              {\nhprime},
\end{align}
\noindent where $\cov_H(\cdot, \cdot)$ is the finite-population unit
variance among hidden population members.

}

\rptstmtonlyproof{res:vhftilde_sampling_sensitivity}

Result~\ref{res:vhftilde_sampling_sensitivity} shows that the bias in the estimator
$\vhabartildehatprime$ with imperfect relative probability weights is the sum of
two terms: one term that arises due to intrinsic bias in any ratio estimator,
and one term that arises due to differences between the imperfect weights and the
true weights. 
A large literature shows that, in many practical situations, 
the intrinsic bias in a ratio estimator will tend to be very small
(see, for example, Online Appendix~\ref{ap:ratio} and also
\citet[][Chap. 5]{sarndal_model_1992}).
When this intrinsic ratio bias is negligible, Result~\ref{res:vhftilde_sampling_sensitivity}
shows that the bias in the estimator for $\vhabartilde$ with imperfect weights can be
approximated by
\begin{align}
    \label{eqn:approxbias-impweight-vhf}
    \text{bias}(\vhabartildehatprime) &\approx 
    \frac{\cov_H(\tilde{v}_{i, \mathcal{A} \cap F}, \wdiff_i)}{\bar{\wdiff}}.
\end{align}

Similar to the discussion of Result~\ref{res:yfh_sampling_sensitivity}, 
we can obtain additional insight into Equation~\ref{eqn:approxbias-impweight-vhf} by
using the fact that
$\cov_H(\tilde{v}_{i, \mathcal{A} \cap F}, \wdiff_i) = 
\correl_H(\tilde{v}_{i, \mathcal{A} \cap F}, \wdiff_i)~
\sd_H(\tilde{v}_{i, \mathcal{A} \cap F})~\sd_H(\wdiff_i)$,
where $\sd_H(\cdot)$ is the unit finite-population standard deviation, and 
$\correl_H(\tilde{v}_{i, \mathcal{A} \cap F}, \wdiff_i)$ 
is the correlation between the $y_i$ and $\wdiff_i$.
Substituting this identity into Equation~\ref{eqn:approxbias-impweight-vhf} yields
\begin{align}
    \label{eqn:approxbias2-impweight-vhf}
    \text{bias}(\vhabartildehatprime) &\approx
    \correl_H(\tilde{v}_{i, \mathcal{A} \cap F}, \wdiff_i)~
          \sd_H(\tilde{v}_{i, \mathcal{A} \cap F})
          \frac{~\sd_H(\wdiff_i)}
         {\bar{\wdiff}}.
\end{align}
\noindent Equation~\ref{eqn:approxbias2-impweight-vhf} provides a qualitative
understanding of factors contributing to bias due to imperfect relative sampling
weights. One term, $\sd_H(\tilde{v}_{i, \mathcal{A} \cap F})$, 
is a property of the population being
studied and will typically be beyond the researcher's control.
The other two terms are related to errors in the weights:  
first, the factor $\frac{\sd_H(\wdiff_i)}{\bar{\wdiff}}$ is the coefficient
of variation in the $\wdiff_i$; it will be minimized when the standard deviation of
the $\wdiff_i$ is small, relative to the mean; that is, it will be minimized
when the errors in the weights are uniform.
Second, the magnitude of
$\correl_H(\tilde{v}_{i, \mathcal{A} \cap F}, \wdiff_i)$
will be minimized when there
is no relationship between the imperfections in the weights, $\wdiff_i$, and the
quantity of interest, $\tilde{v}_{i, \mathcal{A} \cap F}$.

Next, note that $\sd_H(\tilde{v}_{i,\mathcal{A} \cap F}) = \bar{\tilde{v}}_{H,\mathcal{A} \cap F}~
\cv_H(\tilde{v}_{i, \mathcal{A} \cap F})$, where
$\cv_H(\tilde{v}_{i, \mathcal{A} \cap F})$ is the coefficient of variation.
Equation~\ref{eqn:approxbias2-impweight-vhf} can therefore be re-arranged to
yield
\begin{align}
    \text{bias}(\vhabartildehatprime) 
    &\approx 
    \bar{\tilde{v}}_{H, \mathcal{A} \cap F} ~ K_H,
    \label{eqn:approxbias3-impweight-vhf}
\end{align}
\noindent where we have defined 
$K_H = \correl_H(\tilde{v}_{i, \mathcal{A} \cap F}, \wdiff_i) 
\cv_H(\tilde{v}_{i, \mathcal{A} \cap F}) 
\cv_H(\wdiff_i)$
as an index for the amount of error in the imperfect weights.

Using the index $K_H$ helps to clarify the meaning of the $\wdiff_i$ in
Result~\ref{res:vhftilde_sampling_sensitivity}. It may seem unintuitive to
define $\wdiff_i = \frac{\pi_i}{\piprime_i}$, since the result assumes that
neither $\pi_i$ or $\piprime_i$ is known.  But, we note that the $K_H$ in
Expression~\ref{eqn:approxbias3-impweight-vhf} is not impacted if $\epsilon_i$
are multiplied by a constant.  Therefore, if researchers find it more natural
to work with a version of $\wdiff_i$ that involves multiplying all of the
$\piprime_i$ or $\pi_i$ by a constant, then
Result~\ref{res:vhftilde_sampling_sensitivity} still applies.
For example, imagine that a researcher has sampled from the hidden population
using respondent-driven sampling, and then makes estimates under the assumption
that respondents' inclusion probabilities are proportional to their degrees
($\piprime_i \propto d_i$).  This researcher might wonder how her estimate would be
impacted if this sampling assumption was incorrect $(\piprime_i \not\propto d_i)$.
In this case, the researcher could then make the necessary assumptions and
calculate $K_H$ assuming that, for example, $(\piprime_i \propto d_i^0)$, or
$(\piprime_i \propto d_i^2)$.

Finally, since 
$\E[\widehat{\bar{\tilde{v}}}_{H, \mathcal{A} \cap F}] =
 \text{bias}(\widehat{\bar{\tilde{v}}}_{H, \mathcal{A} \cap F}) +
\bar{\tilde{v}}_{H, \mathcal{A} \cap F}$,
we can conclude that
\begin{align}
    \E[\widehat{\bar{\tilde{v}}}_{H, \mathcal{A} \cap F}] &\approx
    \bar{\tilde{v}}_{H, \mathcal{A} \cap F}(1 + K_H).
\end{align}
Therefore, Result~\ref{res:yfh_sampling_sensitivity} directly implies
Corollary~\ref{res:yfh_sampling_sensitivity_estimand}.

\stmt{cor}{res:vhftilde_sampling_sensitivity_estimand}{%
    From Result~\ref{res:yfh_sampling_sensitivity}, we also have
    \begin{align}
        \vhabartildehatprime &\leadsto 
        \bar{\tilde{v}}_{H, \mathcal{A} \cap F} (1 + K_H),
    \end{align}
    where $\leadsto$ means `is consistent and essentially unbiased for,'
    and $K_H = \correl_H(\tilde{v}_{i, \mathcal{A} \cap F}, \wdiff_i) 
               \cv_H(\tilde{v}_{i, \mathcal{A} \cap F})
               \cv_H(\wdiff_i)$ is an index for the amount of error in the
               imperfect relative sampling weights.
}

\subsubsection{Summary and results for all estimators}
\label{sec:sampling_sensitivity_all_table}


\begin{table}
\centering
\begin{tabular}{l >{\compress}p{5cm} p{5cm}}
\toprule
Quantity & Relevant results & Effective estimand \par under imperfect sampling\\
\midrule

$\widehat{y}^{\prime}_{F,\mathcal{A}} = \sum_{i \in s_F} y_{i, \mathcal{A}} / \piprime_i$ &
\begin{enumerate}[(i)]
    \item $\widehat{\bar{d}}_{F,F}$ (Result~\ref{res:kpestimator-dff})
    \item $\widehat{\bar{d}}_{U,F}$ (Result~\ref{res:kpestimator-duf})
    \item $\widehat{\phi}_F$ (Result~\ref{res:goc-phi-estimator})
    \item $\widehat{\delta}_F$ (Result~\ref{res:goc-delta-estimator})
\end{enumerate} 
&
$y_{F, \mathcal{A}}\cdot\bar{\wdiff}\cdot[1 + K_{F_1}]$
\\
$\widehat{y}^{\prime}_{F,H} = \sum_{i \in s_F} y_{i, H} / \piprime_i$ &
\begin{enumerate}[(i)]
    \item $\widehat{y}_{F,H}$ (Result~\ref{res:estimator-yft})
\end{enumerate} 
&
$y_{F, H}\cdot\bar{\wdiff}\cdot[1 + K_{F_2}]$
\\
$\widehat{\bar{\tilde{v}}}^{\prime}_{H,\mathcal{A} \cap F} = 
\frac{\sum_{i \in s_H} \tilde{v}_{i, \mathcal{A} \cap F} / (\cprime \piprime_i)}
{\sum_{i \in s_H} 1 / (\cprime \piprime_i)}$ &
\begin{enumerate}[(i)]
    \item $\widehat{\bar{v}}_{H,F}$ (Result~\ref{res:goc-v-estimator})
\end{enumerate} 
&
$\bar{\tilde{v}}_{H, \mathcal{A} \cap F}\cdot[1 + K_H]$
\\
\midrule

\bottomrule

\end{tabular}
\caption{
    Summary of estimators' sensitivity to imperfect sampling.
    Here, $s_F$ is a probability sample, $s_H$ is a relative probability sample, and
    the $K$s are indices for the magnitude of errors in the imperfect weights; 
    $K_{F_1} = \correl_F(\wdiff_i, y_{i,\mathcal{A}})~\cv_F(\wdiff_i)~
    \cv_F(y_{i, \mathcal{A}})$;
    $K_{F_2} = \correl_F(\wdiff_i, y_{i,H})~\cv_F(\wdiff_i)~\cv_F(y_{i,H})$; and
    $K_{H} = \correl_H(\wdiff_i, \tilde{v}_{i, \mathcal{A} \cap F})~\cv_H(\wdiff_i)~
    \cv_H(\tilde{v}_{i, \mathcal{A} \cap F})$.
    When the weights are exactly correct, each $K$ is equal to 0. 
}
\label{tab:sampling-robustness-all}
\end{table}


Table~\ref{tab:sampling-robustness-all} uses $K$, the index for the magnitude of
errors introduced by imperfect weights, to summarize the
results of our investigation into the impact that imperfect sampling weights will
have on three quantities that play a central role in the estimators we consider
throughout this paper: $\widehat{y}^{\prime}_{F, \mathcal{A}}$,
$\widehat{y}^{\prime}_{F,H}$, and
$\widehat{\bar{\tilde{v}}}^{\prime}_{H,\mathcal{A} \cap F}$.
The results in Table~\ref{tab:sampling-robustness-all} show how the magnitude
of the index $K$ is directly related to the bias that results from imperfect
sampling weights.

\subsection{Combined sensitivity results}
\label{ap:sensitivity-combined}

We now combine our analysis of sensitivity to reporting, network structure, 
and survey construction
(Section~\ref{sec:sampling_sensitivity_prob})
and sensitivity to sampling problems (Section~\ref{sec:sampling_sensitivity_relprob})
to derive results that describe the sensitivity of the
generalized and the modified basic scale-up estimator to all of the
conditions they rely upon.
Roughly, what we show below is that the results about estimators' sensitivity
to nonsampling conditions (such as survey construction and reporting) and
results about estimators' sensitivity to sampling conditions combine
naturally.

\subsubsection{Generalized scale-up}
\label{ap:sensitivity-gnsum}

In this section, we derive an expression
for the sensitivity of the generalized scale-up estimator to all
of the conditions it relies upon.
First, we derive a combined sensitivity result for 
$\widehat{\bar{v}}_{H,F}$ (Result~\ref{res:vhf_combined_sensitivity}).
We then make use of the combined sensitivity result for 
$\widehat{\bar{v}}_{H,F}$ to derive a combined sensitivity result
for the generalized scale-up estimator 
(Result~\ref{res:gnsum_combined_sensitivity_estimand} and 
Corollary \ref{res:gnsum_combined_sensitivity}).

\stmt{result}{res:vhf_combined_sensitivity}{%
    Suppose researchers have obtained a relative probability sample $s_H$
    to estimate $\bar{v}_{H,F}$,
    but that the researchers have imperfect relative sampling weights. 
    Call the imperfect relative sampling weights
    $\wprimeH_i = \frac{1}{\cprime \piprimeH_i}$,
    call the true probabilities of inclusion $\pi_i$, 
    and define 
    $\wdiffH_i = \frac{\piH_i}{\piprimeH_i}$.
    Call the estimator for $\bar{\tilde{v}}_{H, \mathcal{A} \cap F}$ using the
    imperfect relative sampling weights 
    $\widehat{\bar{\tilde{v}}}^{\prime}_{H, \mathcal{A} \cap F}$.

    Suppose also that the researcher's estimate of 
    $N_{\mathcal{A} \cap F}$ 
    is incorrect, so that 
    $\widehat{N}_{\mathcal{A} \cap F} = c_1 \cdot N_{\mathcal{A} \cap F}$.
    Suppose that the reporting condition
    (Equation~\ref{eq:v_tf_reporting}) of Result~\ref{res:goc-v-estimator} is
    incorrect, so that 
    $\tilde{v}_{H, \mathcal{A} \cap F} = c_2 \cdot v_{H,\mathcal{A} \cap F}$. 
    Finally, suppose that the probe alter condition is incorrect, so that
    $\frac{v_{H,{A \cap F}}}{N_{\mathcal{A} \cap F}} = c_3 \cdot \frac{v_{H,F}}{N_F}$. 
    Call the estimator for $\bar{v}_{H,F}$ under these imperfect conditions
    $\widehat{\bar{v}}_{H,F}^{\prime\star}$. 

    Then 
    \begin{align}
        \widehat{\bar{v}}_{H,F}^{\prime\star} &\leadsto
        \bar{v}_{H,F}~\frac{c_3~c_2}{c_1}~(1 + K_H)
    \end{align}
    where $\leadsto$ means `is consistent and essentially unbiased for', and
    $K_H = \correl_H(\tilde{v}_{i, \mathcal{A} \cap F}, \wdiffH_i)
           \cv_H(\tilde{v}_{i, \mathcal{A} \cap F})
           \cv_H(\wdiffH_i)$.
}

\stmtproof{res:vhf_combined_sensitivity}{
First, we note that 
Corollary~\ref{res:vhftilde_sampling_sensitivity_estimand} 
shows that
\begin{align}
\vhabartildehatprime 
\leadsto \vhabartilde (1 + K_H) 
= \frac{\tilde{v}_{H, \mathcal{A} \cap F}}{N_H} (1 + K_H).
\end{align}

The remainder of the proof follows the argument from
Results~\ref{res:vtf_sensitivity} and~\ref{res:goc-v-estimator} very closely.
Under the assumptions listed above, we can write the imperfect estimator 
$\widehat{\bar{v}}_{H,F}^{\prime\star}$ as
\begin{align}
\widehat{\bar{v}}_{H,F}^{\prime\star} &= \frac{1}{c_1}~\frac{N_F}{N_{\mathcal{A} \cap F}}~
\vhabartildehatprime
\end{align}
We follow the same steps as the proof of Results~\ref{res:goc-v-estimator}, but
each time we use one of our assumptions, the associated error is carried with
it.  So our estimator $\widehat{\bar{v}}^{\prime\star}_{H,F}$ is consistent and
essentially unbiased for 
\begin{align}
    \widehat{\bar{v}}^{\prime\star}_{H,F} &\leadsto
    (1 + K_H)\frac{1}{c_1}~\frac{N_F}{N_{\mathcal{A} \cap F}}~
    \frac{\tilde{v}_{H, \mathcal{A} \cap F}}{N_H} \\
    &=
    (1 + K_H)\frac{c_2}{c_1}~\frac{N_F}{N_{\mathcal{A} \cap F}}~
    \frac{v_{H, \mathcal{A} \cap F}}{N_H} \\
    &=
    (1 + K_H)\frac{c_3~c_2}{c_1}~\frac{N_F}{N_{\mathcal{A} \cap F}}~
    \frac{v_{H, F}}{N_H}.
\end{align}
In words, the estimand is now incorrect by $(1 + K_H)\frac{c_3~c_2}{c_1}$. Since
$\widehat{\bar{v}}_{H,F}$ is consistent and essentialy unbiased for
$\bar{v}_{H,F}$, we conclude that $\widehat{\bar{v}}_{H,F}^{\prime\star}$ is
consistent and essentially unbiased for $(1 + K_H)\frac{c_3~c_2}{c_1} \bar{v}_{H,F}$.
Note that if the conditions needed for Result~\ref{res:goc-v-estimator} hold,
then $c_1 = 1$, $c_2=1$, $c_3=1$, and $K_H=0$, then we are left with our original result for
$\widehat{\bar{v}}_{H,F}$ (Result~\ref{res:goc-v-estimator}).  
}

\rptstmtonlyproof{res:vhf_combined_sensitivity}

\stmt{result}{res:gnsum_combined_sensitivity_estimand}{%
    Suppose researchers have obtained a probability sample $s_F$
    to estimate $y_{F,H}$, but that the researchers have imperfect sampling weights. 
    Call the imperfect sampling weights
    $\wprimeF_i = \frac{1}{\piprimeF_i}$,
    call the true weights $\wF_i = \frac{1}{c \piF_i}$, 
    and define 
    $\wdiffF_i = \frac{\piF_i}{\piprimeF_i} = \frac{\wprimeF_i}{\wF_i}$.
    Call the estimator for $y_{F,H}$ under these imperfect conditions
    $y^{\prime}_{F,H}$.

    Suppose also researchers have also obtained a relative probability sample $s_H$
    to estimate $\bar{v}_{H,F}$
    but that the researchers have imperfect relative sampling weights. 
    Call the imperfect relative sampling weights
    $\wprimeH_i = \frac{1}{\cprime \piprimeH_i}$,
    call the true probabilities of inclusion $\pi_i$, 
    and define 
    $\wdiffH_i = \frac{\piH_i}{\piprimeH_i}$.
    Suppose also that the researcher's estimate of 
    $N_{\mathcal{A} \cap F}$ 
    is incorrect, so that 
    $\widehat{N}_{\mathcal{A} \cap F} = c_1 \cdot N_{\mathcal{A} \cap F}$.
    Suppose that the reporting condition
    (Equation~\ref{eq:v_tf_reporting}) of Result~\ref{res:goc-v-estimator} is
    incorrect, so that 
    $\tilde{v}_{H, \mathcal{A} \cap F} = c_2 \cdot v_{H,\mathcal{A} \cap F}$. 
    Finally, suppose that the probe alter condition is incorrect, so that
    $\frac{v_{H,{A \cap F}}}{N_{\mathcal{A} \cap F}} = c_3 \cdot \frac{v_{H,F}}{N_F}$. 
    Call the estimator for $\bar{v}_{H,F}$ under these imperfect conditions
    $\widehat{\bar{v}}_{H,F}^{\prime\star}$. 

    Finally, suppose that there are false positive reports, so that
    $y^{+}_{F,H} = \eta_F y_{F,H}$.
    Let the generalized scale-up estimator for $N_H$ in this situation be
    $\widehat{N}_{H}^{\prime\star} = 
    \frac{y^{\prime}_{F,H}}
      {\widehat{\bar{v}}_{H,F}^{\prime\star}}$.
    Then 
    \begin{align}
        \widehat{N}_{H}^{\prime\star} &\leadsto
        \frac{\wdiffFbar(1 + K_{F_1})}{1 + K_H} \frac{c_1}{c_3~c_2} \frac{1}{\eta_F} N_H,
    \end{align}
    where $\leadsto$ means `is consistent and essentially unbiased for';
    $\wdiffFbar = \frac{1}{N_F} \sum_{i \in F} \wdiffF_i$;
    $K_H = \correl_H(\tilde{v}_{i, \mathcal{A} \cap F}, \wdiffH_i)
           \cv_H(\tilde{v}_{i, \mathcal{A} \cap F})
           \cv_H(\wdiffH_i)$;
    and
    $K_{F_1} = \correl_F(y_{i,H}, \wdiffF_i) \cv_F(y_{i,H}) \cv_F(\wdiffF_i)$.
}

\stmtproof{res:gnsum_combined_sensitivity_estimand}{
    The generalized scale-up estimator is formed from a ratio of estimators, 
    one in the numerator ($\widehat{y}_{F,H}$) and 
    one in the denominator ($\widehat{\bar{v}}_{H,F}$).
    We have already derived results for each of the numerator and the denominator 
    separately; our approach will therefore be to combine them.
    We must account for the fact that,
    in addition to the assumptions required for the estimator of the numerator and
    the denominator, the generalized scale-up estimator also requires the additional
    condition that there are no false positive reports.

    We begin with the denominator, $\widehat{\bar{v}}_{H,F}$. 
    Result~\ref{res:vhf_combined_sensitivity} shows that
    \begin{align}
        \widehat{\bar{v}}_{H,F}^{\prime\star} &\leadsto
        \bar{v}_{H,F}~\frac{c_3~c_2}{c_1}~(1 + K_H),
        \label{eqn:gnsum_combined_denom}
    \end{align}
    where 
    $K_H = \correl_H(\tilde{v}_{i, \mathcal{A} \cap F}, \wdiffH_i)
           \cv(\tilde{v}_{i, \mathcal{A} \cap F})
           \cv(\wdiffH_i)$.
    Thus, Expression \ref{eqn:gnsum_combined_denom} shows the 
    sensitivity of the denominator of the generalized scale-up
    estimator to violations of all of the conditions it relies upon.

    Turning now to the numerator of the generalized scale-up estimator, 
    Corollary~\ref{res:yfh_sampling_sensitivity_estimand} shows that
    \begin{align}
        \yfhhatprime &\leadsto 
        y_{F,H}\cdot\bar{\wdiff}\cdot(1 + K_{F_1}),
        \label{eqn:gnsum_combined_num}
    \end{align}
    where 
    $K_{F_1} = \correl_F(y_{i,H}, \wdiffF_i) \cv_F(y_{i,H}) \cv_F(\wdiffF_i)$.
    Thus, Expression \ref{eqn:gnsum_combined_num} shows
    sensitivity of the numerator of the generalized scale-up
    estimator to violations of all of the conditions it relies upon.

    Using the fact that a ratio estimator is consistent and essentially
    unbiased for the ratio of the estimand of its numerator and denominator
    (see Online Appendix~\ref{ap:ratio} and \citet[][chap. 5]{sarndal_model_1992}), we
    therefore have
    \begin{align}
        \widehat{N}_{H}^{\prime\star} &\leadsto
        \frac{\wdiffFbar(1 + K_{F_1})}{1 + K_H} \frac{~c_1}{c_3~c_2} 
        \frac{y_{F,H}}{\bar{v}_{H,F}}.
        \label{eqn:gnsum_combined_intermed1}
    \end{align}
    Finally, by definition we have $y_{F,H} = y^{+}_{F,H} / \eta_F$,
    which we can substitute into Expression~\ref{eqn:gnsum_combined_intermed1}
    to produce
    \begin{align}
        \widehat{N}_{H}^{\prime\star} &\leadsto
        \frac{\wdiffFbar(1 + K_{F_1})}{1 + K_H} \frac{c_1}{c_3~c_2~\eta_F} 
        \frac{y^{+}_{F,H}}{\bar{v}_{H,F}}.
    \end{align}
    By the argument in Section~\ref{sec:framework} and
    Appendix~\ref{ap:generalized}, $N_H = y^{+}_{F,H} / \bar{v}_{H,F}$.
    Substituting $N_H$ for $y^{+}_{F,H} / \bar{v}_{H,F}$ in 
    the expression above completes the proof.
}

\rptstmtonlyproof{res:gnsum_combined_sensitivity_estimand}

\stmt{cor}{res:gnsum_combined_sensitivity}{%
    From Result~\ref{res:gnsum_combined_sensitivity_estimand}, it follows
    that, for the generalized scale-up estimator,
    \begin{align}
        \widehat{N}_{H}^{\prime\star} \cdot
        \underbrace{\frac{1 + K_H}{\wdiffFbar(1 + K_{F_1})}}_{%
          \substack{\text{sampling} \\ \text{conditions}}   
        } \cdot
        \underbrace{\frac{c_3~c_2}{c_1}}_{%
          \substack{\text{visibility} \\ \text{estimator} \\ \text{conditions}}
        } \cdot
        \underbrace{ \eta_F \phantom{\frac{}{}} }_{%
          \substack{\text{no false} \\ \text{positives} \\ \text{condition}}
        }
        &\leadsto 
        N_H.
    \label{eqn:gnsum_combined_sensitivity}
    \end{align}
}

Researchers who wish to conduct a sensitivity analysis for estimates made using the
generalized scale-up method can therefore (1) assume values or ranges of values for 
$K_H$, $\wdiffFbar$,
$K_{F_1}$, $c_1$, $c_2$, $c_3$, and $\eta_F$ 
and (2) use Corollary~\ref{res:gnsum_combined_sensitivity} to determine 
the resulting values of $N_H$.
Thus, researchers can use this approach to explore the sensitivity of their
estimates to all of the assumptions they had to make.

\subsubsection{Modified basic scale-up}
\label{ap:sensitivity-nsum}

In this section, we develop an expression for the sensitivity of the modified
basic scale-up estimator to all of the conditions it relies upon.
First, we derive a combined sensitivity result for 
$\widehat{\bar{d}}_{F,F}$ (Result~\ref{res:kp_combined_sensitivity}).
We then make use of the combined sensitivity result for 
$\widehat{\bar{d}}_{F,F}$ to derive a combined sensitivity result
for the modified basic scale-up estimator 
(Result~\ref{res:mbnsum_combined_sensitivity_estimand} and 
Corollary~\ref{res:mbnsum_combined_sensitivity}).

\stmt{result}{res:kp_combined_sensitivity}{%
    Suppose researchers have obtained a probability sample $s_F$
    to estimate $\bar{d}_{F,F}$; 
    however, suppose that the researchers have imperfect sampling weights. 
    Call the imperfect sampling weights
    $\wprimeF_i = \frac{1}{\piprimeF_i}$,
    call the true weights $\wF_i = \frac{1}{c \piF_i}$, 
    and define 
    $\wdiffF_i = \frac{\piF_i}{\piprimeF_i}$.
    Let the estimator for $y_{F, \mathcal{A}}$ using these imperfect weights be
    $\widehat{y}^{\prime}_{F, \mathcal{A}}$.

    Suppose also that researchers have chosen a set of probe alters $\mathcal{A}$
    in order to use the known population method
    (Result~\ref{res:kpestimator-dff}).
    However, suppose that the researcher's estimate of 
    $N_{\mathcal{A}}$ is incorrect, so that 
    $\widehat{N}_{\mathcal{A}} = c_1 \cdot N_{\mathcal{A}}$.
    Suppose also that the reporting condition
    (Equation~\ref{eqn:kpcondition-dff-reporting}) of Result~\ref{res:kpestimator-dff} is
    incorrect, so that 
    $y_{F, \mathcal{A}} = c_2 \cdot d_{F,\mathcal{A}}$. 
    Finally, suppose that the probe alter condition 
    (Equation~\ref{eqn:kpcondition-dff})
    of Result~\ref{res:kpestimator-dff} is incorrect, so that
    $\bar{d}_{\mathcal{A}, F} = c_3 \cdot \bar{d}_{F,F}$. 
    Call the estimator for $\bar{d}_{F,F}$ under these imperfect conditions
    $\widehat{\bar{d}}_{F,F}^{\prime\star}$. 

    Let the known population estimator for $\bar{d}_{F,F}$ (Result~\ref{res:kpestimator-dff})
    under these imperfect conditions be $\widehat{\bar{d}}_{F,F}^{\prime\star}$. 
    Then 
    \begin{align}
        \widehat{\bar{d}}_{F,F}^{\prime\star} &\rightarrow 
        \wdiffFbar (1 + K_{F_2})\cdot
       \frac{ c_2~c_3 }{ c_1 }\cdot
       \bar{d}_{F,F},
    \end{align}
    where $\leadstounbiased$ means `is consistent and unbiased for', and
    $K_{F_2} = \correl_F(y_{i,\mathcal{A}}, \wdiffF_i) 
               \cv_F(y_{i,\mathcal{A}}) \cv_F(\wdiffF_i)$.
}

\stmtproof{res:kp_combined_sensitivity}{
    Under the assumptions above, we can write the imperfect estimator
    $\widehat{\bar{d}}^{\prime\star}_{F,F}$ as
    \begin{align}
        \widehat{\bar{d}}^{\prime\star}_{F,F} &=
        \frac{1}{c_1} \cdot \frac{\widehat{y}^{\prime}_{F, \mathcal{A}}}{N_{\mathcal{A}}}
    \end{align}
    Using the exact same argument as Result~\ref{res:yfh_sampling_sensitivity} and
    Corollary~\ref{res:yfh_sampling_sensitivity_estimand}, we have
    \begin{align}
        \widehat{y}^{\prime}_{F,\mathcal{A}} &\leadstounbiased
        \wdiffFbar (1 + K_{F_2}) \cdot y_{F, \mathcal{A}}.
    \end{align}
    Applying this to the imperfect estimator $\widehat{\bar{d}}^{\prime\star}_{F,F}$, 
    we have
    \begin{align}
        \widehat{\bar{d}}^{\prime\star}_{F,F} &\leadstounbiased
        \wdiffFbar (1 + K_{F_2}) \cdot
        \frac{1}{c_1} \cdot \frac{y_{F, \mathcal{A}}}{N_{\mathcal{A}}}
        =
        \wdiffFbar (1 + K_{F_2}) \cdot
        \frac{1}{c_1} \cdot \bar{y}_{F, \mathcal{A}}.
    \end{align}
    We will obtain the rest of the result by following the argument of
    Result~\ref{res:kpestimator-dff} closely, but carrying the errors from the
    conditions that are not met through with each step. 
    First, by assumption, $\bar{y}_{F, \mathcal{A}} = c_2 \bar{d}_{F, \mathcal{A}}$,
    yielding
    \begin{align}
        \widehat{\bar{d}}^{\prime\star}_{F,F} &\leadstounbiased
        \wdiffFbar (1 + K_{F_2}) \cdot
        \frac{c_2}{c_1} \cdot \bar{d}_{F, \mathcal{A}}.
    \end{align}
    Next, again by assumption, $\bar{d}_{F, \mathcal{A}} = c_3 \bar{d}_{F,F}$,
    so we have
    \begin{align}
        \widehat{\bar{d}}^{\prime\star}_{F,F} &\leadstounbiased
        \wdiffFbar (1 + K_{F_2}) \cdot
        \frac{c_2~c_3}{c_1} \cdot \bar{d}_{F, F},
    \end{align}
    which is our result.
}

\rptstmtonlyproof{res:kp_combined_sensitivity}

\stmt{result}{res:mbnsum_combined_sensitivity_estimand}{%
    Suppose researchers have obtained a probability sample $s_F$
    to estimate $y_{F,H}$ and $\bar{d}_{F,F}$ in order to produce estimates
    from the modified basic scale-up method.
    However, suppose that the researchers have imperfect sampling weights. 
    Call the imperfect sampling weights
    $\wprimeF_i = \frac{1}{\piprimeF_i}$,
    call the true weights $\wF_i = \frac{1}{c \piF_i}$, 
    and define 
    $\wdiffF_i = \frac{\piF_i}{\piprimeF_i} = \frac{\wprimeF_i}{\wF_i}$.
    Let the estimator for $y_{F,H}$ using these imperfect weights be
    $y^{\prime}_{F,H}$.

    Suppose also that researchers have chosen a set of probe alters $\mathcal{A}$
    in order to use the known population method
    (Result~\ref{res:kpestimator-dff}).
    However, suppose that the researcher's estimate of 
    $N_{\mathcal{A}}$ is incorrect, so that 
    $\widehat{N}_{\mathcal{A}} = c_1 \cdot N_{\mathcal{A}}$.
    Suppose also that the reporting condition
    (Equation~\ref{eqn:kpcondition-dff-reporting}) of Result~\ref{res:kpestimator-dff} is
    incorrect, so that 
    $y_{F, \mathcal{A}} = c_2 \cdot d_{F,\mathcal{A}}$. 
    Suppose also that the probe alter condition 
    (Equation~\ref{eqn:kpcondition-dff})
    of Result~\ref{res:kpestimator-dff} is incorrect, so that
    $\bar{d}_{\mathcal{A}, F} = c_3 \cdot \bar{d}_{F,F}$. 
    Call the estimator for $\bar{d}_{F,F}$ under these imperfect conditions
    $\widehat{\bar{d}}_{F,F}^{\prime\star}$. 

    Finally, suppose that the basic scale-up conditions do not hold; that is,
    suppose that there are false positive reports, so that
    $y^{+}_{F,H} = \eta_F y_{F,H}$;
    suppose that there are false negative reports, so that
    $\bar{v}_{H,F} = \tau_F \bar{d}_{H,F}$;
    and suppose that the average personal network size of hidden population members
    is not equal to the average personal network size of frame population members,
    so that $\bar{d}_{H,F} = \delta_F \bar{d}_{F,F}$.

    Let the modified basic scale-up estimator for $N_H$ in this situation be
    \begin{align}
       \widehat{N}_{H}^{\prime\star} &= 
       \frac{\widehat{y}^{\prime}_{F,H}}{\widehat{\bar{d}}^{\prime\star}_{F,F}}.
    \end{align}
    Then 
    \begin{align}
       \widehat{N}_{H}^{\prime\star} &\leadsto 
       \frac{(1 + K_{F_1})}{(1 + K_{F_2})}\cdot
       \frac{ c_1 }{ c_2~c_3 }\cdot
       \frac{\tau_F~\delta_F}{\eta_F} \cdot
       N_H,
    \end{align}
    where $\leadsto$ means `is consistent and essentially unbiased for';
    $K_{F_1} = \correl_F(y_{i,H}, \wdiffF_i) \cv_F(y_{i,H}) \cv_F(\wdiffF_i)$; and
    $K_{F_2} = \correl_F(y_{i,\mathcal{A}}, \wdiffF_i) 
               \cv_F(y_{i,\mathcal{A}}) \cv_F(\wdiffF_i)$.
}

\stmtproof{res:mbnsum_combined_sensitivity_estimand}{

    The modified basic scale-up estimator is formed from a ratio of estimators for the
    numerator ($y_{F,H}$) and denominator ($\bar{d}_{F,F}$).
    We have already derived results for each of the numerator and the denominator 
    separately; our approach will therefore be to combine them. 
    We must account for the fact that,
    in addition to the assumptions required for the estimator of the numerator and
    the denominator, the modified basic scale-up estimator also requires the additional
    conditions that there are no false positive reports, that there are no false negative
    reports, and that the degree ratio is one.

    For the numerator, Result~\ref{res:kp_combined_sensitivity} shows that
    \begin{align}
        \widehat{\bar{d}}^{\prime\star}_{F,F} &\leadstounbiased
        \wdiffFbar (1 + K_{F_2}) \cdot
        \frac{c_2~c_3}{c_1} \cdot \bar{d}_{F, F}.
        \label{eqn:mbnsum_combined_denom}
    \end{align}
    Thus, Expression \ref{eqn:mbnsum_combined_denom} shows
    sensitivity of the denominator of the modified basic scale-up
    estimator to violations of all of the conditions it relies upon.

    Turning now to the numerator of the modified basic scale-up estimator, 
    Corollary~\ref{res:yfh_sampling_sensitivity_estimand} shows that
    \begin{align}
        \yfhhatprime &\leadstounbiased 
        y_{F,H}\cdot\bar{\wdiff}\cdot(1 + K_{F_1}),
        \label{eqn:mbnsum_combined_num}
    \end{align}
    where 
    $K_{F_1} = \correl_F(y_{i,H}, \wdiffF_i) \cv_F(y_{i,H}) \cv_F(\wdiffF_i)$.
    Thus, Expression \ref{eqn:mbnsum_combined_num} shows
    sensitivity of the numerator of the modified basic scale-up
    estimator to violations of all of the conditions it relies upon.

    Using the fact that a ratio estimator is consistent and essentially
    unbiased for the ratio of the estimand of its numerator and denominator
    (see Online Appendix~\ref{ap:ratio} and \citet[][chap. 5]{sarndal_model_1992}), we
    therefore have
    \begin{align}
        \widehat{N}_{H}^{\prime\star} &\leadsto
        \frac{(1 + K_{F_1})}{(1 + K_{F_2})} \cdot
        \frac{c_1}{c_2~c_3} \cdot
        \frac{y_{F,H}}{\bar{d}_{F,F}}.
        \label{eqn:mbnsum_combined_intermed1}
    \end{align}

    Finally, by assumption, we have $y_{F,H} = y^{+}_{F,H} / \eta_F$,
    and $\bar{v}_{H,F} = \bar{d}_{F,F} / (\tau_F~\delta_F)$. 
    Substituting these assumptions into Expression~\ref{eqn:mbnsum_combined_intermed1}
    produces
    \begin{align}
        \widehat{N}_{H}^{\prime\star} &\leadsto
        \frac{(1 + K_{F_1})}{(1 + K_{F_2})} \cdot
        \frac{c_1}{c_2~c_3} \cdot
        \frac{\tau_F~\delta_F}{\eta_F} \cdot
        \frac{y^{+}_{F,H}}{\bar{v}_{F,F}}
        \label{eqn:mbnsum_combined_intermed1}
    \end{align}
    By the argument in Section~\ref{sec:framework} and
    Appendix~\ref{ap:generalized}, $N_H = y^{+}_{F,H} / \bar{v}_{H,F}$.
    Substituting $N_H$ for $y^{+}_{F,H} / \bar{v}_{H,F}$ in 
    the expression above completes the proof.
}

\rptstmtonlyproof{res:mbnsum_combined_sensitivity_estimand}

\stmt{cor}{res:mbnsum_combined_sensitivity}{%
    From Result~\ref{res:mbnsum_combined_sensitivity_estimand}, it follows
    that, for the modified basic scale-up estimator,
    \begin{align}
       \widehat{N}_{H}^{\prime\star} \cdot
       \underbrace{\frac{(1 + K_{F_2})}{(1 + K_{F_1})}}_{%
           \substack{\text{sampling} \\ \text{conditions}}   
       } \cdot
       \underbrace{\frac{ c_2~c_3 }{ c_1 }}_{%
           \substack{\text{known} \\ \text{population} \\ \text{conditions}}   
       } \cdot
       \underbrace{\frac{\eta_F}{\tau_F~\delta_F}}_{%
           \substack{\text{basic} \\ \text{scale-up} \\ \text{conditions}}   
       }
       &\leadsto
       N_H.
    \label{eqn:mbnsum_combined_sensitivity}
    \end{align}
}

Researchers who wish to conduct a sensitivity analysis for estimates made using the
generalized scale-up method can therefore (1) assume values or ranges of values for 
$K_{F_1}$, $K_{F_2}$, 
$c_1$, $c_2$, $c_3$, $\delta_F$, $\tau_F$, and $\eta_F$;
and (2) use Corollary~\ref{res:mbnsum_combined_sensitivity} to determine 
the resulting values of $N_H$.
Thus, researchers can use this approach to explore the sensitivity of their
estimates to all of the assumptions they had to make, individually and jointly.

\section{Approximate unbiasedness of compound ratio estimators}
\label{ap:ratio}

\subsection{Overview}

Several of the estimators we propose are nonlinear, which means that they are
not design-unbiased~\citep{sarndal_model_1992}.  While ratio estimators are
common in survey sampling and the bias of these estimators is commonly regarded
as insignificant~\citep{sarndal_model_1992}, several of the estimators we
propose are somewhat more complex than standard ratio estimators.  In fact, all
of our nonlinear estimators turn out to all be special cases of a ratio of
ratios (Table~\ref{tab:nlestimatorform}), which is also known as a double ratio
estimator \citep{rao_double_1968a}.  Any double ratio can be written

\begin{align}
\label{eqn:doubleratio}
R_d &= \frac{R_1}{R_0} = 
\frac{\frac{\bar{y}_1}{\bar{x}_1}}{\frac{\bar{y}_0}{\bar{x}_0}} 
= \frac{\bar{y}_1 \bar{x}_0}{\bar{x}_1 \bar{y}_0}.
\end{align}

\noindent If we have unbiased estimators for each of the four terms, we can
estimate $R_d$ by 
\begin{equation}
\label{eqn:doubleratioestimator}
\widehat{r}_d = \frac{\widehat{\bar{y}}_1 \widehat{\bar{x}}_0}{\widehat{\bar{x}}_1 \widehat{\bar{y}}_0}.
\end{equation}

In this appendix we investigate when we can expect the biases in our estimators
to be small enough to be negligible; we conclude that, in practice, the bias
is typically negligible when compared to sampling and non-sampling error.

{
    \renewcommand{\arraystretch}{1.5}
\begin{sidewaystable}[!ph]
    \fontsize{8}{8}\selectfont
	\centering
    \begin{tabular}{c c l l l l l l l}
    \hline
    Estimator & 
    Reference & 
    Form & 
    $\widehat{\bar{x}}_0$ &
    $\widehat{\bar{y}}_1$ & 
    $\widehat{\bar{x}}_1$ & 
    $\widehat{\bar{y}}_0$ &
    Approx. rel. bias
    \\
    \hline
    $\widehat{\phi}_F$ & 
    Res. \ref{res:goc-phi-estimator} & 
    $K \widehat{\bar{x}}_0 / \widehat{\bar{y}}_0$ &
    $\displaystyle \sum_{i \in s_F} y_{i, \mathcal{A}_{F_1}}/\pi_i$ &    
    - & 
    - & 
    $\displaystyle \sum_{i \in s_F} y_{i, \mathcal{A}_{F_2}} /\pi_i$ &
    $C_{\widehat{\bar{y}}_0}^2 - \relcovbar{y}{0}{x}{0}$
    \\    
    $\widehat{\bar{v}}_{H,F}$ & 
    Res. \ref{res:goc-v-estimator} & 
    $K \widehat{\bar{x}}_0 / \widehat{\bar{y}}_0$ &
    $\displaystyle \sum_{i \in s_H} \widetilde{v}_{i, \mathcal{A}_H \cap F}/ c\pi_i$ &    
    - & 
    - & 
    $\displaystyle \sum_{i \in s_H} 1/c \pi_i$ &
    $C_{\widehat{\bar{y}}_0}^2 - \relcovbar{y}{0}{x}{0}$
    \\
    $\widehat{\bar{d}}_{H,F}$ & 
    Res. \ref{res:goc-df-estimator} & 
    $K \widehat{\bar{x}}_0 / \widehat{\bar{y}}_0$ &
    $\displaystyle \sum_{i \in s_H} y_{i, \mathcal{A}_H \cap F}/ c\pi_i$ &    
    - & 
    - & 
    $\displaystyle \sum_{i \in s_H} 1/c \pi_i$ &
    $C_{\widehat{\bar{y}}_0}^2 - \relcovbar{y}{0}{x}{0}$
    \\
    $\widehat{\delta}_F$ & 
    Res. \ref{res:goc-delta-estimator} & 
    $K \widehat{\bar{x}}_0 / (\widehat{\bar{y}}_0~\widehat{\bar{x}}_1)$ &
    $\displaystyle \sum_{i \in s_H} y_{i, \mathcal{A}_H \cap F} / c\pi_i$ &    
    - & 
    $\displaystyle \sum_{i \in s_F} y_{i, \mathcal{A}_F}/\pi_i$ & 
    $\displaystyle \sum_{i \in s_H} 1 / c\pi_i$ &
    $C_{\widehat{\bar{y}}_0}^2 + C_{\widehat{\bar{x}}_1}^2 - \relcovbar{y}{0}{x}{0}$
    \\
    $\widehat{\TPR}_F$ & 
    Res. \ref{res:goc-tpr-estimator} & 
    $K \widehat{\bar{x}}_0 / (\widehat{\bar{y}}_0 ~ \widehat{\bar{x}}_1) $ &
    $\displaystyle \sum_{i \in s_H} \tilde{v}_{i, \mathcal{A}_H \cap F} / c\pi_i$ &    
    - & 
    $\displaystyle \sum_{i \in s_H} y_{i, \mathcal{A}_H \cap F} / c\pi_i$ &
    $\displaystyle \sum_{i \in s_H} 1/c\pi_i$ & 
    $C_{\widehat{\bar{y}}_0}^2 + C_{\widehat{\bar{x}}_1}^2 - \relcovbar{y}{0}{x}{0}$
    \\    
    $\widehat{N}_H$ & 
    Res. \ref{res:goc-gnsum-new} & 
    $K \widehat{\bar{y}}_1 \widehat{\bar{x}}_0/ \widehat{\bar{y}}_0$ &
    $\displaystyle \sum_{i \in s_H} 1/c\pi_i$ &    
    $\displaystyle \sum_{i \in s_F} y_{i,H} / \pi_i$ &
    - & 
    $\displaystyle \sum_{i \in s_H} \tilde{v}_{i, \mathcal{A}_H \cap F} / c\pi_i$ &
    $C_{\widehat{\bar{y}}_0}^2 - \relcovbar{y}{0}{x}{0}$
    \\
    $\widehat{N}_H$ & 
    Res. \ref{res:goc-gnsum} & 
    $K \widehat{\bar{x}}_0 / \widehat{\bar{y}}_0$ &
    $\displaystyle \sum_{i \in s_F} y_{i,H} / \pi_i$ &
    - & 
    - & 
    $\displaystyle \sum_{i \in s_F} \sum_{j} y_{i, A_j} / \pi_i$&
    $C_{\widehat{\bar{y}}_0}^2 - \relcovbar{y}{0}{x}{0}$
    \\    
    \hline
    \end{tabular}
    \caption{Description of the general form of the nonlinear estimators we propose. $K$ is a constant, $\widehat{\bar{y}}_1$ and $\widehat{\bar{x}}_1$ are taken from $s_F$, while $\widehat{\bar{x}}_0$ and $\widehat{\bar{y}}_0$ are taken from $s_H$. Our nonlinear estimators are all special cases of the double ratio estimator, which we define and discuss below. Note that the estimator for $\widehat{N}_H$ that involves adjusting a basic scale-up estimate (Result \ref{res:goc-gnsum}) would, in practice, take these adjustment factors from other studies; we therefore assume that these adjustment factors are independent of the quantities that go into the scale-up estimate, and treat them as constants. }
    \label{tab:nlestimatorform}    
\end{sidewaystable}}

\subsection{The general case}

We will focus on the relative bias in our estimator, $\widehat{r}_d$.  The
relative bias is given by
\begin{equation}
B_d = \frac{\mathbb{E}[\widehat{r}_d] - R_d}{R_d}.
\label{eqn:relbiasdefn}
\end{equation}
\noindent $B_d$ expresses the bias in our estimator $\widehat{r}_d$ in terms of
the true value; a relative bias of $0.5$, for example, means that our estimator
is typically 0.5 times bigger than the true value. This is a natural quantity
to consider because estimators that have small relative bias have small bias in
substantive terms.

Our approach will be to follow \cite{rao_double_1968a} in using a Taylor series
to form an approximation to the relative bias. This is accomplished in
Result~\ref{res:multiple-ratio}.

\stmt{result}{res:multiple-ratio} {
\citep{rao_double_1968a} If $\widehat{\bar{x}}_0$, $\widehat{\bar{x}}_1$,
$\widehat{\bar{y}}_0$, and $\widehat{\bar{y}}_1$ are unbiased estimators, and
$|(\widehat{\bar{x}}_1 - \bar{x}_1) / \bar{x}_1| < 1$ and
$|(\widehat{\bar{y}}_0 - \bar{y}_0) / \bar{y}_0| < 1$, then the relative bias
of the double ratio estimator, $B_d$, is approximated by
\begin{align}
\label{eqn:bdapprox}
B_d = \frac{\mathbb{E}[\widehat{r}_d] - R}{R} &\approx B_d' = 
\relcovbar{x}{1}{y}{0} - \relcovbar{x}{1}{y}{1} - \relcovbar{y}{0}{y}{1} - \relcovbar{x}{0}{x}{1} - \relcovbar{x}{0}{y}{0} + \relcovbar{y}{1}{x}{0} + C^2_{\bar{y}_0} + C^2_{\bar{x}_1},
\end{align}
\noindent where $C_{\widehat{\bar{x}},\widehat{\bar{y}}} = \frac{\text{cov}(\widehat{\bar{x}},\widehat{\bar{y}})}{\bar{x} \bar{y}}$ is the relative covariance between $\widehat{\bar{x}}$ and $\widehat{\bar{y}}$, and $C^2_{\widehat{\bar{y}}} = \frac{\text{var}(\widehat{\bar{y}})}{\widehat{\bar{y}}^2}$.  
}
\stmtproof{res:multiple-ratio}{
Define
\begin{equation}
\delta_{\widehat{\bar{x}}_0} = \frac{\widehat{\bar{x}}_0 - \bar{x}_0}{\bar{x}_0},
\end{equation}
\noindent with analogous definitions for $\delta_{\widehat{\bar{x}}_1}$, $\delta_{\widehat{\bar{y}}_1}$, and $\delta_{\widehat{\bar{y}}_0}$. We can express $r_d$ as
\begin{equation}
\widehat{r}_d = R \frac{(1 + \delta_{\widehat{\bar{y}}_1})(1 + \delta_{\widehat{\bar{x}}_0})}{(1 + \delta_{\widehat{\bar{y}}_0})(1 + \delta_{\widehat{\bar{x}}_1})}.
\end{equation}
\noindent The relative bias then becomes
\begin{equation}
\label{eqn:relbiasdelta}
B_d = \frac{\mathbb{E}[\widehat{r}_d] - R}{R} = \mathbb{E}\left[\frac{(1 + \delta_{\widehat{\bar{y}}_1})(1 + \delta_{\widehat{\bar{x}}_0})}{(1 + \delta_{\widehat{\bar{y}}_0})(1 + \delta_{\widehat{\bar{x}}_1})} \right] - 1.
\end{equation}
\noindent The strategy is now to expand the two factors in the denominator and to then discard high-order terms. What remains will be an approximation to the true relative bias. 

Recall that if $|x| < 1$ then $\frac{1}{1 - x} = \sum_{i=0}^\infty x^i$ and, in particular, $\frac{1}{1+x} = 1 - x^2 + x^3 - \cdots$. We'll make use of this expansion for the two factors in the denominator of Equation~\ref{eqn:relbiasdelta}; that is, we assume that $|\delta_{\bar{y}_0}| < 1$ and $|\delta_{\bar{x}_1}| < 1$. Then we have
\begin{align}
B_d = \mathbb{E}\left[(1 + \delta_{\widehat{\bar{y}}_1}) (1 + \delta_{\widehat{\bar{x}}_0}) (1 - \delta_{\widehat{\bar{y}}_0} + \delta_{\widehat{\bar{y}}_0}^2 - \cdots)
(1 - \delta_{\widehat{\bar{x}}_1} + \delta_{\widehat{\bar{x}}_1}^2 - \cdots)\right] - 1
\end{align}
\noindent If we multiply this out and retain only terms up to order 2, we obtain the following approximation:
\begin{align}
\label{eqn:expandedapprox}
B_d \approx \mathbb{E}\left[\relerr{x}{1}\relerr{y}{0} + \relerr{x}{0}\relerr{y}{1} - \relerr{x}{0}\relerr{y}{0} - \relerr{x}{0}\relerr{x}{1} - \relerr{x}{1}\relerr{y}{1} - \relerr{y}{0}\relerr{y}{1} + \relerr{x}{0} + \relerr{y}{1} - \relerr{x}{1} - \relerr{y}{0} - \relerr{y}{0}^2 - \relerr{x}{1}^2\right].
\end{align}
Since we assumed that the estimators for the individual components of $r_d$ are
unbiased, we know that
\begin{align}
\mathbb{E}[\relerr{x}{1}] = 0,
\end{align}
\noindent We can also determine that
\begin{align}
\mathbb{E}[\relerr{x}{1}\relerr{y}{1}] = \frac{\text{cov}(\widehat{\bar{x}}_1, \widehat{\bar{y}}_1)}{\bar{x}_1 \bar{y}_1}, 
\end{align}
and, that
\begin{align}
\mathbb{E}[\relerr{x}{1}^2] = \frac{\text{var}(\widehat{\bar{x}}_1)}{\bar{x}_1^2}.
\end{align}
Applying these relationships to Equation~\ref{eqn:expandedapprox}, we find
\begin{align}
\label{eqn:reswithbar}
B_d \approx &
\relcovbar{x}{0}{y}{1} + \relcovbar{x}{1}{y}{0} - \relcovbar{x}{0}{x}{1} - \relcovbar{x}{0}{y}{0} - \relcovbar{x}{1}{y}{1} - \relcovbar{y}{0}{y}{1} + C_{\bar{x}_1}^2 + C_{\bar{y}_0}^2,
\end{align}
which is our result.
}

\rptstmtonlyproof{res:multiple-ratio}


Result~\ref{res:multiple-ratio} is useful because it reveals the behavior of
double ratio estimators in quite general contexts. To understand what it says a
bit more intuitively, note that Result~\ref{res:multiple-ratio} is framed in
terms of the relative covariances and variances of the \emph{estimators}
$\widehat{\bar{x}}_0$, $\widehat{\bar{x}}_1$, $\widehat{\bar{y}}_0$, and
$\widehat{\bar{y}}_1$. In the special case of simple random sampling with
replacement, we can re-write the approximation in terms of the finite
population variances and covariances and a constant, $\kappa$:
\begin{align}
B_d' =
\kappa
\left[ \relcov{x}{1}{y}{0} - \relcov{x}{1}{y}{1} - \relcov{y}{0}{y}{1} - \relcov{x}{0}{x}{1} - \relcov{x}{0}{y}{0} + \relcov{y}{1}{x}{0} + C^2_{y_0} + C^2_{x_1}\right],
\label{eqn:approxrelbiassi}
\end{align}
where $\kappa=\left(\frac{1}{n} - \frac{1}{N} \right)$, $n$ is our sample size,
and $N$ is the size of the population. In the case of
simple random sampling, the relative bias depends upon the finite population
variances of the underlying population values and the size of our sample.

For designs other than simple random sampling, there is no analogous expression
as simple as Equation~\ref{eqn:approxrelbiassi}. However, speaking roughly, if
we have an idea that our sampling plan has a typical design effect
($\text{deff}$) for the quantities inside the square brackets in
Equation~\ref{eqn:approxrelbiassi}, then we can see that we would simply
replace the $\kappa$ in Equation~\ref{eqn:approxrelbiassi} by $(\kappa \cdot
\text{deff})$ in order to get a sense of the approximate relative bias.

Notice, also, that Result~\ref{res:multiple-ratio} is framed largely in terms
of relative covariances. When we apply Result~\ref{res:multiple-ratio}, we will
often make use of the fact that the relative covariances can be expressed in
terms of correlations and coefficients of variation as follows:

\begin{align}
C_{\widehat{\bar{x}},\widehat{\bar{y}}} = 
\frac{\text{cov}(\widehat{\bar{x}},\widehat{\bar{y}})}
     {\bar{x} \bar{y}} &= 
\frac{\rho_{\widehat{\bar{x}},\widehat{\bar{y}}}
      \sqrt{\text{var}(\widehat{\bar{x}})} 
      \sqrt{\text{var}(\widehat{\bar{y}})}}
     {\bar{x} \bar{y}}\\
&= \rho_{\widehat{\bar{x}},\widehat{\bar{y}}}~
\text{cv}(\widehat{\bar{x}})~
\text{cv}(\widehat{\bar{y}}),
\end{align}

\noindent where $\rho_{\widehat{\bar{x}},\widehat{\bar{y}}}$ is the correlation
between the estimators $\widehat{\bar{x}}$ and $\widehat{\bar{y}}$, and
$\text{cv}(\widehat{\bar{x}}) =
\frac{\sqrt{\text{var}(\widehat{\bar{x}})}}{\bar{x}}$ is the coefficient of
variation of the estimator $\widehat{\bar{x}}$. We will also make use of the
fact that $C_{\widehat{\bar{x}}}^2 = \text{cv}(\widehat{\bar{x}})^2$.

\subsection{Applying Result~\ref{res:multiple-ratio} to scale-up}

We now apply Result~\ref{res:multiple-ratio} to understand the biases in the
nonlinear estimators we propose for realistic situations.  For each particular
estimator, we can simplify the expression in Result~\ref{res:multiple-ratio}.
In order to do so, we first remove terms that do not appear in the estimator
itself (for example, in $\widehat{\delta}_F$, there is no
$\widehat{\bar{y}}_1$). Additionally, we assume that the estimates produced
from a sample from the frame population and a sample from the hidden population
will be independent of one another, meaning that their correlation will be 0.
Table~\ref{tab:nlestimatorform} summarizes the nonlinear estimators we propose,
along with the specific version of the approximate relative bias from
Result~\ref{res:multiple-ratio} that applies.

Finally, in order to give a sense of the magnitude of the coefficients of
variation and correlations found in real studies, we estimated the quantities
that go into the approximate relative bias from the studies available to us.
Table~\ref{tab:nsumdcvs} shows the coefficients of variation for the estimated
degree (the values of $\widehat{\bar{x}}_1$ for $\widehat{\delta}_F$) in
surveys from Rwanda, the United States, and Curitiba, Brazil.  Further,
Tables~\ref{tab:gocbiasvals-cv} and \ref{tab:gocbiasvals-cor} show the relevant
coefficients of variation and pairwise correlations for all remaining
quantities using data from Curitiba, Brazil (currently, the only setting where
we have data from a sample of the hidden population).  For all values in these
tables, the estimated variance of the estimators is calculated using the
bootstrap methods presented in Section~\ref{sec:var_nsum}.

Since we have both a sample from the frame population and a sample from the
hidden population in Curitiba, we can compute numerical estimates of the bias
of each nonlinear estimator in the context of that study.  We can see that in
this study bias caused by the nonlinearity of the estimator was not a big
problem: in each case, the estimated approximate bias was less than one percent
of the estimate (Table~\ref{tab:ctbiasvals}).  

To conclude, we derived an expression for the approximate relative
bias in double ratio estimators in general.  We then simplified the
approximation for each specific nonlinear estimator that we propose. Finally,
we used data from a real scale-up study in Curitiba, Brazil to estimate
magnitude of the biases caused by the non-linearity of the estimators in a
specific scale-up study.  From these results, we conclude that theses
estimators are essentially unbiased, and that sampling error and non-sampling
error will dominate any bias introduced by the nonlinear form of the
estimators.

\begin{table}[ht]
\centering
\begin{tabular}{rl}
  \hline
$\widehat{\text{cv}}(\widehat{\bar{d}})$ & source \\ 
  \hline
0.05 & Rwanda \\ 
  0.10 & Curitiba \\ 
  0.02 & US \\ 
   \hline
\end{tabular}
\caption{Estimated coefficients of variation for the average degree from 3 different scale-up surveys. These play a role in the approximate relative bias for the estimate of $\widehat{\delta}_F$. Our approximation tells us that the larger these values are, the worse the relative bias will be.  The estimates were computed using the rescaled bootstrap procedure.} 
\label{tab:nsumdcvs}
\end{table}


\begin{table}[h]
\centering
\begin{tabular}{rr}
  \hline
 & estimated coef. of variation \\ 
  \hline
$\sum_{i \in s_H} y_{i, \mathcal{A} \cap F} / c\pi_i$ & 0.08 \\ 
  $\sum_{i \in s_H} \tilde{v}_{i, \mathcal{A} \cap F} / c\pi_i$ & 0.08 \\ 
  $\sum_{i \in s_H} 1 / c\pi_i$ & 0.06 \\ 
   \hline
\end{tabular}
\caption{Estimated coefficients of variation for quantities derived from a sample from the hidden population. These quantities play a role in the approximate relative bias for the estimate of all of the nonlinear estimators we propose. The estimates were computed using the respondent-driven sampling bootstrap procedure \citep{salganik_variance_2006}.} 
\label{tab:gocbiasvals-cv}
\end{table}

\begin{table}[h]
\centering
\begin{tabular}{rr}
  \hline
 & estimated correlation \\ 
  \hline
$\widehat{\text{cor}}(\sum_{i \in s_H} y_{i, \mathcal{A} \cap F} / c\pi_i, \sum_{i \in s_H} \tilde{v}_{i, \mathcal{A} \cap F} / c\pi_i)$ & 0.92 \\ 
  $\widehat{\text{cor}}(\sum_{i \in s_H} y_{i, \mathcal{A} \cap F} / c\pi_i, \sum_{i \in s_H} 1 / c\pi_i)$ & 0.71 \\ 
  $\widehat{\text{cor}}(\sum_{i \in s_H} \tilde{v}_{i, \mathcal{A} \cap F}/ c\pi_i, \sum_{i \in s_H} 1 / c\pi_i)$ & 0.68 \\ 
   \hline
\end{tabular}
\caption{Estimated pairwise correlations for quantities derived from a sample from the hidden population. These quantities play a role in the approximate relative bias for the estimate of all of the nonlinear estimators we propose.} 
\label{tab:gocbiasvals-cor}
\end{table}

\begin{table}[ht]
\centering
\begin{tabular}{rccc}
  \hline
 & approx. rel. bias, $B_d$ & estimate & estimated absolute bias \\ 
  \hline
$\widehat{\tau}_F$ & 0.0005 & 0.77 & 0.0004 \\ 
  $\widehat{\delta}_F$ & 0.0105 & 0.69 & 0.0073 \\ 
  $\widehat{N}_H$ & 0.0026 & 114498.00 & 298.0000 \\ 
   \hline
\end{tabular}
\caption{Approximate relative bias in the estimates of the nonlinear quantities using data taken from the Curitiba study, the point estimates produced by the Curitiba study, and the estimated implied absolute bias. For each quantity, the bias is very small.} 
\label{tab:ctbiasvals}
\end{table}

\section{Variance estimation and confidence intervals}
\label{ap:variance_estimation}

In addition to producing point estimates, researchers must also produce confidence intervals around their estimates.  The procedure currently used by scale-up researchers begins with the variance estimator proposed in~\citet{killworth_estimation_1998a}: 
\begin{equation}
    \label{eqn:killworth_var}
    \widehat{se}(\widehat{N}_H) = 
    \sqrt{\frac{N \cdot \widehat{N}_H}{\sum_{i \in s_F} \widehat{d}_{i,U}}},
\end{equation}
and then produces a confidence interval:
\begin{equation}
    \label{eqn:killworth_ci}
    \widehat{N}_H \pm z_{1 - \alpha/2} \widehat{se}(\widehat{N}_H),
\end{equation}
\noindent where $1-\alpha$ is the desired confidence level (typically 0.95),
and $z_{\alpha/2}$ is the $\alpha/2$ quantile of the standard Normal distribution.

Unfortunately, the variance estimator (Equation~\ref{eqn:killworth_var}) was derived from the basic scale-up model
(Equation~\ref{eqn:basic-scaleup-model}), and so it suffers from the
limitations of that model.  In particular, it has three main problems, none of which seem to have been appreciated in the scale-up literature and all of
which lead it to underestimate the variance in most situations.  
First, the variance estimator in
Equation~\ref{eqn:killworth_var} does not include any information about the
procedure used to sample respondents, which can lead to problems when complex
sampling designs, such as stratified, multi-stage designs, are used.  Second,
it implicitly assumes that the researchers have learned about $\sum_{i \in s_F}
d_{i,U}$ independent alters, which is not true if there are barrier effects
(i.e., non-random social mixing).  Finally, like virtually all variance
estimators, it only provides a measure of uncertainty introduced by sampling
but not other possible sources of error.  

To address the first two problems but not the third, we propose that
researchers used the rescaled bootstrap variance estimation
procedure~\citep{rao_resampling_1988, rao_recent_1992a, rust_variance_1996} with
the percentile method; a combination that, for convenience, we will refer to as
the rescaled bootstrap.  This procedure, described in more
detail below, has strong theoretical foundations; does not depend on the basic
scale-up model; can handle both simple and complex sample designs; and can be
used for both the basic scale-up estimator and the generalized scale-up
estimator.  

In addition to the theoretical reasons to prefer to rescaled bootstrap,
empirically, we find that the rescaled bootstrap produces intervals with
slightly better coverage properties in three real scale-up studies.  In
particular, using the internal consistency check procedure proposed
in~\citet{killworth_social_1998a} for all groups of known size in three real
scale-up datasets---one collected via simple random
sampling~\citep{mccarty_comparing_2001a} and two collected via complex sample
designs~\citep{salganik_assessing_2011,
rwandabiomedicalcenter_estimating_2012}---we produced a size estimate using the
basic scale-up estimator (Equation~\ref{eqn:basic-scaleup-estimator-intheory}),
and we produced confidence intervals using (i) the current procedure
(Equation~\ref{eqn:killworth_var}); (ii) the simple bootstrap (which does not
account for complex sample designs) with the percentile method; and (iii) the
rescaled bootstrap (which does account for complex sample designs) with the
percentile method.
      
\begin{figure}
  \centering
   \subfigure[United States (simple random sample)]{%
     \label{fig:united_states} 
     \includegraphics[height=.27\textheight]{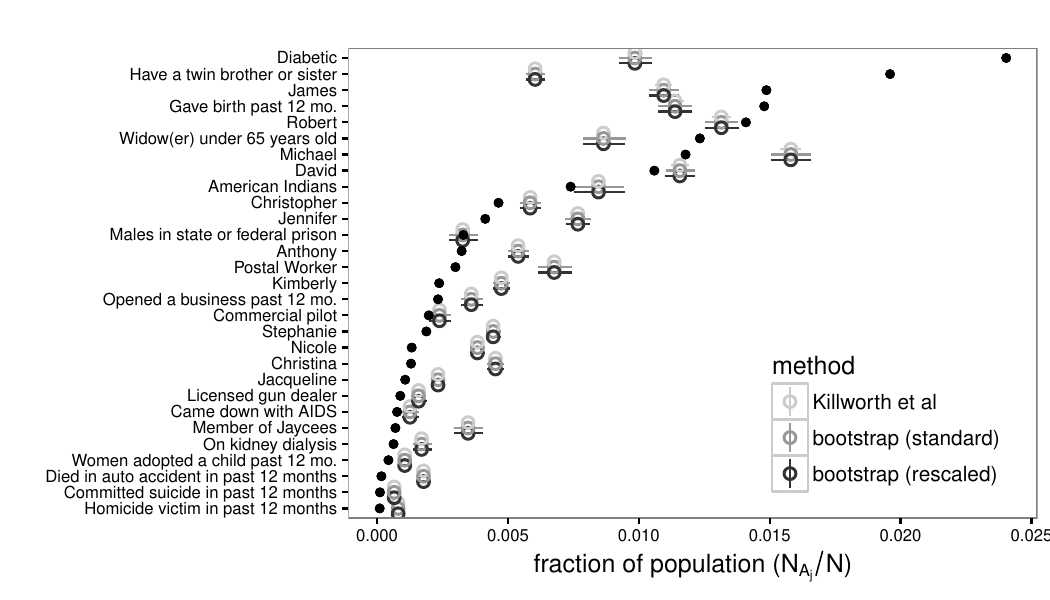}}
  \hspace{0.0in}
    \subfigure[Rwanda (stratified, multi-stage)]{%
     \label{fig:rwanda} 
     \includegraphics[height=.27\textheight]{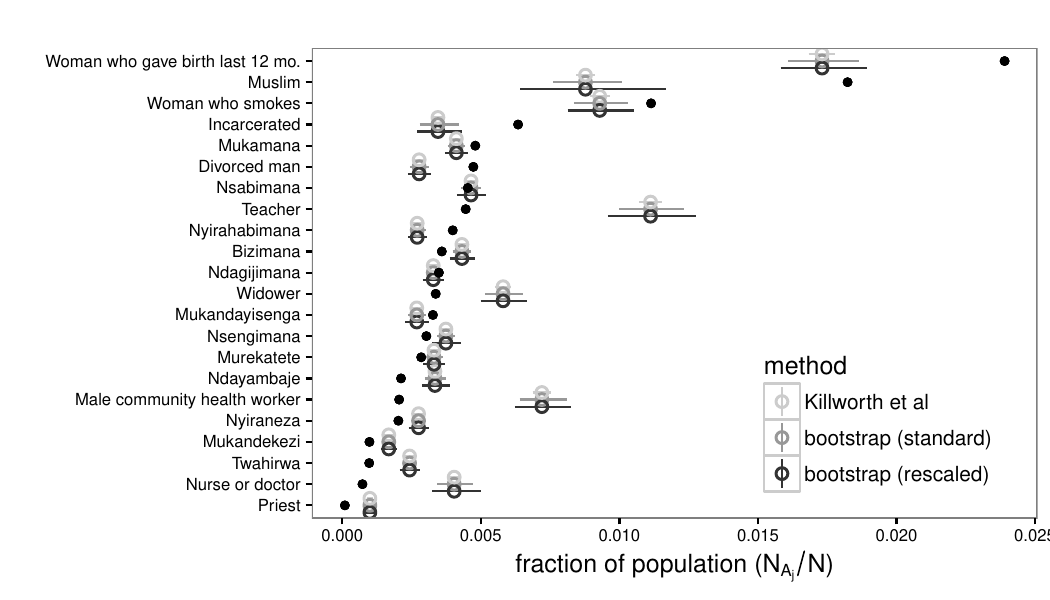}}
  \hspace{0.0in}
    \subfigure[Curitiba, Brazil (multi-stage)]{%
     \label{fig:curitiba} 
     \includegraphics[height=.27\textheight]{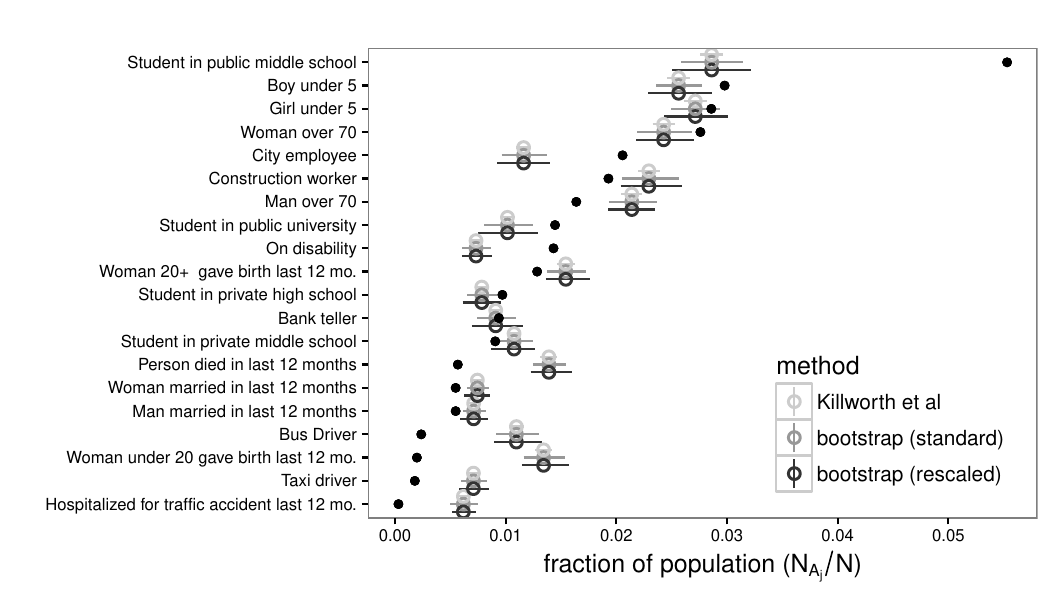}}     
     \vspace{-.1in}
     \caption{Assessing confidence interval procedures using scale-up studies
         in the United States~\citep{mccarty_comparing_2001a},
         Rwanda~\citep{rwandabiomedicalcenter_estimating_2012}, and Curitiba,
         Brazil~\citep{salganik_assessing_2011}.  The true size of each group
         is shown with a black dot.  Estimates made use the basic scale-up estimator are shown with circles.
         The rescaled bootstrap confidence intervals include
         the true group size for 3.4\%, 9.1\%, and 15.0\% of the groups in the US,
         Rwanda, and Curitiba, respectively. 
         The standard bootstrap confidence intervals include the true group size for
         3.4\%, 9.1\%, and 10.0\% of the groups.
         The currently used procedure
         (Equation~\ref{eqn:killworth_var}), contains the true group size for 3.4\%,
         9.1\%, and 5.0\% of the groups.}
     \label{fig:validation_ci} 
\end{figure}

This empirical evaluation (Figure~\ref{fig:validation_ci}) produced three main
results.  First, as expected, we found that the current confidence interval
procedure produces intervals with bad coverage properties: purported 95\%
confidence intervals had empirical coverage rates of about 5\%.  This poor
performance does not seem to have been widely appreciated in the scale-up
literature.  Second, also consistent with expectation, we found that the
rescaled bootstrap produced wider intervals than both the current procedure and
the simple bootstrap, especially in the case of complex sample designs.  Third,
and somewhat surprisingly, the rescaled bootstrap did not work well in an
absolute sense: purported 95\% confidence intervals had empirical coverage
rates of about 10\%, only slightly better than the current procedure.  

We speculate that there are two possible reasons for the surprisingly poor coverage rates of the rescaled bootstrap.   The first is bias in the basic scale-up estimator.  As described in detail in~\citet[Sec 5.2]{sarndal_model_1992}, bias in an estimator can degrade the coverage rates for confidence intervals.  For example, if Native Americans (one of the groups in the study of~\citet{mccarty_comparing_2001a}) have smaller personal networks than other Americans, then there will be a downward bias in the estimated number of Native Americans (Equation~\ref{eqn:addbias}).  This bias will necessarily degrade the coverage properties of any confidence interval procedure, especially if the bias ratio $\left( bias(\hat{N}_H) / se(\hat{N}_H) \right)$ is large (see~\citet[Sec 5.2]{sarndal_model_1992}).  The second possible reason for the surprisingly poor coverage rates could also be some unknown problem with the rescaled bootstrap or the percentile method.  Because (i) the rescaled bootstrap and percentile method have strong theoretical foundations~\citep{rao_resampling_1988, rao_recent_1992a, rust_variance_1996, efron_introduction_1993} and (ii) we expect that the basic scale-up estimates are biased in most situations (see Equation~\ref{eqn:addbias}), we believe that the main reason for the poor coverage is the bias.  However, we also believe that future research should explore the properties of the rescaled bootstrap and percentile method in greater detail.

An additional concern about these empirical results is that they only apply to the basic scale-up estimator and not the generalized scale-up estimator.  Unfortunately, we cannot assess the performance of the rescaled bootstrap procedure when used with the generalized scale-up estimator because the generalized scale-up estimator has not yet been used for populations of known size.  

These empirical results, and the theoretical arguments that follow, lead us to three conclusions.  First, confidence intervals from the rescaled bootstrap are preferable to intervals from the current procedure.  Second, researchers should expect that the confidence intervals from the rescaled bootstrap procedure will be anti-conservative (i.e., they will be too small).  Third, creating confidence intervals around scale-up estimates is an important area for further research.

Next in Section~\ref{sec:var_nsum} we review the standard bootstrap and
rescaled bootstrap; describe how we applied these methods to three real
scale-up datasets; and describe the results in Figure~\ref{fig:validation_ci} in
greater detail.  Finally, in Section~\ref{sec:var_gnsum} we describe how
researchers can use the rescaled bootstrap with the generalized scale-up
estimator.

\subsection{Variance estimation with a sample from $F$}
\label{sec:var_nsum}

\begin{figure}
\centering
\includegraphics[width=0.6\textwidth]{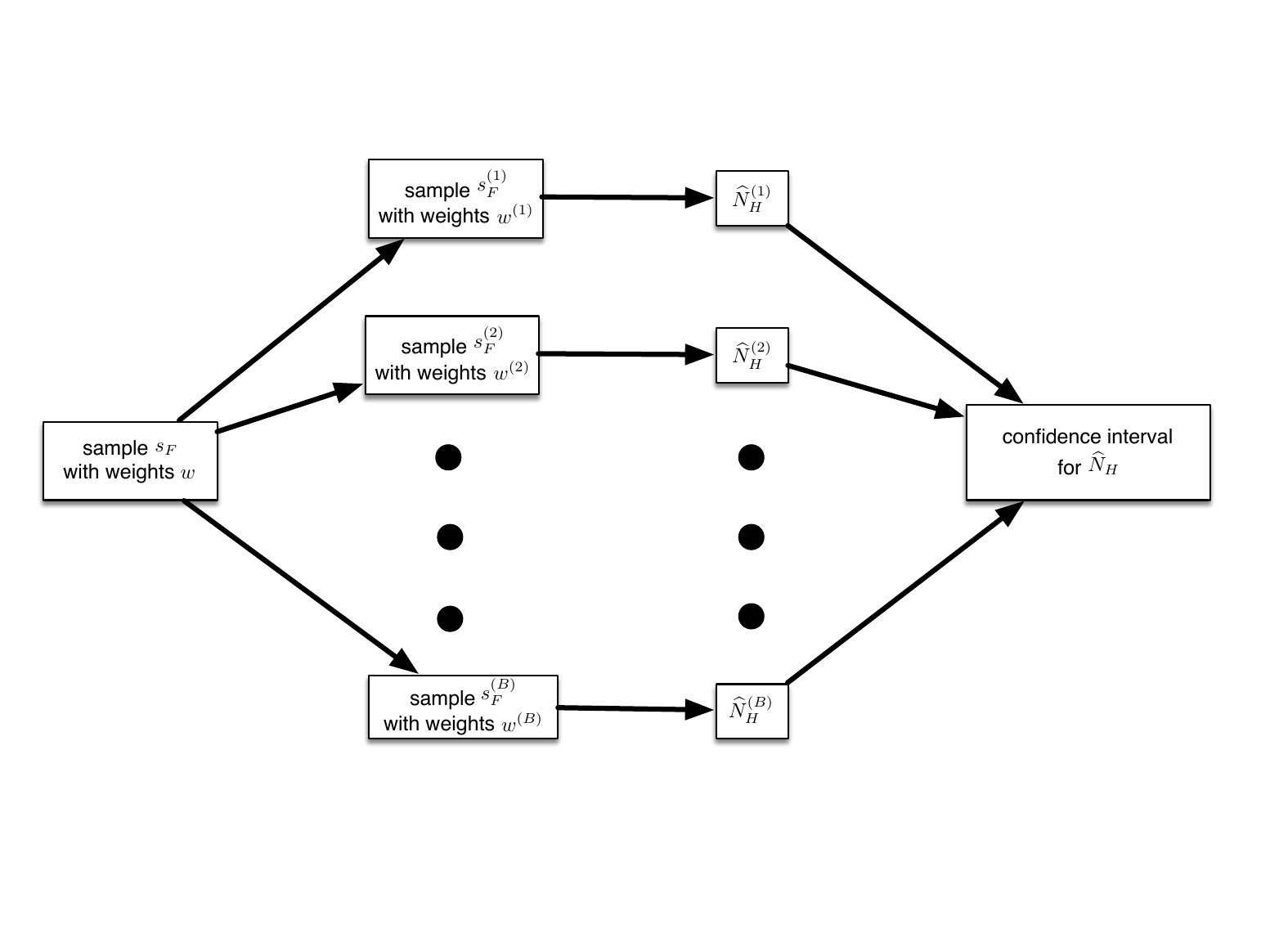}
\caption{Schematic of the bootstrap procedure to put a confidence interval
    around $\widehat{N}_H$ when there is a sample from the frame $s_F$.}
\label{fig:boot_schematic}
\end{figure}

The goal of a bootstrap variance estimation procedure is to put a confidence
interval around an estimate $\widehat{N}_H$ that is derived from a sample
$s_F$.  The most standard bootstrap procedure has three steps.  First,
researchers generate $B$ replicate samples, $s_F^{(1)}, s_F^{(2)}, \ldots,
s_F^{(B)}$ by randomly sampling with replacement from $s_F$.  Second, these
replicate samples are then used to produce a set of replicate estimates,
$\widehat{N}_H^{(1)}, \widehat{N}_H^{(2)}, \ldots, \widehat{N}_H^{(B)}$.
Finally, the replicate estimates are combined to produce a confidence interval;
for example, by the percentile method which chooses the 2.5th and 97.5th
percentiles of the $B$ estimates
(Fig.~\ref{fig:boot_schematic})~\citep{efron_introduction_1993}.  

When the original sample can be modeled as a simple random sample, this
standard bootstrap procedure is appropriate.  For example, consider the
scale-up study of~\citet{mccarty_comparing_2001a} that was based on telephone
survey of 1,261 Americans selected via random digit dialing.%
\footnote{The
    original data file includes 1,375 respondents.  From these cases,  113
    respondents who had missing data for some of the aggregated relational data
    questions and 1 respondent who answered 7 for all questions
    (see~\citet{zheng_how_2006}).  Further, consistent with common practice
    (e.g.,~\citet{zheng_how_2006}), we top coded all responses at 30, affecting
0.26\% of responses.} %
We can approximate the sampling design as simple random sampling, and draw $B =
10,000$ replicate samples of size 1,261.  In this case the bootstrap confidence
intervals are, as expected, larger than the confidence intervals from
Equation~\ref{eqn:killworth_var}, since they account for the clustering of responses
with respondent; on average, they are 2.05 times wider.

This standard bootstrap procedure, however, can perform poorly when the
original data are collected with a complex sample
design~\citep{shao_impact_2003}.  To deal with this problem, Rust and Rao
(1996)\nocite{rust_variance_1996} proposed the rescaled bootstrap procedure that
works well when the data are collected with a general multistage sampling
design, a class of designs that includes most designs that would be used for
face-to-face scale-up surveys.  For example, it includes stratified two-stage
cluster sampling with oversampling (as was used in a recent scale-up study in
Rwanda~\citep{rwandabiomedicalcenter_estimating_2012}) and three-stage
element sampling (as was used in a recent scale-up study in Curitiba,
Brazil~\citep{salganik_assessing_2011}); a full description of the designs
included in this class is presented in~\citet{rust_variance_1996}.

The rescaled bootstrap includes two
conceptual changes from the standard bootstrap.  First, it approximates the
actual sampling design by a closely related one that is much easier to work
with.  In particular, if we assume that primary sampling units (PSUs) are
selected with replacement and that all subsequent stages of sampling are
conducted independently each time a given PSU is selected, then we can use the
with-replacement sampling framework in which variance estimation is much
easier; see~\citet{sarndal_model_1992} Result 4.5.1 for a more formal version
of this claim.  It is important to note that this approximation is generally
conservative because with-replacement sampling usually results in higher
variance than without-replacement sampling.  Therefore, we will be estimating
the variance for a design that has higher variance than the actual design. In
practice, this difference is usually small because the sampling fraction in
each stratum is usually small~\citep{rao_recent_1992a, rust_variance_1996};
see~\citet{sarndal_model_1992} Section 4.6 for a more formal treatment.  To
estimate the variance in this idealized with-replacement design, resampling
should be done independently in each stratum and the units that are resampled
with replacement should be entire PSUs, not respondents.

This change---resampling PSUs, not respondents---introduces the need for a
second change in the resampling procedure.  It is known that the standard
bootstrap procedure is off by a factor of $(n -1)/ n$ where $n$ is the sample
size~\citep{rao_resampling_1988}.  Thus, when the sample size is very small,
the bootstrap will tend to underestimate the variance.  While this issue is
typically ignored, it can become important when we resample PSUs rather than
respondents.  In particular, the number of sampled PSUs in stratum $h$,
$n_{h}$, can be small in complex sample designs.  At the extreme, in a design
with two sampled PSUs per stratum, which is not uncommon, the standard
bootstrap would be expected to produce a 50\% underestimate of the variance.
Therefore,~\citet{rao_recent_1992a} developed the rescaled bootstrap, whereby
the bootstrap sample size is slightly smaller than the original sample size and
the sample weights are rescaled to account for this difference.
\citet{rust_variance_1996} recommend that if the original sample includes
$n_{h}$ PSUs in strata $h$, then researchers should resample $n_{h} - 1$ PSUs
and rescale the respondent weights by $n_{h} / (n_{h} - 1)$.  That is, the
weight for the $j^{th}$ person in PSU $i$ in the $b^{th}$ replicate sample is 
\begin{equation}
w_{ij}^{(b)} = w_{ij} \times \frac{n_{h}}{(n_{h} - 1)} \times r_{i}^{(b)} 
\label{eq:adjusted_weight_simplified_curitiba}
\end{equation}
where $w_{ij}$ is the original weight for the $j^{th}$ unit in the $i^{th}$
PSU, $n_{h}$ is the number of PSUs in strata $h$, and $r_{i}^{(b)}$ is the
number of times the $i^{th}$ PSU was selected in replicate sample $b$.  

In Figure~\ref{fig:reporting-network}, we compared the three different procedures for putting confidence intervals around the basic scale-up estimator: the current procedure~\citep{killworth_estimation_1998a}, the standard bootstrap with the percentile method, and the rescaled bootstrap with the percentile method.  We made this comparison using data from scale-up studies in the United States, Rwanda,%
\footnote{The scale-up study in Rwanda used stratified two-stage cluster
    sampling with unequal probability of selection across strata in order to
    oversample urban areas.  Briefly, the sample design divided Rwanda into
    five strata: Kigali City, North, East, South, and West.  At the first
    stage, PSUs---in this case villages---were selected with probability
    proportional to size and without replacement within each stratum with
    oversampling in the Kigali City stratum.  This approach resulted in a
    sample of 130 PSUs: 35 from Kigali City, 24 from East, 19 from North, 26
    from South, and 26 from West. At the second stage, 20 households were
    selected via simple random sampling without replacement from each PSU in
    Kigali City and 15 households from each PSU in other strata.  Finally, all
    members of the sampled household over the age of 15 were interviewed.  
    The study included a survey experiment which randomized respondents to report
    about one of two different personal networks; to keep things simple, we use 
    responses about only one personal network here.
    For full details see~\citet{rwandabiomedicalcenter_estimating_2012}.  The
    original data file includes 2,406 respondents.  From these cases, we
    removed 2 respondents who had missing data for some of the aggregated
    relational data questions.  Further, consistent with common practice
    (e.g.,~\citet{zheng_how_2006}), we top coded all responses at 30, affecting
0.12\% of responses.%
} and Curitiba, Brazil.%
\footnote{The scale-up study in
    Curitiba, Brazil used two-stage element sampling where 54 primary sampling
    units (PSUs)---in this case census tracks---were selected with probability
    proportional to their estimated number of housing units and without
    replacement.  Then, within each cluster, eight secondary sampling units
    (SSUs)---in this case people---were selected with equal probability without
    replacement.  For full details see~\citet{salganik_assessing_2011}.  The
    original data file includes 500 respondents.  From these cases,  we removed
    no respondents who had missing data for some of the aggregated relational
    data questions.  Further, consistent with common practice
    (e.g.,~\citet{zheng_how_2006}), we top coded all responses at 30, affecting
    0.58\% of responses.%
}  
As expected, the rescaled bootstrap produced confidence intervals that are larger than those from
the standard bootstrap, which in turn are larger than those from the current scale-up variance estimation procedure.  In the study from
Curitiba, the rescaled bootstrap procedure produced confidence intervals 1.17
times larger than the standard bootstrap and 2.84 times larger than the current
procedure.  In the Rwanda case, the rescaled bootstrap procedure
produced confidence intervals 1.35 times larger than the standard bootstrap and
2.65 times larger than the current procedure.

Finally, Figure~\ref{fig:validation_ci} shows the estimated confidence intervals for the
groups of known size in the three studies described above.  
The coverage rates for the bootstrap confidence intervals for the US, Rwanda,
and Curitiba, are 3.4\%, 9.1\%, 15.0\%.  
While this is far from ideal, we note that it is slightly better than the
currently used procedure (Equation~\ref{eqn:killworth_ci}), which produced coverage
rates of 3.4\%, 9.1\%, 5.0\%, and it is also slightly better than the standard
bootstrap, which produced coverage rates of 3.4\%, 9.1\%, and 10.0\%.  

\subsection{Variance estimation with sample from $F$ and $H$}
\label{sec:var_gnsum}

\begin{figure}
\centering
\includegraphics[width=0.6\textwidth]{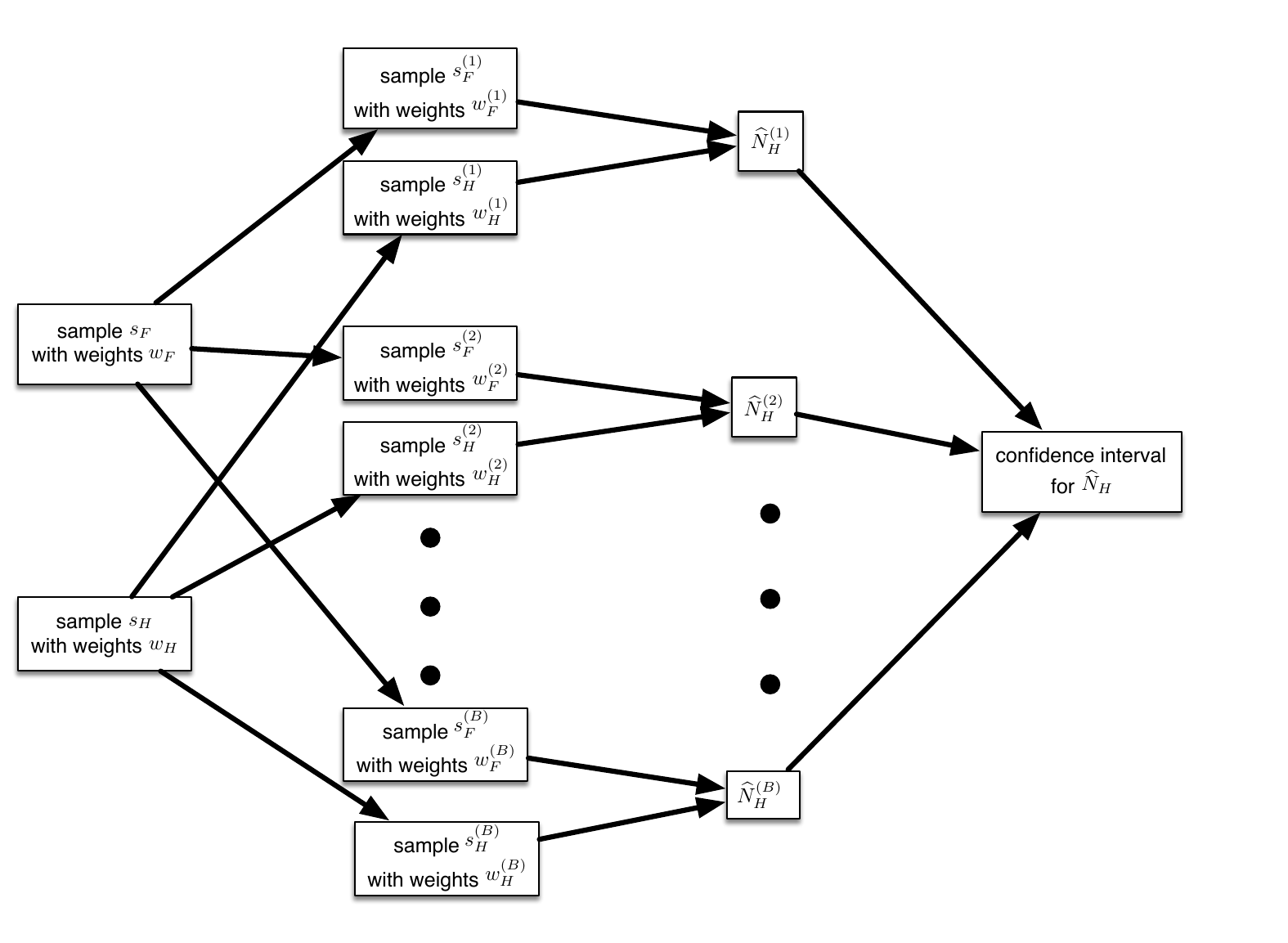}
\caption{Schematic of the bootstrap procedure to put a confidence interval
    around $\widehat{N}_H$ when there is a sample from the frame $s_F$ and a
sample from the hidden population $s_H$.}
\label{fig:bootstrap_two_sample}
\end{figure}

Producing confidence intervals around the generalized scale-up estimator is
more difficult than the basic scale-up estimator because the generalized
estimator has uncertainty from two different samples: the sample from the
hidden population and the sample from the frame population. 
To capture all of this
uncertainty, we propose combining replicate samples from the frame population
with independent replicate samples from the hidden population in order to
produce a set of replicate estimates.  More formally, given $s_F$, a sample
from the frame population, and an independent sample $s_H$ from the hidden
population, we seek to produce a set of $B$ bootstrap replicate samples for
$s_F$ and $s_H$, $s_F^{(1)}, s_F^{(2)}, \ldots, s_F^{(B)}$ and $s_H^{(1)},
s_H^{(2)}, \ldots, s_H^{(B)}$, which are then combined to produce a set of $B$
bootstrap estimates: $\widehat{N}_H^{(1)} = f(s_F^{(1)}, s_H^{(1)})$,
$\widehat{N}_H^{(2)} = f(s_F^{(2)}, s_H^{(2)}), \ldots \widehat{N}_H^{(B)} =
f(s_F^{(B)}, s_H^{(B)})$.  Finally, these $B$ replicate estimates are converted
into a confidence interval using the percentile method
(Fig.~\ref{fig:bootstrap_two_sample}).

Because of the challenges involved in sampling hard-to-reach populations, the
two most likely sampling designs for $s_H$ will be time-location sampling and
respondent-driven sampling.  If $s_H$ was selected with time-location sampling,
we recommend treating the design as a two-stage element sample
(see~\citet{karon_statistical_2012}) and using the procedure of Rust and Rao
(1996). If $s_H$ was selected with respondent-driven sampling, as was done in a
recent study of heavy drug users in Curitiba,
Brazil~\citep{salganik_game_2011a}, we recommend using the best available
bootstrap method for respondent-driven sampling data, which at the present time
is the procedure introduced in~\citet{salganik_variance_2006}.  One
implementation detail of that particular bootstrap procedure is that it requires
researchers to divide the sample of the hidden population into two mutually
exclusive groups.  In this case, we recommend dividing the hidden population
into those who are above and below the median of their estimated visibility
$\widehat{v}_{i,F}$ in order to capture some of the extra uncertainty
introduced if there are strong tendencies for more hidden members of the hidden
population to recruit each other.

Because the generalized scale-up estimator has never been used for groups of known
size, we cannot explore the coverage rate of the proposed procedure. However,
based on experience with respondent-driven sampling, we suspect that variance
estimation procedures for hidden populations will underestimate the actual
uncertainty in the estimates~\citep{goel_respondentdriven_2009,
goel_assessing_2010, yamanis_empirical_2013, verdery_network_2013, rohe_network_2015}. If this is
the case, then the intervals around the generalized scale-up estimates will be
anti-conservative. 

In conclusion, Sec.~\ref{sec:var_nsum} presents a bootstrap procedure for
simple and complex sample designs from the sampling frame, and
Sec.~\ref{sec:var_gnsum} extends these results to account for the sampling
variability introduced by having a sample from the hidden population.  We have
shown that the performance of these procedures on three real scale-up datasets
is consistent with theoretical expectations.  Additional research in this area,
which is beyond the scope of this paper, could adopt a total survey error
approach and attempt to quantify all sources of uncertainty in the estimates,
not just sampling uncertainty.  Additional research could also explore the
properties and sensitivity of these confidence interval procedures though simulation.

\section{Simulation study}
\label{sec:sim_overview}

In this appendix, we describe a simulation study comparing the performance of
the generalized and basic network scale-up estimators. The results of these
simulations confirm and illustrate several of the analytical results in
Section~\ref{sec:relationshiptoscaleup} of the paper. Most importantly, the
simulations show that the generalized network scale-up estimator is unbiased
for all of the situations explored by the simulation, while the basic network
scale-up estimator is biased for all but a few special cases. Moreover, our
analytical results correctly predict the bias of the basic network scale-up
estimator in each case.

Our simulation study is intentionally simple in order to clearly illustrate
our analytical results; it is not designed to be a realistic model of any
scale-up study. Concretely, our simulations compare the performance of generalized and
basic scale-up estimators as three important quantities vary: (1) the size of the
frame population $F$, relative to the size of the entire population, $U$; (2)
the extent to which people's network connections are not formed completely at
random, also called the amount of inhomogenous mixing; and (3) the accuracy of
reporting, as captured by the true positive rate $\tau_F$ (see
Equation~\ref{eqn:tau-defn}). 

We simulate populations consisting of $N=5,000$ people, using a stochastic
block-model \citep{white_social_1976, wasserman_social_1994} to randomly
generate networks with different amounts of inhomogenous mixing. Stochastic
block models assume population members can be grouped into different
\emph{blocks}.  For any pair of people, $i$ and $j$, the probability that there
is an edge between $i$ and $j$ is completely determined by the block
memberships of $i$ and $j$.

In our simulation model, each person can be either in or out of the frame
population $F$ and each person can also be either in or out of the hidden
population $H$, producing four possible blocks: $FH$, $F\lnot H$, $\lnot F\lnot
H$, and $\lnot FH$. (Here, we use the logical negation symbol, $\lnot$, to
denote not being in a group.) The probability of an edge between any two people
$i$ and $j$ is then governed by a Bernoulli distribution whose mean is a
function of the two block memberships:
\begin{equation}
    \text{Pr}(i \leftrightarrow j) \sim \Ber(\mu_{g(i), g(j)}),
\end{equation}
\noindent where $g(i)$ is the block containing $i$, $g(j)$ is the block
containing $j$, $i \leftrightarrow j$ denotes an undirected edge between $i$
and $j$, and $\mu_{g(i),g(j)}$ is the probability of an edge between a member
of group $g(i)$ and a member of group $g(j)$.  In a network with a no
inhomogenous mixing (equivalent to an Erdos-Renyi random graph),
$\mu_{g(i),g(j)}$ will be the same for all $i$ and $j$. On the other hand, in a
network with a high level of inhomogenous mixing, $\mu_{g(i),g(j)}$ will be
relatively small when $g(i)\neq g(j)$ and $\mu_{g(i),g(j)}$ will be relatively
large when $g(i)=g(j)$\footnote{%
    Computer code to perform the simulations was written in
    R~\citep{rcoreteam_r_2014} and used the following packages:
    devtools~\citep{wickham_devtools_2013};
    functional~\citep{danenberg_functional_2013}; 
    ggplot2~\citep{wickham_ggplot2_2009};
    igraph~\citep{csardi_igraph_2006};
    networkreporting~\citep{feehan_networkreporting_2014};
    plyr~\citep{wickham_splitapplycombine_2011};
    sampling~\citep{tille_sampling_2015};
    and stringr~\citep{wickham_stringr_2012}.}. 

Each random network drawn under our simulation model depends on seven parameters.
The first four parameters describe population size and group memberships; they
are:
\begin{itemize}
\item $N$, the size of the population
\item $p_F$, the fraction of people in the frame population
\item $p_H$, the fraction of people in the hidden population
\item $p_{F|H}$, the fraction of hidden population members also in the frame population
\end{itemize}

The next three parameters govern the amount of inhomogenous mixing in the network
that connects people to each other; they are:
\begin{itemize}
\item ${\piw}$, the probability of an edge between two people who are
both in the same block.
\item $\xi$, the relative probability of an edge between two vertices that
differ in frame population membership.  
For example, a value of 0.6 would mean that the chances
of having a connection between a particular person in $F$ and a particular
person not in $F$ is 60\% of the chance of a connection between two members
of $F$ or two members of $\lnot F$.
\item $\rho$, the relative probability of an edge between two vertices that
differ in hidden population membership.  
For example, a value of 0.8 would mean that the chances
of having a connection between a particular person in $H$ and a particular
person not in $H$ is 80\% of the chance of a connection between two members
of $H$ or two members of $\lnot H$.
\end{itemize}
Together, the parameters $\piw$, $\xi$, and $\rho$ are used to construct the
mixing matrix M (Figure~\ref{fig:mixmatrix-fig}). 
Note that varying the parameter $\rho$ will change several structural features
of the network in addition to the amount of inhomogenous mixing; for example,
changing $\rho$ will alter the degree distribution.  Our analytical results
show that the generalized network scale-up estimator is robust to changes in
these structural features.

\begin{figure}
    \label{fig:mixmatrix-fig}
\begin{equation}
    \mathbf{M} =
\bordermatrix{
                    & ~F~H                   & ~F \lnot H              & \lnot F ~H              & \lnot F \lnot H \cr
    ~F~H            & \zeta                  & \rho\cdot\zeta          & \xi\cdot\zeta           & \xi\cdot\rho\cdot\zeta \cr
    ~F \lnot H      & \rho\cdot\zeta         & \zeta                   & \xi\cdot\rho\cdot\zeta  & \xi\cdot\zeta \cr
    \lnot F ~H      & \xi\cdot\zeta          & \xi\cdot\rho\cdot\zeta  & \zeta                   & \rho\cdot\zeta \cr
    \lnot F \lnot H & \xi\cdot\rho\cdot\zeta & \xi\cdot\zeta           & \rho\cdot\zeta          & \zeta
}
\end{equation}
\caption{%
The mixing matrix used to generate a random network using the stochastic block
model. Entry $(i,j$) in the matrix describes the probability of an edge between
two people, one of whom is in group $i$ and one in group $j$. The probabilities
are governed by $\piw$, $\xi$, and $\rho$. In our simulations, we generate
networks with different amounts of inhomogenous mixing between hidden
population members and non-hidden population members by fixing $\piw=0.05$ and
$\xi=0.4$, and then varying $\rho$ from 0.1 (extreme inhomogenous mixing
between hidden and non-hidden population members) to 1 (perfectly random mixing
between hidden and non-hidden population members).
}
\end{figure}

The final parameter, $\tau_F$, is used to control the amount of imperfect reporting. After
randomly drawing a network using the stochastic block model, we generate a reporting
network as follows:
\begin{enumerate}
\item convert all undirected edges $i \leftrightarrow j$ in the social network
    into two directed reporting edges in the reporting network: one $i
    \rightarrow j$ and one $j \rightarrow i$
\item select a fraction, $1-\tau_F$, of the edges that lead from members of the
    frame population to members of the hidden population uniformly at random
    and remove them from the reporting graph.
\end{enumerate}
Given a simple random sample of 500 members of the frame population and a
relative probability sample of 30 members of the hidden population, the
reporting graph is then used to compute the basic and generalized scale-up
estimates for the size of the hidden population.

Across our simulations, we fix five of the parameters at constant values
($N=5,000$; $p_F=0.03$; $p_{F|H} = 1$; $\piw = 0.05$; $\xi=0.4$). We systematically
explore varying the remaining parameters: we investigate $\rho$ for values from 0.1 to 1
in increments of $0.1$; we investigate $p_F$ for values 0.1, 0.5, and 1; and we investigate
$\tau_F$ for 0.1, 0.5, and 1. 
For each combination of the parameter values, we generate 10 random networks.
Within each random network, we simulate 500 surveys.
Each survey consists of two samples: a probability sample from the frame population,
with sample size of 500; and a relative probability from the hidden population of size
30, with inclusion proportional to each hidden population member's personal network size.
For each unique combination of parameters, we averaged the results across the surveys and
across the randomly generated networks.

\singlespacing

\end{document}